\documentclass[12pt]{article}
\usepackage[english]{babel}
\usepackage{bm,epsfig,amssymb,amsfonts,amsmath,cite}

\title{\bf Chiral symmetry and the properties \\of hadrons in the Generalised Nambu--Jona-Lasinio model}
\author{Yu. S. Kalashnikova$^{\rm a,b}$, A. V. Nefediev\thanks{e-mail: nefediev@itep.ru}\hspace*{2mm}$^{\rm a,b,c}$, J. E. F. T. Ribeiro$^{\rm d}$\\[5mm]
${\rm ^a}$ {\small\it Institute for Theoretical and Experimental Physics,}\\
{\small\it 117218, B.Cheremushkinskaya 25, Moscow, Russia}\\
${\rm ^b}$ {\small\it National Research Nuclear University MEPhI, }\\
{\small\it 115409, Kashirskoe highway 31, Moscow, Russia}\\
${\rm ^c}$ {\small\it Moscow Institute of Physics and Technology,}\\
{\small\it 141700, Institutsky lane 9, Dolgoprudny, Moscow Region, Russia}\\
${\rm ^d}$ {\small\it Instituto Superior T{\' e}cnico, Universidade de Lisboa,}\\
{\small\it 1049-001, Av.Rovisco Pais, 1, Lisboa, Portugal}
}
\date{}

\DeclareMathSymbol\hhbar{\mathord}{AMSb}{"7E}

\newcommand{\ds}{\displaystyle}

\newcommand{\toon}{\mathop{\to}\limits_{n\to\infty}}
\newcommand{\vx}{{\bm x}}
\newcommand{\vy}{{\bm y}}
\newcommand{\vz}{{\bm z}}
\newcommand{\vk}{{\bm k}}
\newcommand{\vp}{{\bm p}}
\newcommand{\vP}{{\bm P}}

\newcommand{\vq}{{\bm q}}
\newcommand{\vg}{{\bm \gamma}}
\newcommand{\ld}{\lambda}

\newcommand{\ver}{{\bm r}}

\newcommand{\vpp}{\varphi}

\newcommand{\Tr}{\textrm{Tr}}

\DeclareMathSymbol{\varGamma}{\mathord}{letters}{"00}
\newcommand{\vex}{{\bm x}}
\newcommand{\vesig}{{\bm \sigma}}
\newcommand{\veS}{{\bm S}}
\newcommand{\veL}{{\bm L}}

\newcommand{\vegam}{{\bm \gamma}}

\newcommand{\lan}{\langle}
\newcommand{\ran}{\rangle}

\newcommand{\bp}{\begin{picture}}
\newcommand{\ep}{\end{picture}}

\graphicspath{{Figs.dir/}}

\begin{document}

\maketitle

\begin{abstract}
Various aspects of the Generalised Nambu--Jona-Lasinio model for QCD in four dimensions are reviewed. The properties of mesonic excitations are 
discussed in detail, with special attention paid to the chiral pion. The spontaneous chiral symmetry breaking in the vacuum and the effective chiral 
symmetry restoration in the spectrum of highly excited mesons and baryons are described microscopically.
\end{abstract}

\tableofcontents

\section{Introduction}

Quark models for strong interactions have a long history, starting
from the mid of the previous century when the idea of hadrons, to
be composed of quarks, was commonly accepted. And, as it happens,
not only the number of various quark models, but even the number
of their types turns out to be quite large. For example, the so-called
Coulomb+linear potential model of \cite{Eichten1978} describes
heavy quarkonium spectra with rather good accuracy, which is
clearly due to the heavy quark mass being much larger than the
scale of strong interaction $\Lambda_{\rm QCD}$. A na{\" i}ve kinematical
relativisation \cite{Godfrey1985} of quark model allows one to consider
mesons made of light quarks, though the justification of the potential
model approach is less obvious in this case. The
given approach, as well as similar models, is simple, also for
numerical calculations, however its range of applicability is very
limited, and many phenomena inherent to Quantum Chromodynamics
(QCD), which are of interest for the phenomenology of strong
interactions, cannot be addressed in such a framework. Among those
one should mention the effect of spontaneous breaking of chiral
symmetry in QCD vacuum, its implications in the spectrum of
hadrons, and the effective restoration of chiral symmetry in
excited hadrons.

It is well-known that, in the chiral limit, the $SU(2)_L\times SU(2)_R$ symmetry
of QCD Lagrangian is broken, and this affects the observed
spectrum of hadrons. Thus, the spontaneous symmetry breaking
$SU(2)_L \times SU(2)_R \rightarrow SU(2)$ \cite{Nambu:1961fr} manifests itself through
the absence of low-lying hadrons populating multiplets of the $SU(2)_L \times SU(2)_R$ group, through the Goldstone nature of the
pion, in particular, through its vanishing (beyond the strict chiral limit --- finite but quite small compared to the typical hadronic scale) mass,
through the nonzero value of the chiral condensate in the vacuum, and so on.
Thus, chiral symmetry is realised nonlinearly in the low-lying hadrons.

Meanwhile, there are good reasons to believe
that the aforementioned symmetry is effectively restored both in the spectrum of excited baryons
\cite{Glozman:1999tk,Cohen:2001gb,Cohen:2002st} and excited mesons
\cite{Glozman:2002cp,Glozman:2003bt,Glozman:2002kq}. A nice and convincing
justification of such a restoration in the spectrum of excited hadrons was
suggested in a recent paper \cite{Denissenya:2014poa}, where
the masses of the light hadrons were extracted from the lattice configuration
after the near-zero modes of the Dirac operator, responsible for
the spontaneous chiral symmetry breaking \cite{Banks:1979yr}, had been artificially
removed from the latter. The resulting mass spectrum demonstrated a
remarkably high degeneracy pattern, including formation of the chiral multiplets
\cite{Glozman:2014mka}.

The full solution of QCD would yield a microscopic description of the effect of the spontaneous breaking of chiral symmetry. 
In the absence of a such solution, various approaches were suggested
aimed at identification of gluonic field configurations which could be responsible for chiral symmetry breaking. It is quite natural to relate chiral symmetry breaking to confinement, yet another
prominent feature of QCD. For example, in the approach of \cite{Simonov:2015cza} confining kernel derived in the Vacuum Correlator Method \cite{DiGiacomo:2000irz} gives rise to
the interaction of light quarks with Nambu-Goldstone fields, arriving in such a way at an effective chiral Lagrangian. The subject of the present
review is a
phenomenological approach which employs a simple ansatz for the confining kernel pertinent to the matter in hand. 
The approach gains experience from the 't~Hooft model \cite{'tHooft:1974hx} for
two-dimensional QCD in the limit of the large number of colours ($N_C\to\infty$).

First, we notice that a microscopic description of the effect of the
spontaneous breaking of chiral symmetry requires an intrinsically field
theoretical approach which takes into account, within the very same formalism,
both particles and antiparticles on equal footing. And this necessity lies outside the scope
of constituent quark models
as they merely provide an essentially quantum mechanical approach, even if one considers
relativistic kinematics. Formally, the problems stems from the fact that, working in the formalism
of relativistic quantum-mechanical Hamiltonians, one is stuck
to a particular (positive) sign of the energy while the contributions from the
other (negative) sign of the energy are neglected. Such ``negative''
solutions correspond to antiparticles, so that the interplay of both positive and
negative solutions leads to the Z-like (Zitterbewegung) trajectory
of the particle, that is, to the so-called $Z$-graphs.
The problem can be traced down to the spectrum: the Salpeter equation that emerges for the
bound states is defined with the help of a single-component Hamiltonian
which describes the particle and, therefore, the resulting bound-state equation
is derived curtailed of the full Hamiltonian components related to antiparticles. Such an approximation is well-justified
for heavy particles, however it is obviously misleading for the light quarks and henceforth for
the light hadrons built thereof --- for the chiral pion in the first place.

The proper mechanism to account for the Zitterbewegung motion of particles can
be established in terms of a matrix Hamiltonian and a
two-component wave function.
In \cite{Bars:1977ud} such an approach to the two-dimensional
't~Hooft model was suggested and described in detail. The key
approximations which allowed one to control the pair creation
process is the limit of the large number of colours,
$N_C\to\infty$ (an introduction to this limit in QCD and related
issues can be found in \cite{Donoghue:1992dd}). Also, it has
to be noticed that the limit of the large number of degrees of
freedom allows one to override \cite{Witten:1978qu} the Coleman's
no-go theorem which forbids spontaneous breaking of chiral
symmetry in two dimensions \cite{Coleman:1973ci}. Additional
simplifications in the model arise from the instantaneous type of
the interaction mediated by the two-dimensional gluon. To
establish the latter property, it is sufficient to count the
number of the degrees of freedom for the two-dimensional gluon and
then to arrive straightforwardly at the absence of the gluon
transversal propagating degrees of freedom. Then the 't~Hooft
model in the axial gauge considered in \cite{Bars:1977ud}
describes the interaction of two quark currents taken at equal
time and mediated by the confining potential which depends on the
one-dimensional interquark separation. The terms containing higher
powers of the quark currents do not appear in this Hamiltonian,
which is a reflection of the fact that all correlators of several
gluonic fields either vanish or reduce to the powers of the
bilocal correlator, which is nothing but the gluon propagator. As
a result, there exists only one irreducible field correlator
$\langle\langle A_1A_2\rangle\rangle=\langle A_1A_2\rangle-\langle
A_1\rangle\langle A_2\rangle=\langle A_1A_2\rangle$ with all such
irreducible correlators of higher orders vanishing. This result is
exact in two dimensions and it does not rely on any approximations
or assumptions. A review of the 't~Hooft model in the axial gauge
can be found in \cite{Kalashnikova:2001df}.

In contrast to the two-dimensional case, the instantaneous nature of the interquark
interaction and the absence in the Hamiltonian of the terms with
the product of more than two quark currents are approximations
that allow one to build a realistic quark model which we review in what follows.
Thus, a quark model with quark currents endowed with an instantaneous interaction
was suggested as a model for QCD about 30 years ago in
\cite{Amer:1983qa,LeYaouanc:1983it,LeYaouanc:1983huv,LeYaouanc:1984ntu}
and it was studied in detail in the Hamiltonian formalism in
\cite{Bicudo:1989sh,Bicudo:1989si,Bicudo:1989sj,Bicudo:1993yh,Bicudo:1998mc,Bicudo:2002eu}, as well as in the later works
\cite{Nefediev:2002nw,Bicudo:2003cy,Nefediev:2004by,Nefediev:2009zzb,Antonov:2010qt}.
As was mentioned above, this model can be regarded as the four-dimensional generalisation of the 't~Hooft model in two dimensions.
At the same time, the same model can also be viewed as the generalisation of the four-dimensional Nambu-Jona-Lasinio (NJL) model \cite{Nambu:1961fr}
to a nonlocal interaction of the quark currents. It is important to notice that, in spite
of its long history, the NJL model \cite{Nambu:1961fr}
still remains a useful and convenient tool for various studies in the physics of strong interactions.
An important role for this to come true was played by a detailed study of the connection of the given model with QCD
(see, for example, \cite{Bijnens:1992uz,Arbuzov:2006ia}) and by its further
developments (see the review papers \cite{Volkov:2005kw,Kalinovsky:2016dik}) which
allow one to extend considerably the spectrum of the problems where this model can be successfully employed. An important feature of
the Nambu-Jona-Lasinio-type model, hereafter referred to as the Generalised
Nambu-Jona-Lasinio (GNJL) model, is the presence of the confining interaction which
allows one to employ this model to address the problem of bound states
and which also brings in an intrinsic scale into the model.

The model is defined through the Hamiltonian (for simplicity, only one quark flavour
is considered, generalisation to the multi-flavour case is trivial)
\begin{equation}
\hat{H}=\int d^3 x\bar{\psi}({\bm x},t)\left(-i{\bm \gamma}\cdot
{\bm \bigtriangledown}+m\right)\psi({\bm x},t)+ \frac12\int d^3
xd^3y\;J^a_\mu({\bm x},t)K^{ab}_{\mu\nu}({\bm x}-{\bm y})J^b_\nu({\bm y},t),
\label{HGNJL}
\end{equation}
where, as it was explained above, one has an interaction of the quark currents
$J_{\mu}^a({\bm x},t)=\bar{\psi}({\bm x},t)\gamma_\mu\frac{\lambda^a}{2}\psi({\bm x},t)$
parametrised with the help of the instantaneous kernel
\begin{equation}
K^{ab}_{\mu\nu}({\bm x}-{\bm y})=g_{\mu 0}g_{\nu 0}\delta_{ab}V_0(|{\bm x}-{\bm y}|).
\label{KK}
\end{equation}

Hereinafter the following notations are used:
\begin{itemize}
\item low-case letters from the beginning of the greek (latin) alphabet, that is, $\alpha$, $\beta$ and so on ($a$, $b$ and so on) are used
for the colour indices in the fundamental (adjoint) representation which run over $1,2\ldots N_C$ ($1,2\ldots N_C^2-1$);
\item low-case letters from the middle of the greek alphabet ($\mu$, $\nu$ and so on) are used for the Lorentz indices which take values from 0 to 3;
\item $\psi({\bm x},t)$ is the fermion (quark) field; $\bar{\psi}=\psi^\dag\gamma_0$;
\item $m$ is the mass of the quark (the chiral limit implies that $m=0$);
\item $\gamma^\mu=(\gamma_0,\vegam)$ are the Dirac matrices;
\item $\lambda$ are the colour matrices (generators of the $SU(N_C)$ group);
\item $g_{\mu\nu}$ is the Minkowski metric tensor;
\item $\delta_{ab}$ is the Kronecker symbol.
\end{itemize}

Typically, the confining potential is chosen in a power-like form,
\begin{equation}
V_0(|{\bm x}|)=K_0^{\alpha+1}|{\bm x}|^{\alpha},\quad 0\leqslant\alpha\leqslant 2,
\label{potential}
\end{equation}
where $K_0$ is the parameter of the model which has the dimension of mass.
The qualitative predictions of the model are independent of the particular
form of the potential, provided it should only
be confining for coloured objects, on the one hand, and should demonstrate a
moderate growth with the interquark separation to avoid divergent
integrals, on the other.

The boundary cases with $\alpha=0$ and $\alpha=2$ require a special treatment.
In particular, in the limit $\alpha\to 0$, the potential has to be
re-defined as
\begin{equation}
\left.V_0(|{\bm x}|)\to {\tilde V}_0(|{\bm x}|)=K_0
\frac{(K_0|{\bm x}|)^\alpha-1}{\alpha}
\right|_{\alpha\to 0}=K_0\ln(K_0|{\bm x}|),
\label{ln}
\end{equation}
so that the resulting interaction is logarithmic. Strictly speaking,
the potential can also be defined for negative values of $\alpha$ up to
$\alpha>-1$ (for $\alpha=-1$, that is, for the Coulomb potential
the integrals become divergent again (see \cite{Bicudo:2003cy} for the
details)). However, the negative powers $\alpha$ do not provide confinement
for the quarks, so they will be disregarded in what follows.

In the limit of $\alpha=2$ the Fourier transform of the potential reduces to the
Laplacian of the three-dimensional $\delta$-function,
so that, by taking integrals by parts one can turn all integral
equations into second-order differential equations which are much simpler
to deal with from the
technical point of view. This explains why such a choice is quite popular in the literature (see, for example,
\cite{Amer:1983qa,LeYaouanc:1983it,LeYaouanc:1983huv,LeYaouanc:1984ntu,Bicudo:1989sh,Bicudo:1989si,Bicudo:1989sj,Bicudo:1993yh,Bicudo:1998mc}).
Even larger values of $\alpha$, $\alpha>2$, lead to divergent integrals and are not considered
(a detailed discussion of the problem can be found in
\cite{Amer:1983qa,LeYaouanc:1983it,LeYaouanc:1983huv,Bicudo:2003cy}).
More realistic quantitative predictions can be made with the help of the
linear confinement
\cite{Adler:1984ri,Kalinovsky:1989wx,Bicudo:1995kq,Horvat:1995im,LlanesEstrada:1999uh}.

As was mentioned above, qualitative results are insensitive to the
particular form of the potential, so that in most cases in
what follows it will not be fixed. If, however, a quantitative investigation of
equations is needed, the potential will be chosen in the most appropriate
power-like form as given in equation (\ref{potential}).

The GNJL model meets a wide set of requirements, such as a) the ability to account for relativistic effects; b)
the presence of an explicit confining force and, therefore, it can be employed to address various questions related to bound states of quarks, 
including excited hadrons; c) it is chirally symmetric (for $m=0$),
d) it is able to describe the effect of the spontaneous breaking of chiral symmetry in the vacuum.
The last point above deserves an additional remark. In particular, the given model fulfills all low-energy theorems such as
the Gell-Mann-Oakes-Renner relation \cite{GellMann:1968rz} (see \cite{Amer:1983qa,LeYaouanc:1983it,LeYaouanc:1983huv}), the Goldberger--Treiman
relation \cite{Goldberger:1958vp} (see \cite{Bicudo:2003fp}), the Adler self-consistency condition \cite{Adler:1964um}, and the Weinberg theorem
\cite{Weinberg:1966kf} (see \cite{Bicudo:2001jq}). At the same time, the model possesses an attractive feature to describe microscopically the
phenomenon of the spontaneous breaking of chiral symmetry in the vacuum and its effective restoration in the spectrum of excited hadrons.
These questions are discussed in detail in the review. Furthermore, since the effects of the chiral symmetry breaking and restoration are
closely related to the problem of the Lorentz nature of the confining interaction in quarkonia, the latter issue is also addressed in this review.

\section{BCS approximation, mass-gap equation, and chirally broken vacuum}\label{mgbog}

A convenient approach to studies of the model described by
Hamiltonian (\ref{HGNJL}) is the Bogoliu\-bov-Valatin transformation
which allows one to proceed from the ``bare'' quarks, which are
the relevant degrees of freedom in the chirally symmetric vacuum,
to the ``dressed'' quarks, which are the physical degrees of
freedom in the chirally broken vacuum
\cite{Bicudo:1989sh,Bicudo:1989si,Bicudo:1989sj,Bicudo:1993yh,Bicudo:1998mc}.
The quark field $\psi_{\alpha}({\bm x},t)$ is defined in terms of
annihilation and creation operators $\hat{b}$, $\hat{d}$ and
$\hat{b}^\dagger$, $\hat{d}^\dagger$, and takes the form
\begin{equation}
\psi_{\alpha}({\bm x},t)=\sum_{s=\uparrow,\downarrow}\int\frac{d^3p}{(2\pi)
^3}e^{i{\bm p}{\bm x}} [\hat{b}_{\alpha s}({\bm p},t)u_s({\bm p})+\hat{d}_{\alpha s}^\dagger(-{\bm p},t) v_{-s}(-{\bm p})],
\label{psi}
\end{equation}
\begin{equation}
\left\{
\begin{array}{rcl}
u_s({\bm p})&=&\ds\frac{1}{\sqrt{2}}\left[\sqrt{1+\sin\vpp_p}+
\sqrt{1-\sin\vpp_p}\;({\bm \alpha}\hat{{\bm p}})\right]u_s(0),\\[3mm]
v_{-s}(-{\bm p})&=&\ds\frac{1}{\sqrt{2}}\left[\sqrt{1+\sin\vpp_p}-
\sqrt{1-\sin\vpp_p}\;({\bm \alpha}\hat{{\bm
p}})\right]v_{-s}(0),
\end{array}
\right.
\label{uandv}
\end{equation}
\begin{equation}
\hat{b}_{s}({\bm p},t)=e^{iE_pt}\hat{b}_{s}({\bm p},0),\quad \hat{d}_{s}(-{\bm p},t)=e^{iE_pt}\hat{d}_{s}(-{\bm p},0).
\label{bandd}
\end{equation}
Here the rest-frame bispinors are defined as
\begin{equation}
u_s(0)=\left(
\begin{array}{c}
w_s\\
0\\
\end{array}
\right),
\quad v_{-s}(0)=-i\gamma_2 u_s^*(0)=
\left(
\begin{array}{c}
0\\
i\sigma_2 w_s^*\\
\end{array}
\right),
\label{spinors}
\end{equation}
where $\gamma_2(\sigma_2)$ is the second Dirac(Pauli) matrix, $s=\pm 1$ labels the spin eigenstates, so that $(w_s)_i=\delta_{si}$,
$E_p$ is the dressed-quark energy. The quantity $\vpp_p$ which parametrises the
Bogoliubov-Valatin transformation is known as the chiral angle
and it is defined with the boundary conditions $\vpp_p(p=0)=\pi/2$
and $\vpp_p(p\to\infty)=0$.

Then, after the normal ordering\footnote{In this review, normal ordering of operators is indicated by columns, for example, $:\hat{H}_2:$.} 
in terms of the dressed creation and annihilation operators, Hamiltonian (\ref{HGNJL}) takes the form
\begin{equation}
\hat{H}=E_{\rm vac}+:\hat{H}_2:+:\hat{H}_4:,
\label{H3}
\end{equation}
\begin{equation}
E_{\rm vac}[\vpp_p]=-\frac12gV\int\frac{d^3p}{(2\pi)^3}
\biggl(A_p\sin\vpp_p+B_p\cos\vpp_p\biggr),
\label{Evac1}
\end{equation}
where $V$ is the three-dimensional volume and the factor $g$ counts the total number
of the degrees of freedom for each quark,
$g=(2S+1)N_C$, where $2S+1$ with $S=1/2$ is the number of the quark spin projections
(in the multi-flavour case, $g$ is to be additionally
multiplied by the number of flavours $N_f$). The functions of the momentum $A_p$ and
$B_p$ are given by the formulae
\begin{equation}
A_p=m+\frac{1}{2}\int\frac{d^3k}{(2\pi)^3}V({\bm p}-{\bm k})\sin\vpp_k,\quad
B_p=p+\frac{1}{2}\int \frac{d^3k}{(2\pi)^3}\;(\hat{{\bm p}}\hat{{\bm k}})V({\bm p}-{\bm k})\cos\vpp_k,
\label{ABp}
\end{equation}
where hats in $\hat{{\bm p}}$ and $\hat{{\bm k}}$ denote the unit vectors for the respective momenta (hats over scalar
quantities identify the latter as operators --- see, for example, (\ref{psi})),
$V=C_FV_0$ and $C_F=(N_C^2-1)/(2N_C)$ is the eigenvalue of the fundamental
Casimir operator. To ensure that the potential takes finite values
in the limit $N_C\to\infty$, its strength is subject to an appropriate rescaling,
that is, $K_0^{\alpha+1}N_C\underset{N_C\to\infty}{\to}\mbox{const}$.

The explicit form of the chiral angle $\vpp_p$ is determined from the requirement that
the vacuum energy is kept to the minimum.
For a qualitative investigation of the properties of the
corresponding functional (\ref{Evac1}) it is convenient to use the following
trick \cite{Bicudo:2002eu}. Suppose the given functional has a minimum at a
particular function
$\vpp_0(p)$. Then, if evaluated at a rescaled function $\vpp_0(p/\xi )$,
with $0\leqslant \xi <\infty$, it must take larger values
for all $\xi \neq 1$, and it should reproduce the above minimum at $\xi =1$.
Finally, taking the limit
$\xi\to 0$ is equivalent to taking an infinitely large argument of the chiral angle
and, since
$\vpp_p(p\to\infty)\to 0$, such a limit is equivalent to the evaluation of the
energy functional for the trivial, chirally symmetric solution. Thus,
it proves instructive to study the behaviour of the function $E_{\rm vac}(\xi)$
which should have a minimum at $\xi =1$.
For simplicity, consider the chiral limit and set $m=0$. In this case, the only
remaining dimensional parameter is the potential strength $K_0$.
Then, by the redefinition of the integration variable in the functions $A_p$ and $B_p$, $p\to p/\xi $, one readily arrives at
\begin{equation}
E_{\rm vac}(\xi )=C_1\xi ^{d+1}+C_2K_0^{\alpha+1}\xi ^{d-\alpha},
\label{Evac11}
\end{equation}
where $D$ is the dimension of the space-time, $d=D-1$, while $C_1$ and $C_2$ are two $\xi$-independent constants. For convenience, let us count the
energy from the chirally symmetric solution $\vpp_p\equiv 0$ which corresponds
to $\xi=0$, that is, we set $E_{\rm vac}(0)=0$.
Whether or not there exists a minimum with a negative energy at $\xi =1$ depends
on the relation between the coefficients and the powers
of the two contributions in expression (\ref{Evac11}).
An interesting case is given by the limit $\alpha=d$. The 't~Hooft model
for two-dimensional
QCD constitutes an example of such a limit for which $\alpha=d=1$. Naively, one could expect that, in this limit, the second term in (\ref{Evac11})
turns to a constant, so that no nontrivial minimum can exist. However, this is not
the case. It is important to notice that, for $\alpha=d$, the
integrals in momentum are logarithmically divergent in the infrared and, as such,
they need a regulator, hereinafter denoted as $\lambda$.
Then, in the given limit, the second term in formula (\ref{Evac11}) contains a
logarithmic dependence on $\xi$,
\begin{equation}
E_{\rm vac}^{(\alpha=d)}(\xi )=C_1\xi ^{d+1}+C_2K_0^{d+1}\ln\left(\xi\frac{K_0}{\lambda}\right),
\label{Evac12}
\end{equation}
that entails two consequences: (i) a nontrivial minimum is possible, if the
coefficients $C_1$ and $C_2$ have different signs and
(ii) the vacuum energy grows as one approaches the trivial solution at $\xi=0$.
In other words, the chirally symmetric phase of the theory
ceases to exist \cite{Bicudo:2002eu}. A similar conclusion for the 't~Hooft model
is made in \cite{Gogokhia:2002ui}.

\begin{figure}[t]
\begin{center}
\epsfig{file=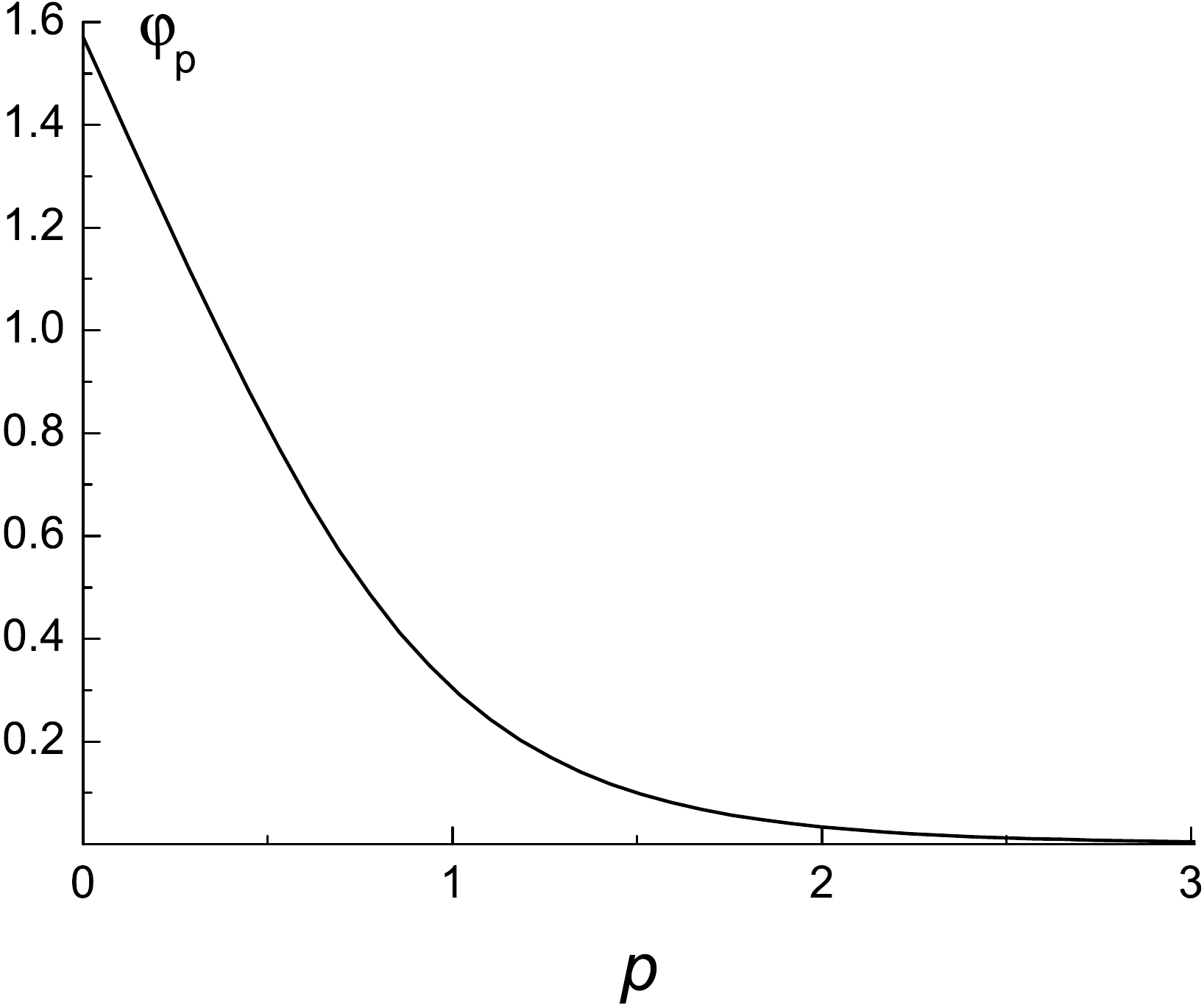,width=8cm}
\caption{Solution of the mass-gap equation (\ref{mge}) for $m=0$ and for the linear confining potential $V(r)=\sigma r$б where the parameter $\sigma$
has the dimension of the mass squared. The momentum $p$ is shown in the units of $\sqrt{\sigma}$.}\label{figvp}
\end{center}
\end{figure}

For the Generalised Nambu-Jona-Lasinio model $d=3$, and
in view of the restrictions for the value of the power
$\alpha$ --- see
(\ref{potential}) --- one always has $\alpha<d$. This ensures, for a particular choice of the signs of the coefficients
$C_1$ and $C_2$ in (\ref{Evac11}), the existence of a nontrivial energetically favourable solution, as compared to the trivial
vacuum. By a straightforward check, one can ensure that, indeed, the needed signs take place.

It should be noticed that the requirement that the vacuum energy should be a minimum guarantees at the same time that the quadratic part of the
Hamiltonian, $:\hat{H}_2:$, is diagonal (that is, the anomalous terms of the form $\hat{b}^\dagger
\hat{d}^\dagger-\hat{d}\hat{b}$ are missing), and the corresponding equation is known as the mass-gap equation
\cite{Amer:1983qa,LeYaouanc:1983it,LeYaouanc:1983huv,LeYaouanc:1984ntu,Bicudo:1989sh,Bicudo:1989si,Bicudo:1989sj,Bicudo:1993yh,Bicudo:1998mc},
\begin{equation}
A_p\cos\vpp_p=B_p\sin\vpp_p.
\label{mge}
\end{equation}
Then the dressed-quark dispersive law is
\begin{equation}
E_p=A_p\sin\vpp_p+B_p\cos\vpp_p.
\label{Ep}
\end{equation}

It is easy to verify that the solution of the mass-gap equation for a free particle
takes the form $\vpp_p=\arctan(m/p)$, and
then the free dispersive law $E_p=\sqrt{p^2+m^2}$ is readily reproduced.
It is also worthwhile mentioning that the same angle defines the
Foldy-Wouthuysen transformation that brings the free Dirac Hamiltonian
$H={\bm \alpha}\vp+\beta m$ to the diagonal form
$H'=\beta E_p$. Such a deep connection between the chiral angle and the
Foldy-Wouthuysen transformation persists for the nontrivial confining
interaction and for the chiral angle given by the solution for the
corresponding mass-gap equation (see, for example, \cite{Kalashnikova:2001df,Nefediev:2004by}).

For an arbitrary power-like confining potential (\ref{potential}), the
mass-gap
equation takes the form (in the chiral limit, that is, for $m=0$):
\begin{equation}
p^3\sin\vpp_p=\frac12K_0^3\left[p^2\vpp''_p+2p\vpp_p'+\sin2\vpp_p\right],
\label{diffmge}
\end{equation}
for $\alpha=2$
\cite{Amer:1983qa,LeYaouanc:1983it,LeYaouanc:1983huv,LeYaouanc:1984ntu,Bicudo:1989sh,Bicudo:1989si,Bicudo:1989sj,Bicudo:1993yh,Bicudo:1998mc}, and
\begin{equation}
p^3\sin\vpp_p=K_0^{\alpha+1}\Gamma(\alpha+1)\sin\frac{\pi\alpha}{2}
\int_{-\infty}^{\infty}
\frac{dk}{2\pi}\left\{\frac{pk\sin[\vpp_k-\vpp_p]}{|p-k|^{\alpha+1}}+
\frac{\cos\vpp_k\sin\vpp_p}{(\alpha-1)|p-k|^{\alpha-1}}\right\},
\label{mg2}
\end{equation}
for $0\leqslant\alpha<2$ \cite{Bicudo:2003cy}, where $\Gamma(\alpha+1)$ is the Euler Gamma function. 
For convenience and to make the formulae more compact, the absolute value of the
momentum $p$ is
formally prolonged to the domain $p<0$ according to the rule:
$\cos\vpp_{-p}=-\cos\vpp_p$, $\sin\vpp_{-p}=\sin\vpp_p$.
As was mentioned above, the mass-gap equation for the Harmonic
Oscillator potential
reduces to a second-order differential equation.

In Fig.~\ref{figvp}, the behaviour of the chiral angle as a function of the momentum is exemplified by the solution of the mass-gap equation with the
linear potential. Qualitatively, the shape of the curve does not depend on
the particular form of the interquark potential. Further
details of the formalism of the chiral angle can be found in
\cite{Amer:1983qa,LeYaouanc:1983it,LeYaouanc:1983huv,LeYaouanc:1984ntu,Bicudo:1989sh,Bicudo:1989si,Bicudo:1989sj,Bicudo:1993yh,Bicudo:1998mc,Bicudo:2002eu,Nefediev:2004by},
whereas the details of various studies of the mass-gap
equation can be found in \cite{Bicudo:2003cy} (for the four-dimensional theory) and in
\cite{Bars:1977ud,Bicudo:2003mw} (for the two-dimensional theory). In particular,
in some of the works mentioned above it was
pointed out that the
mass-gap equation supports the existence of ``excited'' solutions, with the
chiral angle possessing knots. Attempts to prescribe a physical
meaning to such solutions can be found in \cite{Bicudo:2002eu,Nefediev:2002nw,Antonov:2010qt}. In what follows, the problem of excited solutions
(replicas) is not discussed, and we always refer to the chiral
angle of the form depicted in Fig.~\ref{figvp} as to the nontrivial
solution of the
mass-gap equation.

For the chiral angle --- solution of the mass-gap equation (\ref{mge})
Hamiltonian (\ref{H3}) takes a diagonal form
\cite{Bicudo:1989sh,Bicudo:1989si,Bicudo:1989sj,Bicudo:1993yh,Bicudo:1998mc},
\begin{equation}
\hat{H}=E_{\rm vac}+\sum_{\alpha=1}^{N_C}\sum_{s=\uparrow,\downarrow}\int
\frac{d^3 p}{(2\pi)^3} E_p[\hat{b}^\dagger_{\alpha s}({\bm p}) \hat{b}_{\alpha
s}({\bm p})+\hat{d}^\dagger_{\alpha s}(-{\bm p}) \hat{d}_{\alpha
s}(-{\bm p})],
\label{H2diag}
\end{equation}
and the contribution of the omitted term $:\hat{H}_4:$ is suppressed as $1/\sqrt{N_C}$ in the large-$N_C$ limit. 
In the literature, such an approximation is
often referred to as the BCS approximation, in analogy with the similar approach by
Bardeen, Cooper, and Schrieffer to the theory of superconductivity.
The new, dressed, operators $b$ and $d$ annihilate the vacuum $|0\rangle$
which is related to the trivial vacuum $|0\rangle_0$,
annihilated by the bare operators, through the following relations
\cite{Bicudo:1989sh,Bicudo:1989si,Bicudo:1989sj,Bicudo:1993yh,Bicudo:1998mc}:
\begin{equation}
|0\rangle=e^{Q-Q^\dagger}|0\rangle_0,\quad Q^\dagger=\frac12\sum_{\vp}\vpp_p C_p^\dagger,\quad
C_p^\dagger=\sum_{\alpha=1}^{N_C}\sum_{s,s'=\uparrow,\downarrow}b^{\dagger}_{\alpha s}(\vp)[({\bm \sigma}\hat{{\bm
p}})i\sigma_2]_{ss'}d^{\dagger}_{\alpha s'}(\vp),
\label{S00}
\end{equation}
where $\vesig$ is given by the standard Pauli matrices and the operator
$C_p^\dag$ creates quark-antiquark pairs with the quantum numbers of
the vacuum, $J^{PC}=0^{++}$, that is, $^3P_0$ pairs.
With the help of the (anti)commutation relations between the quark and the
antiquark operators, one can
arrive at the following representation for the chirally broken (BCS) vacuum
\cite{Bicudo:1989sh,Bicudo:1989si,Bicudo:1989sj,Bicudo:1993yh,Bicudo:1998mc},
\begin{equation}
|0\rangle=\mathop{\prod}\limits_{p}\left[\sqrt{w_{0p}}+
\frac{1}{\sqrt{2}}\sqrt{w_{1p}}C^\dagger_p
+\frac12\sqrt{w_{2p}}C^{\dagger 2}_p\right]|0\rangle_0,
\label{nv}
\end{equation}
where the coefficients take the form
\begin{equation}
w_{0p}=\cos^4\frac{\vpp_p}{2},\quad
w_{1p}=2\sin^2\frac{\vpp_p}{2}\cos^2\frac{\vpp_p}{2},\quad
w_{2p}=\sin^4\frac{\vpp_p}{2},
\label{ws}
\end{equation}
and they obey the condition $w_{0p}+w_{1p}+w_{2p}=1$. It should be noticed that the coefficients (\ref{ws}) support a natural interpretation
in terms of probabilities to find in the new vacuum one ($w_{1p}$) or two ($w_{2p}$)
quark-antiquark pairs with the given relative momentum $2p$, or to find no such pairs at all ($w_{0p}$)
\cite{Nefediev:2009zzb}.
The Fermi statistics for the quark and the antiquarks makes it impossible
to create more pairs with the same relative momentum.

It is straightforward to ensure, with the help of equations (\ref{nv}) and (\ref{ws}),
that the wave function of the BCS vacuum is normalised
(the trivial vacuum is assumed to be normalised as well),
\begin{equation}
\label{norm00}
\langle 0|0\rangle=\prod_p(w_{0p}+w_{1p}+w_{2p})=1,
\end{equation}
and that the two vacua are orthogonal in the limit of an infinite volume $V$,
\begin{equation}
\langle 0|0\rangle_0=\exp\left[\sum_p \ln\left(\cos^2\frac{\vpp_p}{2}\right)\right]
=\exp\left[V\int\frac{d^3p}{(2\pi)^3}\ln\left(\cos^2\frac{\vpp_p}{2}\right)\right]
\mathop{\longrightarrow}\limits_{V\to \infty}0.
\end{equation}

It is easy to see that the BCS vacuum describes a cloud of strongly correlated
quark-antiquark pairs at each point of the configuration space
that is created by the operator $\exp[Q-Q^\dagger]$,
and this fact ensures the appearance of a nonzero quark-antiquark
condensate in the vacuum,
\begin{equation}
\langle \bar{\psi}\psi\rangle=-\frac{N_C}{\pi^2}\int_0^{\infty}dp\;p^2\sin\vpp_p,
\label{chircond}
\end{equation}
which vanishes at the trivial solution $\vpp_p\equiv 0$ but which takes
nonzero values for the nontrivial solution depicted in Fig.~\ref{figvp}.
Therefore, spontaneous breaking of chiral symmetry takes place: the Hamiltonian of
the theory is chirally symmetric while the BCS vacuum is not.
The large-momentum asymptotic of the chiral angle is related to the chiral condensate as
\begin{equation}
{\vpp_p}_{|{m=0}}\mathop{\approx}\limits_{p\to\infty}-
\frac{\pi}{N_C}\Gamma(\alpha+2)K_0^{\alpha+1}\sin\left(\frac{\pi\alpha}{2}\right)
\frac{\langle\bar{\psi}\psi\rangle}{p^{\alpha+4}}.
\label{asym}
\end{equation}

It is instructive to notice that, by a substitution $\varphi(p)\to\varphi(p/\xi)$ and a
subsequent variable change $p=\xi p'$ in formula
(\ref{chircond}), it is easy to demonstrate that the chiral condensate scales as $\xi^3$.
Then, one can rewrite (\ref{Evac11}) in the form of the function
$E_{\rm vac}(\langle \bar{\psi}\psi\rangle)$ which,
therefore, supports
the interpretation as an effective potential which
reaches the minimum at a nonzero
value of the chiral condensate.

\begin{figure}[t]
\begin{center}
\epsfig{file=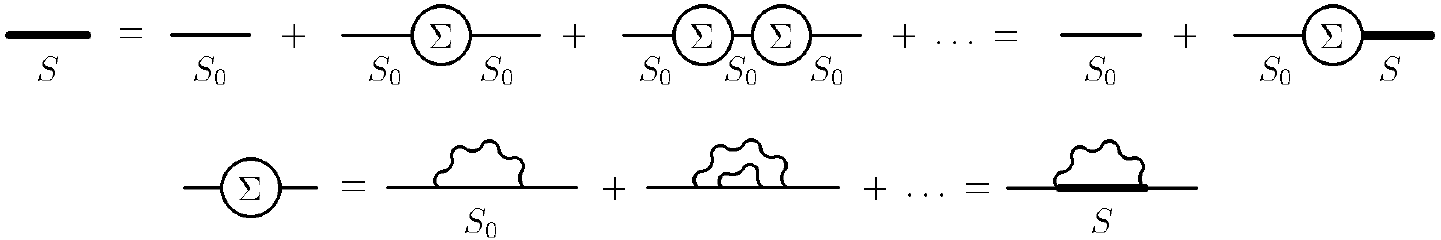,width=0.8\textwidth}
\caption{Graphical representation of the equation for the propagator and for the mass operator
of the dressed quark.}\label{diagrams}
\end{center}
\end{figure}

An alternative approach to the derivation of the mass-gap equation is related to
the Dyson equation for the dressed quark propagator that is
shown graphically in Fig.~~\ref{diagrams}. Schematically, this equation
can be represented as a sum of the infinite series of loops,
\begin{equation}
S=S_0+S_0\Sigma S_0+S_0\Sigma S_0\Sigma S_0+\ldots=S=S_0+S_0\Sigma S,
\label{Ds}
\end{equation}
with the mass operator given by the integral from the dressed propagator,
\begin{equation}
i\Sigma({\bm p})=\int\frac{d^4k}{(2\pi)^4}V({\bm p}-{\bm k})\gamma_0 S(k_0,{\bm k})\gamma_0,\quad
V(\vp)=C_FV_0(\vp),\quad C_F=\frac{N_C^2-1}{2N_C}.
\label{Sigma01last}
\end{equation}

The propagator $S(p_0,{\bm p})$ can be written with the help of the projectors on the positive- and negative-energy solutions of the Dirac equation,
\begin{equation}
S(p_0,{\bm p})=\frac{\Lambda^{+}({\bm p})\gamma_0}{p_0-E_p+i0}+
\frac{{\Lambda^{-}}({\bm p})\gamma_0}{p_0+E_p-i0},
\label{Feynman}
\end{equation}
where
\begin{equation}
\Lambda^\pm({\bm p})=\frac12[1\pm\gamma_0\sin\vpp_p\pm({\bm \alpha}\hat{{\bm p}})\cos\vpp_p].
\label{Lpm}
\end{equation}

The pole of the dressed quark is given by the value $E_p$ ($-E_p$ for the antiquark)
which, in turn, depends on the mass operator, so that one arrives
at a closed system of equations,
\begin{equation}
i\Sigma({\bm p})=\int\frac{d^4k}{(2\pi)^4}V({\bm p}-{\bm k})\gamma_0\frac{1}{S_0^{-1}(k_0,\vk)-\Sigma(\vk)}\gamma_0,
\quad S_0(p_0,\vp)=\frac{1}{\gamma_0 p_0-\vegam\vp-m+i0}.
\label{Sigma03}
\end{equation}

Since the Fourier transform of the potential does not depend on the energy
(this is a consequence of the instantaneous form of the interaction),
the integral on the temporal component of the momentum in the mass operator
(\ref{Sigma01last}) only touches upon the propagator (\ref{Feynman})
and, therefore, it can be evaluated explicitly, which, in turn,
allows one to parametrise
the mass operator in the form
\begin{equation}
\Sigma({\bm p})=[A_p-m]+({\bm \gamma}\hat{{\bm p}})[B_p-p],\quad
E_p=A_p\sin\vpp_p+B_p\cos\vpp_p,
\end{equation}
and gives for the propagator
\begin{equation}
S^{-1}(p_0,{\bm p})=\gamma_0p_0-({\bm \gamma}\hat{{\bm p}})B_p-A_p.
\end{equation}
The self-consistency condition for such a parametrisation is nothing
but the mass-gap equation for the chiral angle (\ref{mge}).

\section{Beyond the BCS level. Mesonic states}\label{app:BS}

In the previous chapter, the Generalised Nambu--Jona-Lasinio model was studied
in the BCS approximation having the dressed
quarks as the physical degrees of freedom. This approximation allows one to describe
microscopically the phenomenon of the spontaneous breaking of chiral symmetry
in the vacuum. Notice that the model contains
confinement and, therefore, it does not support the existence of free quarks.
Then, a natural next step is to proceed beyond the BCS approximation, with the
inclusion of the interaction between
the dressed quarks and thus with the building of colourless objects thereof --- the hadrons.
In this chapter this problem is addressed in the framework of two approaches:
in the matrix formalism (see
\cite{Amer:1983qa,LeYaouanc:1983it,LeYaouanc:1983huv,LeYaouanc:1984ntu,Bicudo:1989sh,Bicudo:1989si,Bicudo:1989sj,Bicudo:1993yh,Bicudo:1998mc,Nefediev:2004by}
for the details) and with the help of the generalised Bogoliubov-Valatin transformation
(the relevant details can be found in \cite{Nefediev:2004by}).

\subsection{Bethe-Salpeter equation}

\begin{figure}[t]
\begin{center}
\epsfig{file=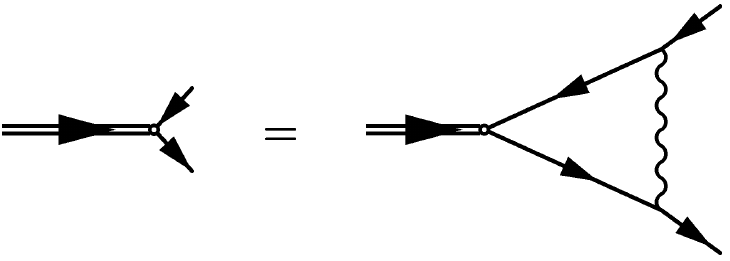,width=0.5\textwidth}
\caption{Graphical representation for the Bethe-Salpeter equation for the amplitude $\chi({\bm p};M)$.}\label{fig:bseq}
\end{center}
\end{figure}

In the framework of the matrix formalism, proceeding beyond the BCS approximation is
done by considering the Bethe-Salpeter equation for the bound states
of quarks and antiquarks which is written as an equation for the mesonic
amplitude $\chi({\bm p};M)$ in the meson rest frame (here
$\vp$ is the momentum of the quark and $M$ is the mass of the meson) --- see Fig.~\ref{fig:bseq},
\begin{equation}
\chi({\bm p};M)=-i\int\frac{d^4q}{(2\pi)^4}V({\bm p}-{\bm q})\;
\gamma_0 S({\bm q},q_0+M/2)\chi({\bm q};M)S({\bm q},q_0-M/2)\gamma_0.
\label{GenericSal}
\end{equation}

The instantaneous form of the interaction allows one to simplify this equation considerably. In particular, once the integral in the energy in
equation (\ref{GenericSal}) only depends on the position of the poles of the propagators,
it is easy to see that, when the propagators are substituted
in the form of equations (\ref{Feynman}), only two terms of the four survive,
with the poles in the $q_0$ complex plane located on different sides from the
real axis. The corresponding integrals are straightforwardly evaluated then and give
\begin{equation}
\int_{-\infty}^{\infty}\frac{dq_0}{2\pi i}\left[\frac{1}{q_0\pm
M/2-E_q+i0}\right] \left[\frac{1}{q_0\mp
M/2+E_q-i0}\right]=-\frac{1}{2E_q\mp M},
\end{equation}
so that equation (\ref{GenericSal}) turns to a system of coupled equations,
\begin{equation}
\left\{
\begin{array}{l}
[2E_p-M]\chi^{[+]}=-\ds\int\frac{d^3q}{(2\pi)^3}V({\bm p}-{\bm q})\;
\gamma_0\left[(\Lambda^+\gamma_0)\chi^{[+]}(\Lambda^-\gamma_0)
+(\Lambda^-\gamma_0)\chi^{[-]}(\Lambda^+\gamma_0)\right]\gamma_0\\[0mm]
[2E_p+M]\chi^{[-]}=-\ds\int\frac{d^3q}{(2\pi)^3}V({\bm p}-{\bm q})\;
\gamma_0\left[(\Lambda^+\gamma_0)\chi^{[+]}(\Lambda^-\gamma_0)
+(\Lambda^-\gamma_0)\chi^{[-]}(\Lambda^+\gamma_0)\right]\gamma_0,
\end{array}
\label{Salpeterlev3}
\right.
\end{equation}
where we introduced the amplitudes
$$
\chi^{[+]}({\bm q};M)=\frac{\chi({\bm q};M)}{2E_q-M},\quad
\chi^{[-]}({\bm q};M)=\frac{\chi({\bm q};M)}{2E_q+M}.
$$

In order to proceed, we
\begin{itemize}
\item multiply the first equation in the system
(\ref{Salpeterlev3}) by $\bar{u}_{s_1}$ from the left and by
$v_{s_2}$ from the right, and do the same for the second equation,
however, with $\bar{v}_{s_3}$ and $u_{s_4}$, respectively;
\item
represent the projectors $\Lambda^\pm$ through the bispinors,
\begin{equation}
\Lambda^+({\bm p})=\sum_{s}u_{s}({\bm p})\otimes u^\dagger_{s}
({\bm p}),\quad \Lambda^-({\bm p})=\sum_{s}v_{-s}(-{\bm p})\otimes
v^\dagger_{-s} (-{\bm p}); \label{Lambdas}
\end{equation}
\item define matrix amplitudes $\phi_{s_1s_2}^+=[\bar{u}_{s_1}\chi^{[+]}v_{-s_2}]$ and $\phi_{s_1s_2}^-=[\bar{v}_{-s_1}\chi^{[-]}u_{s_2}]$.
\end{itemize}

As a result, the Bethe-Salpeter equation takes the form
\begin{equation}
\left\{
\begin{array}{l}
[2E_p-M]\phi_{s_1s_2}^+=-\ds\sum_{s_3s_4}\int\frac{d^3q}{(2\pi)^3}V({\bm p}-{\bm q})
\left\{[v^{++}]_{s_1s_3s_4s_2}\phi_{s_3s_4}^+ +[v^{+-}]_{s_1s_3s_4s_2}\phi_{s_3s_4}^-\right\}\\[1mm]
[2E_p+M]\phi_{s_1s_2}^-=-\ds\sum_{s_3s_4}\int\frac{d^3q}{(2\pi)^3}V({\bm p}-{\bm q})
\left\{[v^{-+}]_{s_1s_3s_4s_2}\phi_{s_3s_4}^+ +[v^{--}]_{s_1s_3s_4s_2}\phi_{s_3s_4}^-\right\},
\label{Salpeterlev5}
\end{array}
\right.
\end{equation}
where we defined the quantities $v^{\pm\pm}$,
\begin{equation}
\begin{array}{c}
[v^{++}({\bm p},{\bm q})]_{s_1s_3s_4s_2}=[\bar{u}_{s_1}({\bm p})\gamma_0 u_{s_3}({\bm q})]
[\bar{v}_{-s_4}(-{\bm q})\gamma_0 v_{-s_2}(-{\bm p})],\\[0cm]
[v^{+-}({\bm p},{\bm q})]_{s_1s_3s_4s_2}=[\bar{u}_{s_1}({\bm p})\gamma_0 v_{-s_3}(-{\bm q})]
[\bar{u}_{s_4}({\bm q})\gamma_0 v_{-s_2}(-{\bm p})],\\[0cm]
[v^{-+}({\bm p},{\bm q})]_{s_1s_3s_4s_2}=[\bar{v}_{-s_1}(-{\bm p})\gamma_0 u_{s_3}({\bm q})]
[\bar{v}_{-s_4}(-{\bm q})\gamma_0 u_{s_2}({\bm p})],\\[0cm]
[v^{--}({\bm p},{\bm q})]_{s_1s_3s_4s_2}=[\bar{v}_{-s_1}(-{\bm
p})\gamma_0 v_{-s_3}(-{\bm q})][\bar{u}_{s_4}({\bm q})\gamma_0
u_{s_2}({\bm p})],
\end{array}
\label{ampls}
\end{equation}
with
\begin{equation}
\begin{array}{c}
[\bar{u}_{s}({\bm k_1})\gamma_0 u_{s'}({\bm
k_2})]=\left[C_{k_1}C_{k_2}+S_{k_1}S_{k_2} ({\bm \sigma}\hat{{\bm
k_1}})({\bm \sigma}\hat{{\bm k_2}})\right]_{s,s'},\\[0cm]
[\bar{v}_{-s}({-\bm k_1})\gamma_0 v_{-s'}({-\bm
k_2})]=\left[(-i\sigma_2)(C_{k_1}C_{k_2}+S_{k_1}S_{k_2} ({\bm
\sigma}\hat{{\bm
k_1}})({\bm \sigma}\hat{{\bm k_2}}))(i\sigma_2)\right]_{s,s'},\\[0cm]
[\bar{v}_{-s}({-\bm k_1})\gamma_0 u_{s'}({\bm
k_2})]=\left[(S_{k_1}C_{k_2}({\bm \sigma}\hat{{\bm
k_1}})-S_{k_2}C_{k_1}({\bm \sigma}\hat{{\bm
k_2}}))(i\sigma_2)\right]_{s,s'},\\[0cm]
[\bar{u}_{s}({\bm k_1})\gamma_0 v_{-s'}({-\bm
k_2})]=-\left[(i\sigma_2)(S_{k_1}C_{k_2}({\bm \sigma}\hat{{\bm
k_1}})-S_{k_2}C_{k_1}({\bm \sigma}\hat{{\bm k_2}}))\right]_{s,s'},
\end{array}
\label{uv}
\end{equation}
and where the following shorthand notations are used:
\begin{equation}
C_p=\cos\frac12\left(\frac{\pi}{2}-\vpp_p\right)=\sqrt{\frac{1+\sin\vpp_p}{2}},\quad
S_p=\sin\frac12\left(\frac{\pi}{2}-\vpp_p\right)=\sqrt{\frac{1-\sin\vpp_p}{2}}.
\label{SandC}
\end{equation}

It also proves convenient to include the potential into the definition of the amplitudes, thus writing
\begin{equation}
\begin{array}{l}
[T^{++}({\bm p},{\bm q})]_{s_1s_3s_4s_2}=[\bar{u}_{s_1}({\bm p})\gamma_0 u_{s_3}({\bm q})]
[-V({\bm p}-{\bm q})][\bar{v}_{-s_4}(-{\bm q})\gamma_0 v_{-s_2}(-{\bm p})],\\[0cm]
[T^{+-}({\bm p},{\bm q})]_{s_1s_3s_4s_2}=[\bar{u}_{s_1}({\bm
p})\gamma_0 v_{-s_3}(-{\bm q})]
[-V({\bm p}-{\bm q})][\bar{u}_{s_4}({\bm q})\gamma_0 v_{-s_2}(-{\bm p})],\\[0cm]
[T^{-+}({\bm p},{\bm q})]_{s_1s_3s_4s_2}=[\bar{v}_{-s_1}(-{\bm
p})\gamma_0 u_{s_3}({\bm q})]
[-V({\bm p}-{\bm q})][\bar{v}_{-s_4}(-{\bm q})\gamma_0 v_{-s_2}({\bm p})],\\[0cm]
[T^{--}({\bm p},{\bm q})]_{s_1s_3s_4s_2}=[\bar{v}_{-s_1}(-{\bm
p})\gamma_0 v_{-s_3}(-{\bm q})] [-V({\bm p}-{\bm
q})][\bar{u}_{s_4}({\bm q})\gamma_0 u_{s_2}({\bm p})],
\end{array}
\label{ampls2}
\end{equation}
or, symbolically,
\begin{equation}
\begin{array}{c}
T^{++}=[\bar{u}\gamma_0 u][-V][\bar{v}\gamma_0 v],\quad T^{+-}=[\bar{u}\gamma_0 v][-V][\bar{u}\gamma_0 v],\\
T^{-+}=[\bar{v}\gamma_0 u][-V][\bar{v}\gamma_0 u],\quad T^{--}=[\bar{v}\gamma_0 v][-V][\bar{u}\gamma_0 u].
\end{array}
\end{equation}
Equations (\ref{Salpeterlev5}) comprise the Bethe-Salpeter equation in the so-called energy-spin formalism of \cite{Bicudo:1989sh,Bicudo:1989si,Bicudo:1989sj,Bicudo:1993yh,Bicudo:1998mc}.

In \cite{Bars:1977ud}, the approach of the matrix wave functions is suggested
for the two-dimensional QCD which is convenient in various
applications. Below, this approach is generalised to the
four-dimensional Generalised Nambu--Jona-Lasinio model \cite{Nefediev:2004by}.

To begin with, we notice that it is convenient to define the Foldy operator
$T_p$ and to re-write the Dirac projectors
$\Lambda^\pm$ (\ref{Lpm}) with its help,
\begin{equation}
\Lambda^\pm({\bm p})=T_pP_\pm T_p^\dagger,\quad
P_\pm=\frac{1\pm\gamma_0}{2},\quad
T_p=\exp{\left[-\frac12({\bm \gamma}\hat{{\bm p}})\left(\frac{\pi}{2}-\vpp_p\right)\right]}.
\end{equation}

As a nest step, equation (\ref{GenericSal}) for the mesonic
amplitude is re-written through the matrix wave function \begin{equation}
\tilde{\phi}({\bm p};M_\pi)=\int\frac{dp_0}{2\pi}S({\bm
p},p_0+M)\chi({\bm p};M)S({\bm p},p_0-M), \label{Psiqq} \end{equation} which
is subject to the rotation with the Foldy operator $T_p$ both from
the left and from the right, thus defining $\phi({\bm
p};M)=T^\dagger_p\tilde{\phi}({\bm p};M)T^\dagger_p$. For such a
matrix wave function, the Bethe-Salpeter equation
(\ref{GenericSal}) takes the form 
\begin{equation} 
\phi({\bm p};M)=-\int\frac{d^3q}{(2\pi)^3}V({\bm p}-{\bm q})\left[
P_+\frac{T_p^\dagger T_q\phi({\bm q};M)T_qT_p^\dagger}{2E_p-M}P_-
+P_-\frac{T_p^\dagger T_q\phi({\bm
q};M_\pi)T_qT_p^\dagger}{2E_p+M}P_+ \right]. 
\label{PHI} 
\end{equation} 
It is
easy to see that the solution of equation (\ref{PHI}) has the form
\begin{equation} 
\phi({\bm p};M)=P_+{\cal A}P_-+P_-{\cal B}P_+, 
\label{AB} 
\end{equation}
where ${\cal A}$ and ${\cal B}$ are two unknown matrix functions
which can be expanded in the complete set of the $4\times 4$
matrices,
$\{1,\gamma_\mu,\gamma_5,\gamma_\mu\gamma_5,\sigma_{\mu\nu}\}$. It
should be noticed however that, due to the orthogonality
properties of the projectors $P_+P_-=P_-P_+=0$ and also due to the
fact that the matrix $\gamma_0$ can always be absorbed into their
definition, the actual set of matrices is reduced to just two,
$\{\gamma_5,{\bm \gamma}\}$, so that the wave function (\ref{AB})
can be represented as \begin{equation} \phi({\bm p};M)=\left(
\begin{array}{cc}
0&\varphi^+({\bm p})\\
\varphi^-({\bm p})&0\\
\end{array}
\right), \end{equation} where $\varphi^{\pm}({\bm p})$ are $2 \times 2$
matrices. It is a straightforward exercise to demonstrate that the
eigenvalue problem given by (\ref{Salpeterlev5}) is equivalent to the one given by
(\ref{PHI}), with
 \begin{equation}
 \phi^+_{s_1s_2}=i(\varphi^+\sigma_2)_{s_1s_2},\quad \phi^-_{s_1s_2}=i(\sigma_2
 \varphi^-)_{s_1s_2}.
 \end{equation}

The further transformations correspond to projecting the matrix amplitudes onto the states with the
given total momentum and spatial and charge parities.

\subsection{The chiral pion}\label{subsec:pion}

Consider first the case of the chiral pion. For the corresponding
matrix amplitude one has
\begin{equation}
\phi_{s_1s_2}^{\pm}({\bm
p})=\left[\frac{i}{\sqrt{2}}\sigma_2\right]_{s_1s_2}
Y_{00}(\hat{{\bm p}})\vpp_\pi^\pm(p),
\label{vppi}
\end{equation}
where
$Y_{00}(\hat{{\bm p}})=1/\sqrt{4\pi}$ is the normalised to unity
lowest spherical harmonic. Then, if the amplitudes
$T_\pi^{\pm\pm}(p,q)$ are introduced according to equation
(\ref{ampls2}) and all spin traces are taken explicitly, then one
arrives at the following system of equations for the scalar wave
functions $\vpp_\pi^\pm$: 
\begin{equation} 
\left\{
\begin{array}{l}
[2E_p-M_\pi]\vpp_\pi^+(p)=\ds\int\frac{\ds q^2dq}{\ds (2\pi)^3}
[T^{++}_\pi(p,q)\vpp_\pi^+(q)+T^{+-}_\pi(p,q)\vpp_\pi^-(q)]\\[0cm]
[2E_p+M_\pi]\vpp_\pi^-(p)=\ds\int\frac{\ds q^2dq}{\ds (2\pi)^3}
[T^{-+}_\pi(p,q)\vpp_\pi^+(q)+T^{--}_\pi(p,q)\vpp_\pi^-(q)],
\end{array}
\right.
\label{bsp}
\end{equation}
where
\begin{eqnarray}
T_\pi^{++}(p,q)=T_\pi^{--}(p,q)&=&-\ds\int d\Omega_q V({\bm
p}-{\bm q})
\left[\cos^2\frac{\vpp_p-\vpp_q}{2}-\frac{1-(\hat{{\bm p}}\hat{{\bm q}})}{2}\cos\vpp_p\cos\vpp_q\right],\nonumber\\[-2mm]
\label{pia}\\[-2mm]
T_\pi^{+-}(p,q)=T_\pi^{-+}(p,q)&=&-\ds\int d\Omega_q V({\bm
p}-{\bm q}) \left[\sin^2\frac{\vpp_p-\vpp_q}{2}+\frac{1-(\hat{{\bm
p}}\hat{{\bm q}})}{2}\cos\vpp_p\cos\vpp_q\right].\nonumber
\end{eqnarray}

The resulting system of equations (\ref{bsp}) can be interpreted as a bound-state equation
for a quark-antiquark pair in the channel with the quantum numbers of
the pion. The phy\-sical interpretation of the two amplitudes used to describe one meson
comes from the observation that the quark-antiquark pair in it can move
both forward and backward in time, and each type of the motion is described by an
independent amplitude
\cite{Amer:1983qa,LeYaouanc:1983it,LeYaouanc:1983huv,LeYaouanc:1984ntu,Bicudo:1989sh,Bicudo:1989si,Bicudo:1989sj,Bicudo:1993yh,Bicudo:1998mc,Bars:1977ud}.
Thus, the Hamiltonian turns out to be a matrix in the space
of the so-called energy spin, and the bound-state equation takes the form of a
system of two coupled equations.

One can explicitly verify that, in the strict chiral limit $m=0$,
the function
\begin{equation}
\vpp_\pi^+(p)=\vpp_\pi^-(p)=\sin\vpp_p,
\label{masslesspion}
\end{equation}
is a solution of system (\ref{bsp})
with the eigenvalue $M_{\pi}=0$. Indeed, substituting function
(\ref{masslesspion}) and $M_{\pi}=0$ into system (\ref{bsp})
one arrives at the single equation
\begin{equation}
2E_p\vpp_\pi(p)=\int\frac{q^2dq}{(2\pi)^3}[T_\pi^{++}(p,q)+T_\pi^{+-}(p,q)]\vpp_\pi(q)=
-\int\frac{d^3q}{(2\pi)^3}V({\bm p}-{\bm q})\vpp_\pi(q),
\end{equation}
which holds true due to the mass-gap equation (\ref{mge}) and dispersive
law (\ref{Ep}). The resulting equation looks especially simple
and instructive in the coordinate space, 
\begin{equation}
[2E_p+V(r)]\vpp_\pi=0, 
\label{mgesalp} 
\end{equation} 
that is, formally, it
takes the form of the simple Salpeter equation with equal masses
and with the eigenvalue $M=0$; however, the form of the quantity
$E_p$ is very different from the simple kinetic energy of the free
quark $\sqrt{p^2+m^2}$ that guarantees the existence of the
vanishing eigenvalue.

We show in such a way that in the chiral limit the pion Bethe-Salpeter equation is
equivalent to the mass-gap equation for the chiral angle, which, in turn,
demonstrates the celebrated dualism of the pion: as a Goldstone boson, it appears
already at the BCS level while, beyond the BCS, the same pion emerges
from the Bethe-Salpeter equation, as the lowest level in the
spectrum of the quark-antiquark states.

The system of equations (\ref{bsp}) allows one to study the
behaviour of the pionic solution near the chiral limit. In
particular, one can demonstrate that, for $M_\pi\to 0$, the
solution of this system has the form (higher-order terms in the
pion mass are neglected)
\begin{equation}
\vpp_\pi^\pm(p)=\frac{\sqrt{2\pi
N_C}}{f_\pi}\left[\frac{1}{\sqrt{M_\pi}}\sin\vpp_p \pm
\sqrt{M_\pi}\Delta_p\right],\quad
f_\pi^2=\frac{N_C}{\pi^2}\int_0^\infty p^2dp\Delta_p\sin\vpp_p,
\label{vppm}
\end{equation}
where the function $\Delta_p$ obeys an equation
which does not contain $M_\pi$ any more (see also
\cite{Bicudo:1989sh,Bicudo:1989si,Bicudo:1989sj,Bicudo:1993yh,Bicudo:1998mc}):
\begin{equation}
2E_p\Delta_p=\sin\vpp_p+\int\frac{d^3k}{(2\pi)^3}V(\vp-\vk)\Bigl(\sin\vpp_p\sin\vpp_k+(\hat{\vp}\hat{\vk}
)\cos\vpp_p\cos\vpp_k\Bigr)\Delta_k. \end{equation} It is easy to verify that
the normalisation condition for the wave functions $\vpp_\pi^\pm$
takes the form \begin{equation}
\int\frac{p^2dp}{(2\pi)^3}\left[\vpp_\pi^{+2}(p)-\vpp_\pi^{-2}(p)\right]=1.
\label{norm1} \end{equation} The physical interpretation of such a
normalisation will become clear from the generalised
Bogoliubov-Valatin transformation for the mesonic operators.

Let us consider now the matrix structure of the pionic wave
function. In case of the pion, it is obvious that only $\gamma_5$
contributes, so that one can extract the matrix structure of the
quantities ${\cal A}$ and ${\cal B}$ explicitly and introduce the
scalar wave functions $\vpp_\pi^\pm$ as
\begin{equation}
{\cal A}_\pi=\gamma_5\vpp^+_\pi(p),\quad {\cal B}_\pi=\gamma_5\vpp_\pi^-(p),
\label{AB2}
\end{equation}
where the signs and the coefficients are chosen to comply with definition
(\ref{vppi}) used before. Thus, with the help of equations
(\ref{AB}) and (\ref{AB2}), it is easy to see that the pion wave
function takes the form
\begin{equation}
\tilde{\phi}({\bm p};M_\pi)=T_p\left[P_+\gamma_5\vpp_\pi^++P_-\gamma_5\vpp_\pi^-\right]T_p=
\gamma_5G_\pi+\gamma_0\gamma_5T_p^2F_\pi,
\label{exp}
\end{equation}
where $G_\pi=\frac12(\vpp_\pi^++\vpp_\pi^-)$ and $F_\pi=\frac12(\vpp_\pi^+-\vpp_\pi^-)$, and the Bethe-Salpeter
equation (\ref{PHI}) can be re-written in the form
$$
M_\pi\tilde{\phi}({\bm p};M_\pi)=[({\bm \alpha}{\bm p})+\gamma_0m]\tilde{\phi}({\bm p};M_\pi)+
\tilde{\phi}({\bm p};M_\pi)[({\bm \alpha}{\bm p})-\gamma_0m]\hspace*{5cm}
$$
\begin{equation}
+\int\frac{d^3q}{(2\pi)^3}V({\bm p}-{\bm q})\left\{\Lambda^+({\bm q})\tilde{\phi}({\bm p};M_\pi)\Lambda^-(-{\bm q})-
\Lambda^+({\bm p})\tilde{\phi}({\bm q};M_\pi)\Lambda^-(-{\bm p})\right.
\label{matrix}
\end{equation}
$$
\hspace*{5cm}\left.-\Lambda^-({\bm q})\tilde{\phi}({\bm p};M_\pi)\Lambda^+(-{\bm q})+
\Lambda^-({\bm p})\tilde{\phi}({\bm q};M_\pi)\Lambda^+(-{\bm p})\right\}.
$$

On multiplying the latter equation by $\gamma_0\gamma_5$, integrating it in the
momentum $\vp$, and taking the trace in the spin
matrices, one arrives at the relation
\begin{equation}
M_\pi\int\frac{d^3p}{(2\pi)^3}F_\pi\sin\vpp_p=2m\int\frac{d^3p}{(2\pi)^3}G_\pi,
\label{GMOR2}
\end{equation}
which can be easily identified as the celebrated Gell-Mann--Oakes--Renner
relation, if the explicit form of the pion wave function (\ref{vppm})
is used together with the quantities $G_\pi$ and $F_\pi$
defined as
\begin{equation}
G_\pi=\frac{\sqrt{2\pi N_C}}{f_\pi\sqrt{M_\pi}}\sin\vpp_p,\quad F_\pi=\frac{\sqrt{2\pi M_\pi N_C}}{f_\pi}\Delta_p,
\end{equation}
where the pion decay constant $f_\pi$ and the function $\Delta_p$ were introduced
in equation (\ref{vppm}). Then the
conventional form of the Gell-Mann--Oakes--Renner relation \cite{GellMann:1968rz}
is readily restored as soon as formula (\ref{chircond})
for the chiral condensate is used,
\begin{equation}
f_\pi^2M_\pi^2=-2m\langle\bar{\psi}\psi\rangle.
\label{GMOR}
\end{equation}

\subsection{Bogoliubov transformation for mesonic operators}

In \cite{Kalashnikova:1999wt}, an alternative approach to
mesonic states in the two-dimensional model for QCD was proposed allowing one to study
mesonic states in this theory with the help of the generalised
Bogoliubov-Valatin transformation for the mesonic sector.
This approach can be naturally
generalised to the four-dimensional Generalised Nambu--Jona-Lasinio model.
Such a generalisation suggested in \cite{Nefediev:2004by} is described in detail below.

Let us define four operators quadratic in the quark operators.
Among those, the first two,
\begin{equation}
\begin{array}{c}
\hat{B}_{ss'}({\bm p},{\bm p}')=\ds\frac{\ds
1}{\ds\sqrt{N_C}}\sum_\alpha \hat{b}_{\alpha
s}^\dagger({\bm p})\hat{b}_{\alpha s'}({\bm p}'),\quad
\hat{D}_{ss'}({\bm p},{\bm p}')=\frac{\ds
1}{\ds\sqrt{N_C}}\sum_\alpha \hat{d}_{\alpha
s}^\dagger(-{\bm p})\hat{d}_{\alpha s'}(-{\bm p}'),
\end{array}
\label{operatorsBD}
\end{equation}
``count'' the number of quarks and antiquarks while the other two,
\begin{equation}
\begin{array}{c}
\hat{M}^\dagger_{ss'}({\bm p},{\bm p}')=\ds\frac{\ds
1}{\ds\sqrt{N_C}}\sum_\alpha \hat{b}^\dagger_{\alpha
s'}({\bm p}')\hat{d}^\dagger_{\alpha s}(-{\bm p}),\quad
\hat{M}_{ss'}({\bm p},{\bm p}')=\ds\frac{\ds
1}{\ds\sqrt{N_C}}\sum_\alpha \hat{d}_{\alpha s}(-{\bm p})\hat{b}_{\alpha
s'}({\bm p}'),
\end{array}
\label{operatorsMM}
\end{equation}
create and annihilate quark-antiquark pairs. In the limit $N_C\to\infty$,
the introduced operators obey the standard bosonic commutation relations.
In particular, the only nonvanishing commutator reads
\begin{equation}
[\hat{M}_{ss'}({\bm p},{\bm p}')\;\hat{M}_{\sigma\sigma'}^\dagger({\bm q},{\bm q}')]=
(2\pi)^3\delta^{(3)}({\bm p}-{\bm q})
(2\pi)^3\delta^{(3)}({\bm p}'-{\bm q}')\delta_{s\sigma}\delta_{s'\sigma'}.
\label{MMcom}
\end{equation}

It is easy to see that, at the BCS level, Hamiltonian (\ref{H2diag}) is expressed entirely in terms of the first pair of the above operators,
\begin{equation}
\hat{H}=E_{\rm
vac}+\sqrt{N_C}\sum_{s=\uparrow,\downarrow}\int\frac{d^3
p}{(2\pi)^3}E_p[\hat{B}_{ss}({\bm p},{\bm p})+\hat{D}_{ss}({\bm p},{\bm p})],
\label{H2diag2}
\end{equation}
while the omitted (at the BCS level, suppressed in the large-$N_C$ limit)
part of the Hamiltonian $:\hat{H}_4:$ contains all four operators.
The key observation of the approach is the statement that, in the presence
of confinement, quarks and antiquark cannot be created or annihilated
as isolated objects --- this is only possible for quark-antiquark pairs.
Therefore, beyond the BCS approximation, operators
(\ref{operatorsBD}) cannot be independent, but they must be expressed through
operators (\ref{operatorsMM}). In the large-$N_C$ limit, it is
sufficient to stick to the minimal number of the quark-antiquark pairs, that is,
to retain only one accompanying antiquark for each created quark
and vice versa and not to consider the entire quark-antiquark cloud. Then, the
sought relation between the operators reads
\begin{equation}
\left\{
\begin{array}{c}
\hat{B}_{ss'}({\bm p},{\bm p}')=\frac{\ds
1}{\ds\sqrt{N_C}}\ds\sum_{s''} \int\frac{\ds d^3p''}{\ds
(2\pi)^3}\hat{M}_{s''s}^\dagger({\bm p}'',{\bm p})
\hat{M}_{s''s'}({\bm p}'',{\bm p}')\\
\hat{D}_{ss'}({\bm p},{\bm p}')=\frac{\ds
1}{\ds\sqrt{N_C}}\ds\sum_{s''} \int\frac{\ds d^3p''}{\ds
(2\pi)^3}\hat{M}_{ss''}^\dagger({\bm p},{\bm p}'')
\hat{M}_{s's''}({\bm p}',{\bm p}'').
\end{array}
\right.
\label{anzatz}
\end{equation}

It is easy to verify that, in the limit $N_C\to\infty$, substitution (\ref{anzatz})
reproduces the commutation relations between the operators
(\ref{operatorsBD}), so that it can be interpreted as an
independent solution for the equations given by these commutation relations.

If relations (\ref{anzatz}) are substituted in Hamiltonian (\ref{H3}),
the terms $:\hat{H}_2:$ and $:\hat{H}_4:$
appear to be of the same order of magnitude, while all other terms,
suppressed in the limit
$N_C\to\infty$, can be neglected. Then the centre-of-mass Hamiltonian of the
quark-antiquark cloud takes the form
\begin{equation}
\hat{H}=E_{\rm vac}'+\int \frac{d^3P}{(2\pi)^3}
\hat{\cal H}({\bm P}),
\label{HH1}
\end{equation}
where (for simplicity, the Hamiltonian density $\cal H$ is taken in the rest
frame, with ${\bm P}=0$)
$$
\hat{\cal H}\equiv\hat{\cal H}({\bm P}=0)=\sum_{s_1s_2}\int\frac{d^3
p}{(2\pi)^3}
2E_p\hat{M}_{s_1s_2}^\dagger({\bm p},{\bm p})\hat{M}_{s_2s_1}({\bm p},{\bm p})
+\frac12\sum_{s_1s_2s_3s_4}\int\frac{d^3p}{(2\pi)^3}\frac{d^3q}{(2\pi)^3}
V({\bm p}-{\bm q})
$$
\begin{equation}
\times \left\{[v^{++}({\bm p},{\bm q})]_{s_1s_3s_4s_2}
\hat{M}^\dagger_{s_2s_1}({\bm p},{\bm p})\hat{M}_{s_4s_3}({\bm q},{\bm q})\right.
+[v^{+-}({\bm p},{\bm q})]_{s_1s_3s_4s_2}
\hat{M}^\dagger_{s_2s_1}({\bm q},{\bm q})\hat{M}^\dagger_{s_3s_4}({\bm p},{\bm p})
\label{HH2}
\end{equation}
$$
\left.+[v^{-+}({\bm p},{\bm q})]_{s_1s_3s_4s_2}
\hat{M}_{s_1s_2}({\bm p},{\bm p})\hat{M}_{s_4s_3}({\bm q},{\bm q})
+[v^{--}({\bm p},{\bm q})]_{s_1s_3s_4s_2}
\hat{M}_{s_3s_4}({\bm p},{\bm p})\hat{M}_{s_1s_2}^\dagger({\bm q},{\bm q})\right\},
$$
and the amplitudes $v$ are given by the expressions from equation (\ref{ampls}).

Strictly speaking, only two amplitudes of the four in equation (\ref{ampls}),
for example, $v^{++}$ and $v^{+-}$, are independent while the others,
$v^{--}$ and $v^{-+}$, are related to them through the operation of Hermitian
conjugation. Nevertheless, we prefer to keep all four amplitudes explicitly
in order to preserve the most symmetric form of the equations.

\subsubsection{The case of the chiral pion}
Before we come to the diagonalisation of the full Hamiltonian (\ref{HH2}), we treat
the case of the chiral pion separately.
For the pion, $J=L=S=0$, so that the operator $\hat{M}_{ss'}({\bm p},{\bm p})$ can be written in the form
\begin{equation}
\hat{M}_{ss'}({\bm p},{\bm p})=\left[\frac{i}{\sqrt{2}}\sigma_2Y_{00}(\hat{{\bm p}})\right]_{ss'}\hat{M}(p),
\label{po}
\end{equation}
where the spin-angular structure is equivalent to the one in the matrix
wave function of the pion (\ref{vppi}).

On substituting expression (\ref{po}) into Hamiltonian (\ref{HH2}), one can find
$$
\hat{\cal H}_\pi=\int\frac{p^2d
p}{(2\pi)^3}2E_p\hat{M}^\dagger(p)\hat{M}(p)-\frac12\int\frac{p^2
dp}{(2\pi)^3}\frac{q^2dq}{(2\pi)^3}
\left\{T^{++}_\pi(p,q)\hat{M}^\dagger(p)M(q)\right.
$$
\begin{equation}
\left.+T^{+-}_\pi(p,q)\hat{M}^\dagger(q)\hat{M}^\dagger(p)
+T^{-+}_\pi(p,q)\hat{M}(p)\hat{M}(q)+T^{--}_\pi(p,q)
\hat{M}^\dagger(q)\hat{M}(p)\right\},
\label{HHpi}
\end{equation}
where the amplitudes $T_\pi^{\pm\pm}(p,q)$ are nothing but combinations of the amplitudes
$v^{\pm\pm}({\bm p},{\bm q})$ and the potential $V({\bm p}-{\bm q})$, integrated in the angle --- see equation (\ref{pia}).

Expression (\ref{HHpi}) is a typical Hamiltonian requiring
diagonalisation through the bosonic Bogoliubov-Valatin transformation of the form
\begin{equation}
\left\{
\begin{array}{l}
\hat{M}(p)=\hat{m}_\pi\vpp_\pi^+(p)+\hat{m}_\pi^\dagger\vpp_\pi^-(p)\\
\hat{M}^\dagger(p)=\hat{m}^\dagger_\pi\vpp_\pi^+(p)+\hat{m}_\pi\vpp_\pi^-(p),
\end{array}
\right.
\label{Mmpi}
\end{equation}
that can be inverted as
\begin{equation}
\left\{
\begin{array}{l}
\hat{m}_\pi=\ds\int\frac{p^2dp}{(2\pi)^3}\left[\hat{M}(p)\vpp_\pi^+(p)-
\hat{M}^\dagger(p)\vpp_\pi^-(p)\right]\\[2mm]
\hat{m}_\pi^\dagger=\ds\int\frac{p^2dp}{(2\pi)^3}\left[\hat{M}^\dagger
(p)\vpp_\pi^+(p)- \hat{M}(p)\vpp_\pi^-(p)\right].
\end{array}
\right.
\label{mMpi}
\end{equation}

The operators $\hat{m}_\pi^\dagger$ and $\hat{m}_\pi$ support a clear physical
interpretation: they create and annihilate the pion in its rest frame.
Then, with the help of the commutator
\begin{equation}
[\hat{M}(p),\;\hat{M}^\dagger
(q)]=\frac{(2\pi)^3}{p^2}\delta(p-q),
\end{equation}
which follows directly from equation (\ref{MMcom}), it is straightforward to find that
\begin{equation}
[\hat{m}_\pi ,\; \hat{m}_\pi^\dagger]=
\int\frac{p^2dp}{(2\pi)^3}\left[\vpp_\pi^{+2}(p)-\vpp_\pi^{-2}(p)\right].
\end{equation}
Therefore, the requirement of the canonical commutation relation between
the bosonic creation and annihilation operators for the pion,
$[\hat{m}_\pi,\hat{m}_\pi^\dagger]=1$, leads to the normalisation condition
(amplitudes $\vpp_\pi^\pm(p)$ are chosen real) of the form
(\ref{norm1}) which is just the standard one for the Bogoliubov amplitudes.
At the same time, the equation which guarantees cancellation
of the anomalous Bogoliubov terms in Hamiltonian (\ref{HHpi}), that is, that
$\langle\Omega|\hat{\cal H}_\pi|\pi\pi\rangle =0$ and $\langle \pi\pi|\hat{\cal H}_\pi|\Omega\rangle=0$ (here $|\Omega\rangle$ is the vacuum
annihilated by the mesonic operators, for example, $\hat{m}_\pi$),
takes the form of the bound-state equation for the amplitudes
$\vpp_\pi^\pm(p)$ --- see equation (\ref{bsp}).

It is important to note that the vacuum $|\Omega\rangle$, annihilated by
the operator $\hat{m}_\pi$, differs from the BCS vacuum
$|0\rangle$ and both vacua are related through a unitary transformation,
$$
\hat{m}_\pi|\Omega\rangle=\hat{m}_\pi U^\dagger|0\rangle
=U^\dagger(U\hat{m}_\pi U^\dagger)|0\rangle \propto
U^\dagger\hat{M}(p)|0\rangle=0.
$$
Since the quark-antiquark pair creation is suppressed in the large-$N_C$ limit,
then the deviation of the operator
$U^\dagger$ from unity demonstrates the same suppression pattern. Similarly, the vacuum energy $E_{\rm vac}'$ in equation (\ref{HH1})
differs from the vacuum energy $E_{\rm vac}$ in the BCS Hamiltonian (\ref{H2diag})
and it contains
contributions from the commutators of the operators $\hat{M}$ and $\hat{M}^\dagger$
(suppressed in the limit $N_C\to\infty$).
Finally, the chiral condensate evaluated in the BCS approximation provides
the leading-order term in the expansion of the exact condensate in the inverse powers
of $N_C$.

Hamiltonian (\ref{HHpi}) diagonalised in the given order in $N_C$ takes the form
\begin{equation}
\hat{\cal H}_\pi=M_\pi \hat{m}_\pi^\dagger \hat{m}_\pi,\quad M_\pi=\langle\pi|\hat{\cal H}_\pi|\pi\rangle,
\label{diagtot}
\end{equation}
where $M_\pi$ is the pion mass, and the omitted (suppressed in $N_C$) terms describe the pion-pion scattering.

\subsubsection{The general case}

Now we diagonalise the full Hamiltonian (\ref{HH2}) in terms of compound mesonic states.
With a trivial generalisation of equations (\ref{Mmpi}) and (\ref{mMpi}),
\begin{equation}
\left\{
\begin{array}{l}
\hat{M}_{ss'}({\bm p},{\bm p})=\ds\sum_n[\hat{m}_{n}\phi^+_{n,ss'}({\bm p})+\hat{m}_{n}^\dagger\phi^{-}_{n,ss'}({\bm p})]\\[2mm]
\hat{M}_{ss'}^\dagger({\bm p},{\bm p})=\ds\sum_n [\hat{m}^\dagger_{n}\phi^{+\dagger}_{n,ss'}({\bm p})+\hat{m}_{n}\phi^{-\dagger}_{n,ss'}({\bm p})]
\end{array}
\right.
\label{Mmgen}
\end{equation}
and
\begin{equation}
\left\{
\begin{array}{l}
\hat{m}_{n}=\ds\int\frac{d^3p}{(2\pi)^3}\Tr\left[\hat{M}({\bm p},{\bm p})\phi^{+\dagger}_{n}({\bm p})-
\hat{M}^\dagger({\bm p},{\bm p})\phi^-_{n}({\bm p})\right]\\
\hat{m}^\dagger_{n}=\ds\int\frac{d^3p}{(2\pi)^3}\Tr\left[\hat{M}^\dagger({\bm p},{\bm p})\phi^{+}_{n}({\bm p})-
\hat{M}({\bm p},{\bm p})\phi^{-\dagger}_{n}({\bm p})\right],
\end{array}
\right.
\label{mMgen}
\end{equation}
it is straightforward to find for the commutators $\hat{m}_n$ and $\hat{m}_m^\dagger$ the following expressions:
\begin{equation}
\begin{array}{l}
[\hat{m}_{n},\;\hat{m}_{m}^\dagger]=\ds
\int\frac{d^3p}{(2\pi)^3}\Tr\left[\phi^{+\dagger}_{n}({\bm p})\phi^+_{m}({\bm p})-
\phi^{-\dagger}_{m}({\bm p})\phi^-_{n}({\bm p})\right],\\[3mm]
[\hat{m}_{n},\;\hat{m}_{m}]=\ds
\int\frac{d^3p}{(2\pi)^3}\Tr\left[\phi^{+\dagger}_{n}({\bm p})\phi^-_{m}({\bm p})-
\phi^{+\dagger}_{m}({\bm p})\phi^-_{n}({\bm p})\right].
\end{array}
\end{equation}
Here the subscripts $n$ and $m$ denote the complete set of quantum numbers describing mesonic states.

A natural requirement that $[\hat{m}_{n},\;\hat{m}_{m}^\dagger]=\delta_{mn}$
and $[\hat{m}_{n},\;\hat{m}_{m}]=0$ leads to the orthogonality condition for the wave functions in the
form
\begin{equation}
\begin{array}{l}
\ds
\int\frac{d^3p}{(2\pi)^3}\Tr\left[\phi^{+\dagger}_{n}({\bm p})\phi^+_{m}({\bm p})-
\phi^{-\dagger}_{m}({\bm p})\phi^-_{n}({\bm p})\right]=\delta_{nm},\\[3mm]
\ds
\int\frac{d^3p}{(2\pi)^3}\Tr\left[\phi^{+\dagger}_{n}({\bm p})\phi^-_{m}({\bm p})-
\phi^{+\dagger}_{m}({\bm p})\phi^-_{n}({\bm p})\right]=0.
\end{array}
\label{normgen}
\end{equation}

It is easy to verify then that representation (\ref{Mmgen}), together with the orthogonality and normalisation condition
(\ref{normgen}), guarantees that the Hamiltonian is diagonal, that is,
\begin{equation}
\hat{\cal H}=\sum_{n}M_{n}m^\dagger_{n}m_{n}+
O\left(\frac{1}{\sqrt{N_C}}\right),
\label{hdgen}
\end{equation}
provided the mesonic wave functions $\phi^{\pm}_{n,s_1s_2}$ obey the system of equations (\ref{Salpeterlev5}) with the eigenvalue $M=M_n$.

In the leading order in $N_C$, Hamiltonian (\ref{hdgen}) describes stable mesons
while the neglected ($N_C$-suppressed) terms include quark exchanges and, therefore, they describe decays and scattering of the
mesons --- see review \cite{Kalashnikova:2001df} where such suppressed terms are
restored for two-dimensional QCD.

In practical applications one diagonalises the Hamiltonian in the $J^{PC}$ basis. To this end one should have in mind that, 
while the Hamiltonian commutes with the
sum ${\bm J}=\veS+\veL$, it does not commute either with the operator
of the total
quark spin ${\bm S}$ or with the operator of the
angular momentum ${\bm L}$ separately.

The case of spin-singlets, $P=(-1)^{J+1}$, $C=(-1)^{J}$, is trivial in this respect as the wave functions are given by the expression
\begin{equation}
\phi_{n,s_1s_2}^{\pm}({\bm p})=\left[\frac{i}{\sqrt{2}}\sigma_2\right]_{s_1s_2}Y_{Jm}(\hat{{\bm p}})\vpp_n^\pm(p),
\label{singlet}
\end{equation}
where $Y_{Jm}(\hat{{\bm p}})$ is the spherical harmonic with the momentum $J$ and magnetic quantum number $m$.
Spin-triplets with $J=L$, $P=(-1)^{J+1}$, $C=(-1)^{J+1}$ are described by
\begin{eqnarray}
\phi_{n,s_1s_2}^{+}({\bm p})=\left[({\bm \sigma} {\bm Y}_{JJm}(\hat{{\bm p}}))\frac{i}{\sqrt{2}}\sigma_2\right]_{s_1s_2}\vpp_n^+(p),\nonumber\\[-2mm]
\label{triplet2}\\[-2mm]
\phi_{n,s_1s_2}^{-}({\bm p})=\left[\frac{i}{\sqrt{2}}\sigma_2({\bm \sigma} {\bm Y}_{JJm}(\hat{{\bm p}}))\right]_{s_1s_2}\vpp_n^-(p),\nonumber
\end{eqnarray}
where ${\bm Y}_{Jlm}(\hat{{\bm p}})$ is the spherical vector with the total momentum $J$, orbital momentum $l$, and magnetic quantum number $m$.

The case of $L=J\pm 1$, $P=(-1)^{J}$, $C=(-1)^{J}$ is more elaborated, as it has to be described by four scalar amplitudes $\vpp_{J \pm 1,n}^{\pm}(p)$
(with an obvious exception of the $0^{++}$ scalar meson with $J=0$ and $l=1$):
\begin{eqnarray}
&&\phi_{n,s_1s_2}^{+}({\bm p})=\left[({\bm \sigma} {\bm Y}_{JJ-1m}(\hat{{\bm p}}))\frac{i}{\sqrt{2}}\sigma_2\right]_{s_1s_2}
\vpp_{J-1,n}^+(p)+\left[({\bm \sigma} {\bm Y}_{JJ+1m}(\hat{{\bm p}}))\frac{i}{\sqrt{2}}\sigma_2\right]_{s_1s_2}
\vpp_{J+1,n}^+(p),\nonumber\\[-2mm]
\label{triplet4}\\[-2mm]
&&\phi_{n,s_1s_2}^{-}({\bm p})=\left[\frac{i}{\sqrt{2}}\sigma_2({\bm \sigma} {\bm Y}_{JJ-1m}(\hat{{\bm p}}))\right]_{s_1s_2}
\vpp_{J-1,n}^-(p)+\left[\frac{i}{\sqrt{2}}\sigma_2({\bm \sigma} {\bm Y}_{JJ+1m}(\hat{{\bm p}}))\right]_{s_1s_2}
\vpp_{J+1,n}^-(p),\nonumber
\end{eqnarray}
and the interaction in the system of equations (\ref{Salpeterlev5}) mixes all four amplitudes from (\ref{triplet4}) thus giving rise,
after projection onto spin-angular states, to four coupled equations for the scalar amplitudes $\vpp_{J \pm 1,n}^{\pm}(p)$.
This can be exemplified by the $\rho$-meson: its quantum numbers $1^{--}$ correspond to two terms --- $^3S_1$ and $^3D_1$ --- so
that the $\rho$-meson has to be described by four amplitudes,
rather than the only two needed for, say, a $0^{-+}$ meson. It is instructive to notice that system (\ref{Salpeterlev5}) would
describe not only the $\rho$-meson, but also a heavier vector meson which is defined by the orthogonal
combination of the $S$ and $D$ waves. Thus, the doubling of the number of scalar functions
is nothing but a mere consequence of the situation in which
the wave function of the $\rho$-meson is ``entangled'' with the wave function
of its heavier partner.

Wave functions (\ref{singlet})-(\ref{triplet4}) are spelled out in the $LS$ basis, however,
the problem of the adequate choice of the basis cannot be solved in general terms since the mixing pattern of different
partial waves with the same quantum numbers is a dynamical problem. The $LS$ basis is quite suitable for heavy quarkonia
where partial-wave mixing can be treated as a relativistic correction. Another notable exception
is provided by the regime of the effective restoration of chiral symmetry in the
spectrum of excited mesons (see Section~\ref{chirrest} below)
filling chiral multiplets and, as a result, possessing the wave functions strictly fixed by chiral symmetry --- see
review \cite{Glozman:2007ek} and references therein.
In paper \cite{Glozman:2007at} a chiral basis is discussed in detail which provides a much
more convenient framework for studies of the spectrum of mesons in the regime of the
effectively restored chiral symmetry. However, it has to be noticed that
this chiral basis {\it per se} cannot solve the problem of the dynamical
mixing of different waves; it only refers to particular
combinations of such waves corresponding to the multiplets with the restored chiral symmetry.

One more final remark is in order here. The Bethe-Salpeter equation (\ref{Salpeterlev5})
derived above describes the spectrum of the genuine quark-antiquark
states. In the limit $N_C\to\infty$, that is, in the limit inherent to the
model under study, such states possess well-known properties.
In particular, as the number of colours grows, the mass of a genuine
$\bar{q}q$ state remains nearly constant while its width tends to zero
since the effects of the light-quark pair creation from the vacuum are suppressed
in this limit. As it can be seen from equation (\ref{hdgen}),
the leading suppressed terms describing the amplitudes of the two-body decays of the mesons
behave as $O(1/\sqrt{N_C})$ thus yielding for the width of the mesons the well-known typical behaviour $1/N_C$. This property allows one to tell genuine
quarkonia from dynamically
generated objects, for example, from the scalar state $f_0(500)$.
Thus, in \cite{Pelaez:2006nj}, in the framework of the unitarised
chiral perturbation theory, it was demonstrated that, in the limit $N_C\to\infty$,
the poles which describe the genuine quarkonia indeed behave as it was
explained above. In the meantime, the pole responsible for the $f_0(500)$ (in the cited paper an obsolete notation $f_0(600)$ is used)
demonstrates a severely different behaviour: its real part (the mass) grows with $N_C$ while the $N_C$-dependence on its width is rather nontrivial
and it does not follow the law $1/N_C$. This observation confirms the common belief that the $f_0(500)$ is a result of the strong interaction between
mesons in the final state, so that to describe this state one needs to proceed beyond the formalism used above.

\section{The Lorentz nature of confinement}\label{DShl}

One of the important problems of the phenomenology of strong interactions is related to the Lorentz nature of the confining interaction.
For example, spin-dependent interactions in the quark-antiquark system are very sensitive to the relations between the potentials added to the mass
(scalar interaction) and the potentials added to the energy or to the momentum (vector interaction) --- see, for instance, the key papers
\cite{Eichten:1980mw,Gromes:1984ma} as well as a series of later works, like
\cite{Brambilla:1997kz,Simonov:1997et,Kalashnikova:1997bu,Simonov:2002er,Simonov:2000hd}
and others.
Phenomenology of heavy quarkonia and lattice calculations \cite{Koma:2006fw} are better
compatible with the spin-dependent potentials which
stem from the scalar confinement. Meanwhile, in a theory with scalar confinement,
chiral symmetry would be broken explicitly and that would contradict
the idea
of its spontaneous breaking (see \cite{Biernat:2013fka} for the discussion
of a possibility of the co-existence of the scalar confinement and the
spontaneous breaking of chiral symmetry). In order to investigate this problem
we consider a heavy-light quarkonium with the heavy quark treated as
a static centre. This will allow us to study the Lorentz nature of the confining potential
and some properties of the quark-antiquark mesons
avoiding unnecessary technical complications. The spectrum of the heavy-light system
should be described by the system of equations
(\ref{Salpeterlev5}) generalised to the case of two quark flavours. Later,
the limit of the static antiquark will be taken explicitly in equation
(\ref{Salpeterlev5}). But first, it would be helpful to stick to a
different approach to the heavy-light quarkonium based on the
Vacuum Background Correlators Method
(see review \cite{DiGiacomo:2000irz} and references therein) and to investigate
the Lorentz nature of confinement in such a system
\cite{Brambilla:1997kz,Simonov:1997et,Kalashnikova:1997bu,Simonov:2002er,Simonov:2000hd}.

The motion of the light quark in the field of the static antiquark should be described
by a single-particle Dirac-like equation with the interaction
with the static centre given by an effective potential. The Lorentz nature of
this potential can be investigated this way.

We start from the Green's function of such a heavy-light quarkonium $S_{q\bar{Q}}$
taken in the form
\cite{Simonov:1997et,Simonov:2002er,Simonov:2000hd} (until stated otherwise, all expressions are written in Euclidean space)
\begin{equation}
S_{q\bar Q}(x,y)=\frac{1}{N_C}\int D{\psi}D{\psi^\dagger}DA_{\mu}\exp{\left\{-\frac14\int d^4x F_{\mu\nu}^{a2}-\int d^4x
\psi^\dagger(-i\hat \partial -im -\hat A)\psi \right\}}
\label{SqQ}
\end{equation}
$$
\times\psi^\dagger(x) S_{\bar Q} (x,y|A)\psi(y),
$$
where $S_{\bar Q} (x,y|A)$ is the propagator of the static antiquark placed at the origin.
To proceed it is convenient to stick to a particular
version of the Fock-Schwinger gauge allowing
to express the vector potential through the field tensor \cite{Balitsky:1985iw},
\begin{equation}
{\bm x}{\bm A}(x_4,{\bm x})=0,\quad A_4(x_4,{\bm 0})=0.
\label{5}
\end{equation}
This particular gauge condition proves convenient because the gluonic field vanishes
at the trajectory of the static antiquark, so that its Green's function
takes a particularly simple form,
\begin{equation}
S_{\bar Q}(x,y|A)=S_{\bar Q}(x,y)= i\frac{1-\gamma_4}{2} \theta (x_4-y_4)e^{-M(x_4-y_4)}+i\frac{1+\gamma_4}{2}\theta(y_4-x_4)e^{-M(y_4-x_4)},
\label{SQ}
\end{equation}
where $\theta$ is the step-like function.

It is easy to notice then that equation (\ref{SqQ}) takes the form
\begin{equation}
S_{q\bar Q}(x,y)=\frac{1}{N_C}\int D{\psi}D{\psi^\dagger}\exp{\left\{-\int d^4x L_{\rm
eff}(\psi,\psi^\dagger)\right\}}\psi^\dagger(x) S_{\bar Q} (x,y)\psi(y),
\end{equation}
that is, the antiquark is completely decoupled from the system and the dynamics of the light quark is defined by the effective Lagrangian
$L_{\rm eff}(\psi,\psi^\dagger)$, such that
$$
\int d^4x L_{\rm eff}(\psi,\psi^\dagger)=\int d^4x\psi^\dagger_{\alpha}(x)(-i\hat\partial -im)\psi^{\alpha}(x)+
\int d^4x\psi^\dagger_{\alpha}(x)\gamma_{\mu}\psi^{\beta}(x)\langle\lan {A_{\mu}}^{\alpha}_{\beta}\rangle\ran
$$
\begin{equation}
+\frac{1}{2}\int d^4x_1d^4x_2\psi^\dagger_{\alpha_1}(x_1)\gamma_{\mu_1}\psi^{\beta_1}(x_1)
\psi^\dagger_{\alpha_2}(x_2)\gamma_{\mu_2}\psi^{\beta_2}(x_2)\langle\lan
{A_{\mu_1}}_{\beta_1}^{\alpha_1}(x_1){A_{\mu_2}}_{\beta_2}^{\alpha_2}(x_2)\rangle\ran+
\ldots,
\label{8}
\end{equation}
where $\alpha$ and $\beta$ are the colour indices in the fundamental representation,
and the gluonic field enters in the form of the
irreducible correlators $\langle\lan {A_{\mu_1}}_{\beta_1}^{\alpha_1}(x_1)\ldots {A_{\mu_n}}_{\beta_n}^{\alpha_n}(x_n)\rangle\ran$
of all orders, as was already mentioned in the Introduction.
Retaining only the first nonvanishing, that is, the Gaussian, correlator is an approximation (here it is taken into account that
$\langle\lan {A_{\mu}}^{\alpha}_{\beta}\rangle\ran=\langle {A_{\mu}}^{\alpha}_{\beta}\rangle=0$).
Discussions on the justification for this approximation can be found,
for example, in review paper \cite{DiGiacomo:2000irz}. It is also important to mention here the results of the
lattice calculations \cite{Bali:2000un} and their relation to the Casimir
scaling in QCD traced in papers
\cite{Shevchenko:2000du,Shevchenko:2003ks}.

Then, defining the interaction kernel of the two quark currents through the bilocal
correlator of the gluonic fields in the vacuum,
\begin{equation}
\lan\langle{A_{\mu}}_{\beta}^{\alpha}(x){A_{\nu}}_{\delta}^{\gamma}(y)\rangle\ran=
\langle{A_{\mu}}_{\beta}^{\alpha}(x){A_{\nu}}_{\delta}^{\gamma}(y)\rangle
\equiv
2(\lambda_a)_{\beta}^{\alpha}(\lambda_a)_{\delta}^{\gamma}K_{\mu\nu}(x,y),
\end{equation}
making use of the Fierz identity $(\lambda_a)_{\beta}^{\alpha}(\lambda_a)_{\delta}^{\gamma}=\frac12
\delta_{\delta}^{\alpha}\delta_{\beta}^{\gamma}-\frac{1}{2N_C}\delta_{\beta}^{\alpha}\delta_{\delta}^{\gamma}$
and taking the limit of the infinite number of colours, we can
write
\begin{equation}
L_{\rm eff}(\psi,\psi^\dagger)=\psi^\dagger_{\alpha}(x)(-i\hat\partial -im)\psi^{\alpha}(x)
+\frac{1}{2}\int d^4y\;
\psi^\dagger_{\alpha}(x)\gamma_{\mu}\psi^{\beta}(x)\psi^\dagger_{\beta}(y)\gamma_{\nu}\psi^{\alpha}(y)
K_{\mu\nu}(x,y),
\end{equation}
that entails the Schwinger-Dyson equation for the light quark in the form
\cite{Simonov:1997et,Simonov:2002er,Simonov:2000hd}
\begin{eqnarray}
&\ds (-i\hat{\partial}_x-im)S(x,y)-i\int d^4zM(x,z)S(z,y)=\delta^{(4)}(x-y),&\nonumber\\[-3mm]
\label{DS4}\\[-3mm]
&\ds -iM(x,z)=K_{\mu\nu}(x,z)\gamma_{\mu}S(x,z)\gamma_{\nu}.&\nonumber
\end{eqnarray}
Here $S(x,y)=\frac{1}{N_C}\langle\psi^{\beta}(x)\psi^\dagger_{\beta}(y)\rangle$. It is instructive to notice that, although (\ref{DS4})
looks like a single-particle equation, it nevertheless contains the information about the heavy antiquark since the kernel
$K_{\mu\nu}$ is evaluated in gauge (\ref{5}) which is closely related to the static antiquark placed at the origin.

Making use of the aforementioned property of the gauge condition (\ref{5}), we can express the vector potential of the gluonic field through the field
tensor \cite{Balitsky:1985iw},
\begin{eqnarray}
\ds A^a_4(x_4,\vex)&=&\int_0^1 d\alpha x_i F^a_{i4}(x_4,\alpha{\bm x})\nonumber\\[-3mm]
\label{AF}\\[-3mm]
\ds A^a_i(x_4,\vex)&=&\int_0^1\alpha x_k F^a_{ki}(x_4,\alpha{\bm x}) d\alpha,\quad i=1,2,3,\nonumber
\end{eqnarray}
and, therefore, the interaction kernel $K_{\mu\nu}$ can be expressed through the field correlator
$\langle F^a_{\mu\nu}(x)F^b_{\lambda\rho}(y)\rangle$. Then, with the help of the Vacuum Background Correlators Method (see review
\cite{DiGiacomo:2000irz}) and retaining only the confining part of the interaction, one can arrive at the kernel
$K_{\mu\nu}(x,y)=K_{\mu\nu}(x_4-y_4,\vx,\vy)$ ($\tau=x_4-y_4$) in the form (for a detailed derivation see papers
\cite{Simonov:1997et,Kalashnikova:1997bu,Nefediev:2007pc})
\begin{equation}
\left\{
\begin{array}{l}
K_{44}(\tau,\vx,\vy)=\ds(\vx\vy)\int_0^1d\alpha\int_0^1 d\beta D(\tau,|\alpha\vx-\beta\vy|),\\[1mm]
K_{i4}(\tau,\vx,\vy)=\ds K_{4i}(\tau,\vx,\vy)=0,\\[1mm]
K_{ik}(\tau,\vx,\vy)=\ds((\vx\vy)\delta_{ik}-y_ix_k)\int_0^1\alpha d\alpha\int_0^1 \beta d\beta
D(\tau,|\alpha\vx-\beta\vy|),
\end{array}
\right.
\label{kern1}
\end{equation}
where the function $D(\tau,\lambda)$ decreases in all directions and describes the profile of the bilocal correlator of the nonperturbative
gluonic fields in the QCD vacuum --- see review \cite{DiGiacomo:2000irz}.

Equation (\ref{DS4}) is essentially nonlinear. It can, however, be linearised if the free Green's function is substituted,
$S(x,z)\to S_0(x,z)$, in the mass operator $M(x,z)$. Such an approach, appropriate in the heavy-quark limit, was used in papers
\cite{Brambilla:1997kz,Kalashnikova:1997bu} to derive the effective potentials and the spin-dependent corrections to it.
The leading correction due to the proper dynamics of the string was found in \cite{Nefediev:2003mx}.
Meanwhile, the above linearisation is only possible if $mT_g\gg 1$ \cite{Kalashnikova:1997bu},
where $m$ is the mass of the quark and $T_g$ is the correlation length of the vacuum which governs the decrease of the correlator $D$
(see papers \cite{Badalian:2008kc,Badalian:2008sv} and references therein for the extraction of the correlation length from the
interquark potentials). In the opposite limit of $mT_g\ll 1$ such a linearisation procedure is misleading and it results in a divergent series
\cite{Kalashnikova:1997bu}, so that, in this limit, the nonlinear equation (\ref{DS4}) has to be studied in the full form.

Once the question discussed in this chapter is related
to the spontaneous breaking of chiral symmetry, we must have a small quark mass and,
therefore, it is exactly the nonpotential regime with $mT_g\ll 1$
which is adequate for the situation. Thus, we need to use a different simplification
scheme for the equation. To begin with, we neglect the
spatial part of kernel (\ref{kern1}), $K_{ik}$, which does not affect the
qualitative result.
Then we take the Fourier transform of $K_{44}$ in time,
\begin{equation}
K_{44}(\omega,\vx,\vy)\equiv K(\omega,\vx,\vy)= K(\vx,\vy)=
(\vx\vy)\int_0^1d\alpha\int_0^1 d\beta \int_{-\infty}^{\infty} d\tau D(\tau,|\alpha\vx-\beta\vy|).
\label{kern2}
\end{equation}

To proceed further we notice that the vacuum correlation length extracted from the
lattice data is very small compared to the other scales of the
problem ($T_g\lesssim 0.1$~fm \cite{Badalian:2008kc,Badalian:2008sv}).
It is therefore natural to take the so-called string limit
$T_g\to 0$ which, given the normalisation condition,
\begin{equation}
\sigma=2\int_0^\infty d\nu\int_0^\infty d\lambda D(\nu,\lambda),
\label{sigma}
\end{equation}
where the parameter $\sigma$ defines the tension of the QCD string
\cite{DiGiacomo:2000irz}, yields for the correlator a
$\delta$-function-like profile,
\begin{equation}
D(\tau,\ld)=2\sigma\delta(\tau)\delta(\lambda),
\end{equation}
so that
\begin{equation}
K(\vx,\vy)=2\sigma(\vx\vy)\int_0^1d\alpha\int_0^1 d\beta\;\delta(|\alpha\vx-\beta\vy|).
\label{K4}
\end{equation}

The fact that kernel (\ref{K4}) does not vanish only for
collinear vectors $\vx$ and $\vy$ is a consequence of
the infinitely thin
(in the limit $T_g\to 0$) string connecting the quark and the antiquark.
Then, the integral in (\ref{K4}) can be taken exactly to yield
\begin{equation}
K(\vx,\vy)=2\sigma {\rm min}(|\vx|,|\vy|)=
\left\{
\begin{array}{ll}
\sigma (|\vx|+|\vy|-|\vx-\vy|),&\vx\parallel\vy\\
0,&\vx\nparallel\vy.
\end{array}
\right.
\label{K4v4}
\end{equation}
The expression arrived can be viewed as the three-dimensional
generalisation of the one-dimensional kernel derived in
\cite{Kalashnikova:2001df} for the 't~Hooft model. Notice that the condition of collinearity for the vectors $\vx$ and $\vy$
is trivial in case of only one spatial dimension, however, for
kernel (\ref{K4v4}) it leads to technical complications not important for the
mechanisms of the spontaneous breaking of chiral symmetry. Therefore,
it is natural to relax this condition and to consider, for any $\vx$ and $\vy$, the interaction kernel in the form
\begin{equation}
K(\vx,\vy)=\sigma(|\vx|+|\vy|-|\vx-\vy|).
\label{K4v3}
\end{equation}
This kernel possesses a number of attractive features, such as
\begin{itemize}
\item it allows one to pass over trivially from Euclidean space to Minkowski space --- from now on only Minkowski space is considered;
\item it admits a simple physical interpretation: the part $-\sigma|\vx-\vy|$
describes the self-interaction of the light quark
while the term $\sigma(|\vx|+|\vy|)$ is responsible for the interaction of the quark
with the static antiquark. The fact that both interactions are
encoded in the same kernel is a consequence of gauge condition (\ref{5})
which results in the static antiquark decoupling from the system. Then,
once the gauge condition violates translational invariance, then kernel (\ref{K4v3})
does not demonstrate such an invariance either;
\item it admits a natural generalisation to an arbitrary profile of the interquark
interaction potential $V(r)$, so that the generic form of the kernel reads
\begin{equation}
K(\vx,\vy)=V(|\vx|)+V(|\vy|)-V(|\vx-\vy|);
\end{equation}
\item it establishes a natural relation between the Vacuum Background Correlators Method
and the Generalised Nambu--Jona-Lasinio model, since
from now on any equation can be derived with the help of the either approach of the two.
\end{itemize}

Although the above consideration cannot be treated as a true derivation of the Generalised
Nambu--Jona-Lasinio model from QCD, it nevertheless
allows one to establish a close relation between the fundamental theory and
this model. In the literature, one can find a similar derivation of the Hamiltonian in the form of equation
(\ref{HGNJL}) in the Gaussian approximation for the QCD vacuum (see \cite{Bicudo:1998bz}) 
as well as attempts of a more rigorous derivation of the classic Nambu--Jona-Lasinio model from QCD --- see, in particular, 
papers \cite{Bijnens:1992uz,Arbuzov:2006ia}.

We are now in a position to return to the equation for the heavy-light quarkonium.
In particular, Schwinger-Dyson equation (\ref{DS4})
for the light quark can be written in the form
\begin{equation}
\left(-i\gamma_0\frac{\partial}{\partial t}+i\vg\frac{\partial}{\partial \vx}-m\right)
S(t,\vx,\vy)-\int d^3z M(\vx,\vz)S(t,\vz,\vy)=\delta(t)\delta^{(3)}(\vx-\vy),
\label{DS5}
\end{equation}
where
\begin{equation}
M(\vx,\vz)=-\frac{i}{2}K(\vx,\vz)\gamma_0\Lambda(\vx,\vz),\quad\Lambda(\vx,\vz)=2i\int\frac{d\omega}{2\pi}S(\omega,\vx,
\vz)\gamma_0.
\label{Mop01}
\end{equation}
The Lorentz nature of the interaction described by the kernel $K$ depends on
the matrix structure of the mass operator
$M(\vx,\vy)$. Thus, if $M(\vx,\vy)$ acquires a contribution proportional to
the unity matrix, it gives rise to the interaction added to
the mass, that is, to scalar confinement. For a detailed study of
this problem we make use of the natural separation of kernel
(\ref{K4v3}) on the local and nonlocal part. As was explained above,
the local part of the kernel is responsible for the light quark
self-interaction
and, therefore, it defines ``dressing'' of the quark. Indeed, it is easy to see that, omitting the nonlocal contribution $\sigma(|\vx|+|\vy|)$,
one can proceed from equation (\ref{DS4}) to the Dyson equation
\begin{equation}
(\gamma_0 p_0-\vg\vp-m-\Sigma(\vp))S(p_0,\vp)=1,
\label{eqS0}
\end{equation}
where the mass operator for the light quark $\Sigma(\vp)$ takes the form
\begin{equation}
\Sigma({\bm p})=-i\int\frac{d^4k}{(2\pi)^4}V({\bm p}-{\bm k})\gamma_0 S(k_0,{\bm k})\gamma_0,
\label{Sigma01}
\end{equation}
and, due to the instantaneous nature of the interaction, it does not depend on the
energy. It is easy to verify that expression (\ref{Sigma01})
for the mass operator reproduces equation (\ref{Sigma01last}) which was derived
above through the summation of the Dyson series for the dressed-quark
propagator --- see Fig.~\ref{diagrams}.

Once the Green's function $S(p_0,\vp)$ is defined from equation (\ref{eqS0})
then its substitution to (\ref{Sigma01}) results in the
self-consistence condition which is nothing but the mass-gap equation (\ref{Sigma03}) in the Generalised Nambu--Jona-Lasinio model
\cite{Amer:1983qa,LeYaouanc:1983it,LeYaouanc:1983huv,LeYaouanc:1984ntu,Bicudo:1989sh,Bicudo:1989si,Bicudo:1989sj,Bicudo:1993yh,Bicudo:1998mc}
which can be conveniently written as equation (\ref{mge}) for the chiral angle $\vpp_p$.

For the function $\Lambda(\vp,\vq)$ parametrised through the chiral angle, which is
the double Fourier transform of the quantity $\Lambda(\vx,\vy)$
introduced in equation (\ref{Mop01}), it is straightforward to find
\begin{equation}
\Lambda(\vp,\vq)=2i\int\frac{d\omega}{2\pi}S(\omega,\vp,\vq)\gamma_0=(2\pi)^3\delta^{(3)}(\vp-\vq)U_p,
\label{iS4}
\end{equation}
where
\begin{equation}
U_p=\beta\sin\vpp_p+({\bm \alpha}\hat{\vp})\cos\vpp_p,\quad \beta=\gamma_0,\quad{\bm \alpha}=\gamma_0\vg.
\label{Up4}
\end{equation}

Let us revisit equation (\ref{DS5}) and rewrite it in the form of the bound-state
equation for the wave function $\tilde{\Psi}(\vex)$,
\begin{equation}
({\bm \alpha}\hat{\vp}+\beta m)\tilde{\Psi}(\vx)+\beta\int d^3z M(\vx,\vz)\tilde{\Psi}(\vz)=E\tilde{\Psi}(\vx),
\label{DS6}
\end{equation}
where now both local and nonlocal parts of the kernel are taken into account.
Then, passing over to the momentum space and employing the
mass-gap equation in the form
\begin{equation}
E_pU_p={\bm \alpha}\vp+\beta
m+\frac12\int\frac{d^3k}{(2\pi)^3} V(\vp-\vk)U_k,
\end{equation}
one can write equation (\ref{DS6}) as
\begin{equation}
E_pU_p\tilde{\Psi}(\vp)+\frac12\int\frac{d^3k}{(2\pi)^3}V(\vp-\vk)[U_p+U_k]\tilde{\Psi}(\vk)=E\tilde{\Psi}(\vp).
\label{Se10}
\end{equation}

Equation (\ref{Se10}) admits an exact Foldy-Wouthuysen transformation\footnote{This possibility is closely related to the instantaneous
nature of the interaction and to the presence of an infinitely heavy particle in the system \cite{Bars:1977ud}.} \cite{Kalashnikova:2005tr}
\begin{equation}
\tilde{\Psi}({\bm p})=T_p\Psi({\bm p}),\quad\Psi({\bm p})={\psi({\bm p})\choose 0},\quad T_p=\exp{\left[-\frac12({\bm
\gamma}\hat{{\bm p}})\left(\frac{\pi}{2}-\vpp_p\right)\right]},
\label{Tpop}
\end{equation}
which brings it to the Shr{\"o}dinger-like equation for the two-component spinor
for the light quark $\psi(\vp)$,
\begin{equation}
E_p\psi({\bm p})+\int\frac{d^3k}{(2\pi)^3}V(\vp-\vk)\left[C_pC_k+
(\vesig\hat{\vp})(\vesig\hat{\vk})S_pS_k\right]\psi({\bm k})=E\psi({\bm p}),
\label{FW4}
\end{equation}
where $C_p$ and $S_p$ are defined in (\ref{SandC}).

Before we study in detail the properties of equation (\ref{FW4}), let us derive it
directly in the framework of the Generalised Nambu--Jona-Lasinio model.
First of all, we notice that the bound-state equation for the quark-antiquark
system (\ref{PHI}) is symmetric with respect to the change
\begin{equation}
\{M_{n},\varphi_{n}^\pm({\bf p})\}\leftrightarrow\{-M_{n},\varphi_{n}^\mp({\bm p})\}.
\label{sym1}
\end{equation}

As was explained in Subsec.~\ref{subsec:pion}, the two components of the wave function, $\varphi_n^+$
and $\varphi_n^-$, describe the forward and backward
in time motion of the quark-antiquark pair in the meson and, what is more, because of the instantaneous form of the interaction kernel (\ref{KK}), the quark and the antiquark can only move forth and 
back in time in unison.
Therefore, once the static antiquark can never move back in time,
the other quark is forced to do the same.
Thus one expects that, in the limit of the static antiquark, system (\ref{PHI}) splits into two disentangled equations.

Indeed, equation (\ref{GenericSal}) is generalised to the heavy-light system as
\begin{equation}
\chi({\bm p};M)=-i\int\frac{d^4k}{(2\pi)^4}V({\bm p}-{\bm k})\; \gamma_0
S_q({\bm k},k_0+M/2)\chi({\bm k};M)S_{\bar Q}({\bm k},k_0-M/2)\gamma_0,
\label{GenericSalHL}
\end{equation}
where, similarly to equation (\ref{Feynman}),
\begin{equation}
S_q(p_0,{\bm p})=\frac{\Lambda^{+}({\bm p})\gamma_0}{p_0-E_p+i0}+\frac{{\Lambda^{-}}({\bm
p})\gamma_0}{p_0+E_p-i0},
\label{FeynmanHL}
\end{equation}
\begin{equation}
\Lambda^\pm({\bm p})=T_pP_\pm T_p^\dagger,\quad P_\pm=\frac{1\pm\gamma_0}{2},
\end{equation}
while the chiral angle for the static antiquark is simply
$\vpp_{\bar Q}(p)\equiv\frac{\pi}{2}$, so that the positive- and negative-energy projectors
take a simpler form and so does the Green's function of the antiquark,
\begin{equation}
S_{\bar Q}(p_0,{\bm p})=\frac{P_+\gamma_0}{p_0-m_{\bar Q}+i0}+ \frac{P_-\gamma_0}{p_0+m_{\bar Q}-i0}.
\end{equation}

Similarly to the generic case (see equation (\ref{Psiqq})), it proves convenient
to define the matrix wave function,
\begin{equation}
\tilde{\phi}({\bm p})=\int\frac{dp_0}{2\pi}S_q({\bm p},p_0+M/2)\chi({\bm p};M)S_{\bar Q}({\bm p},p_0-M/2),
\end{equation}
which is subject to the Foldy-Wouthuysen transformation with the help of
the operator $T_p$ (see definition (\ref{Tpop})) from the left
(for the light quark) and with the help of the
operator $T_p(\vpp_p\equiv\pi/2)=\hat{1}$ from the right (for the static antiquark),
\begin{equation}
\tilde{\phi}({\bm p})=T_p\phi({\bm p})\hat{1}.
\label{Fop}
\end{equation}
Then it is easy to arrive at the following equation:
\begin{equation}
(E-E_p)\phi({\bm p})=P_+\left[\int\frac{d^3k}{(2\pi)^3}V({\bm p}-{\bm k})T_p^\dagger
T_k\phi({\bm k})\right]P_-
\label{psip},
\end{equation}
where $E$ is the excess of the energy over the mass of the static antiquark,
$E=M-m_{\bar Q}$.
The form of the solution of equation (\ref{psip}) follows from the projectors on
the right-hand side (r.h.s),
\begin{equation}
\phi({\bm p})= \left(
\begin{array}{cc}
0&\psi({\bm p})\\
0&0
\end{array}
\right)={\psi({\bm p})\choose 0}\otimes(0\;1)=\Psi(\vp)\otimes\Psi^T_{\bar{Q}}(\vp),
\label{qbQ}
\end{equation}
where the r.h.s. is written in the form of the tensor product of the components describing the light (see equation (\ref{Tpop})) and
the heavy degree of freedom. Substituting the explicit form of the operators
$T_p$ and $T_k$ into (\ref{psip}), its is easy to re-arrive at equation
(\ref{FW4}).

Due to symmetry (\ref{sym1}) of system (\ref{PHI}), the solution
for the meson with the energy $M_n=-m_{\bar{Q}}-E_n$ can be obtained with the
help of the same (inverse) Foldy-Wouthuysen transformation (\ref{Fop}), now applied to the wave function $(0,\psi({\bm p}))^T$.
As a result, one can reproduce equation (\ref{DS6}) with the propagator given by \cite{Kalashnikova:2005tr}
\begin{equation}
S(\omega,\vp,\vk)=\sum_{E_n>0}\frac{\tilde{\Psi}_n(\vp)\tilde{\Psi}^\dagger_n(\vk)\gamma_0}
{\omega-E_n+i0}+\sum_{E_n<0}\frac{\tilde{\Psi}_n(\vp)\tilde{\Psi}^\dagger_n(\vk)\gamma_0}
{\omega+E_n-i0},
\end{equation}
while for the quantity $\Lambda(\vp,\vk)$ result (\ref{iS4}) is reproduced with
\begin{equation}
U_p=T_p\gamma_0T_p^\dagger.
\end{equation}
Equations (\ref{DS5}) and (\ref{FW4}) allow one to answer the
question on the Lorentz nature of the confining interaction in the heavy-light
quarkonium. For the low-lying states with the small relative momentum between
the quarks, the chiral angle $\vpp_p$ takes values close to
$\pi/2$ (see Fig.~\ref{figvp}). Then, in the limit $\vpp_p=\frac{\pi}{2}$, it is easy
to find that $C_p=1$, $S_p=0$, so that it is straightforward
to pass over to the coordinate space in equation (\ref{FW4}), and the interaction
reduces to the linear potential $\sigma r$.
If, in addition, the kinetic term $E_p$ is substituted by
the energy of the free particle,\footnote{This procedure is definitely
ill-defined for the chiral pion, however for the other mesonic states it provides a
rough but rather adequate approximation.}
then the resulting equation reproduces the Salpeter equation,
\begin{equation}
[\sqrt{\vp^2+m^2}+\sigma r]\psi=E\psi,
\label{Salp}
\end{equation}
which is commonly used in the literature in regard to the hadronic spectroscopy
(see, for example, \cite{Allen:2003wz,Allen:2004pt}).

On the other hand, for $\vpp_p=\frac{\pi}{2}$, one has $U_p=\gamma_0$ and, therefore,
\begin{equation}
\Lambda(\vx,\vy)=\gamma_0\delta^{(3)}(\vx-\vy),\quad
M(\vx,\vy)=\sigma|\vx|\delta^{(3)}(\vx-\vy),
\end{equation}
so that the entire potential $\sigma |\vex|$ in equation (\ref{DS5}) is added to the
mass, that is, the interquark interaction is purely scalar.
It is important to notice that this scalar has essentially dynamical origins and
it appears entirely due to the chiral angle deviation from the trivial
solution which, in turn, is closely related to the
effect of chiral symmetry breaking in the vacuum.

In the opposite limit of large interquark momenta, when the chiral angle decreases and tends to zero, the contribution of the scalar interaction
also decreases while, on the contrary, the contribution of the (spatial) vectorial interaction increases. This regime is realised for highly excited
states in the spectrum of hadrons --- see a detailed discussion of this problem in Sect.~\ref{chirrest}. It has to be noticed that the matrix
$\Lambda(\vp,\vk)$ does not contain contributions proportional to the unity matrix which
could have brought about the temporal component of the
rising-with-distance vectorial interaction and 
which would, therefore, be potentially dangerous from the point of view of the Klein paradox.

In short, we used the heavy-light quark-antiquark system to demonstrate, at
the microscopic level, the emergence of the effective scalar
interquark interaction as a result of the phenomenon of the spontaneous breaking
of chiral symmetry in the vacuum. Besides that we traced the
connection between the Generalised Nambu--Jona-Lasinio model and QCD in
the Gaussian approximation for the gluonic fields in the vacuum.

\section{Effective chiral symmetry restoration in the spectrum of hadrons}\label{chirrest}

\subsection{Introductory comments}\label{subsec:gencom}

In the previous chapters, the Generalised Nambu--Jona-Lasinio model was used to address
microscopically the phenomenon of the spontaneous
breaking of chiral symmetry in the vacuum. Besides, the properties of the chiral pion --- the lowest state in the spectrum of hadrons which
also plays the role of the pseudo Goldstone boson --- were described in detail.
Meanwhile, there are good reasons to expect that the effects of
the spontaneously broken chiral symmetry are not manifest in the spectrum of excited hadrons, so that it is relevant to discuss its effective restoration
and how it comes about --- see review \cite{Glozman:2007ek} and references therein.
It is important to emphasise that the discussion in this chapter concerns the way chiral symmetry is realised in the spectrum of excited hadrons
and, in particular, it will be demonstrated that the properties of highly excited hadrons are only weakly sensitive to the
phenomenon of the spontaneous chiral symmetry breaking in the vacuum. This entails various observable consequences which will also be discussed below.

In papers
\cite{Nowak:2003ra,Beane:2001uj,Golterman:2002mi,Afonin:2004yb,Swanson:2003ec,DeGrand:2003sf,Kalashnikova:2005tr}
this phenomenon was described in the framework of various approaches to QCD.
Meanwhile, regardless of the particular model used, such an effective
chiral symmetry restoration implies the emergence of multiplets of hadronic states
approximately degenerate in mass. An important comment
is in order here. It is well-known that the spectrum of mass of the quark-antiquark
mesons bound by the linear potential shows a
Regge
behaviour, that is, $M_{n,l}^2\propto n$ and $M_{n,l}^2\propto l$ for $n,l\gg 1$. Here $n$ and $l$ are the radial quantum number and the angular
momentum, respectively. It is easy to see that the states with the opposite parity which
form approximate degenerate doublets possess the angular momenta
different by one unit (for example, the scalar ${}^3P_0$ and the pseudo-scalar ${}^1S_0$).
Therefore, for a given angular momentum $l_0$, the splitting
in such a pair is
\begin{equation}
\Delta M^{+-}_{n,l_0}\equiv M^+_{n,l_0+1}-M^-_{n,l_0}\sim \frac{1}{M^+_{n,l_0+1}+M^-_{n,l_0}}\sim\frac{1}{\sqrt{n}},
\end{equation}
that is, it decreases with the growth of the radial quantum number.
Clearly, such a decrease does not imply the effective chiral symmetry restoration.
Indeed, exactly the same dependence takes place for the splitting between
the same-parity neighbours,
\begin{equation}
\Delta M^{\pm\pm}_{n,l_0}\equiv M^\pm_{n,l_0+1}-M^\pm_{n,l_0}\sim
\frac{1}{M^\pm_{n,l_0+1}+M^\pm_{n,l_0}}\sim\frac{1}{\sqrt{n}},
\end{equation}
which has nothing to do with chiral symmetry. Therefore, it is necessary to define
the quantity which would allow one to
judge whether or not the effective restoration of chiral symmetry in the spectrum occurs.
For such a quantity one can choose the splitting between the
masses squared, $\Delta (M^{+-})^2=(M^+)^2-(M^-)^2$,
\cite{Kalashnikova:2005tr}\footnote{For the generic power-like potential (\ref{potential}) the power of the masses to be considered is
$(\alpha+1)/\alpha$.}
or, equivalently, the ratio of the splittings $\Delta M^{+-}/\Delta M^{\pm\pm}$ within the same chiral multiplet \cite{Shifman:2007xn}.

Thus, it would be natural to take advantage of the microscopic approach
to chiral symmetry breaking provided by the Generalised Nambu--Jona-Lasinio
model and to use it to study
the influence of chiral symmetry breaking over the spectrum of excited hadrons.

\subsection{Quantum fluctuations and the quasiclassical regime in the spec\-trum of excited hadrons}\label{subsec:quasi}

The phenomenon of the effective restoration of chiral symmetry in the spectrum
of excited hadrons has a simple qualitative explanation.
Once the spontaneous breaking of chiral symmetry is a consequence of quantum fluctuations
(loops) then it must be a quantum effect itself.
The parameter defining the role played by such
fluctuations is provided by the ratio
$\hbar/{\cal S}$, where ${\cal S}$ is the classical action responsible for the internal
degrees of freedom in the hadron.
For large values of the quantum numbers, that is, in the quasiclassical region of the
spectrum, one has ${\cal S}\gg\hbar$ and, therefore,
the effect of the spontaneous breaking of chiral symmetry cannot affect the properties
of the highly excited hadrons \cite{Glozman:2004gk}.

Below, we exemplify this qualitative picture with the help of the Generalised
Nambu--Jona-Lasinio mode. As before, we take the large-$N_C$ limit
that allows us to consider only planar (ladder and rainbow) diagrams and, in addition,
for illustrative purposes, we stick to the simplest structure of the confining potential,
$\gamma_0\times\gamma_0$, --- see equation (\ref{KK}).

Consider Dyson equation (\ref{Sigma03}) for the mass operator.
Similarly to many nonlinear equations, this equation possesses several solutions. One
of them is perturbative and it is given by the series
\begin{equation}
\Sigma=\int d^4k\;V\gamma_0 S_0\gamma_0+\int d^4k\;d^4q\;V^2\gamma_0 S_0\gamma_0 S_0 \gamma_0
S_0\gamma_0+\ldots,
\label{Sser}
\end{equation}
which converges fast in the limit of a weak interaction. It is easy to
demonstrate that this solution is nothing but a series in the powers of the Planck
constant $\hbar$. To this end let us restore the latter explicitly
in formula (\ref{Sigma03}).

The confining potential is defined by the averaged Wilson loop,
\begin{equation}
\langle W(C)\rangle=\exp\left[-\frac{\sigma A}{\hbar c}\right],
\end{equation}
where $\sigma$ is the string tension and $A$ is the area of the minimal surface in Euclidean
space which is bounded by the contour $C$.
For convenience, the speed of light $c$ is also explicitly shown, to be omitted later when appropriate.
Then, for a rectangular loop one has
\begin{equation}
\frac{\sigma A}{\hbar c}=\frac{\sigma R\times (cT)}{\hbar c}=\frac{1}{\hbar}\int_0^T \sigma R dt=
\frac{1}{\hbar}\int_0^T V(R)dt.
\label{Wc}
\end{equation}
From equation (\ref{Wc}), it is easy to find the linear confinement
$V(r)=\sigma r$ as the interquark potential which is assumed to be a classical
quantity to survive in the formal classical limit of $\hbar\to 0$.
Then, for its Fourier transform one finds
\begin{equation}
V(\vp)=\int d^3 x e^{i\vp\vx/\hbar} \sigma
|\vx|=-\frac{8\pi\sigma\hbar^4}{p^4}=\hbar^4 \tilde{V}(\vp),
\label{FV}
\end{equation}
where the quantity $\tilde{V}(\vp)$ does not contain $\hbar$.
As a result, it is easy to arrive at
\begin{equation}
i\Sigma({\bm p})=\hbar\int\frac{d^4k}{(2\pi)^4}\tilde{V}({\bm p}-{\bm k})
\gamma_0\frac{1}{S_0^{-1}(k_0,\vk)-\Sigma(\vk)} \gamma_0,
\label{Sigma02}
\end{equation}
where the factor $\hbar^4$ from the potential in the numerator cancels the factor $\hbar^4$ from the differential
$d^4k/(2\pi\hbar)^4$ in the denominator.
The remaining $\hbar$ is easily restored to provide the correct dimension of the r.h.s.
Therefore, the perturbative expansion in powers of potential (\ref{Sser})
is the expansion in the loops, and each power of the potential
(each loop) brings $\hbar$.

Consider now mass-gap equation (\ref{mge}) with the Planck constant $\hbar$
and the speed of light $c$ restored explicitly,
\begin{equation}
pc\sin\vpp_p-mc^2\cos\vpp_p=\frac{\hbar}{2}\int\frac{d^3k}{(2\pi)^3}
\tilde{V}(\vp-\vk)\left[\cos\vpp_k\sin\vpp_p-
(\hat{{\bm p}}\hat{{\bm k}})\sin\vpp_k\cos\vpp_p\right].
\label{mgh}
\end{equation}
Let us study the limit $m=0$ first. It is easy to see that, in the formal
classical limit $\hbar\to 0$, the r.h.s of equation
(\ref{mgh}) vanishes and the only solution to this equation is given by the trivial
chirally symmetric solution $\vpp_p=0$. This result is
quite natural because, for $m=0$, the chiral angle parametrises the contribution of
the loops which is a purely quantum effect and which, therefore,
must vanish in the classical limit. Any attempt to find a solution of equation (\ref{mgh})
in the form of the series in the powers of $\hbar$,
$\vpp_p=\hbar\times f_1(p)+\hbar^2\times f_2(p)+\ldots$, has to fail as all
coefficients in such a series vanish. This should
not come as a surprise either, for it has a simple qualitative explanation. Indeed,
the full form of the given expansion of the chiral angle in powers
of $\hbar$ should look like
\begin{equation}
\vpp_p=\frac{\hbar}{\cal S}\times f_1\left(\frac{p}{\mu c}\right)+\frac{\hbar^2}{{\cal S}^2}\times f_2\left(\frac{p}{\mu
c}\right)+\ldots,
\label{hSexp}
\end{equation}
where the Planck constant enters divided by the quantity ${\cal S}$ which has the dimension of the action while the momentum is measured
in the units of some mass parameter $\mu$. It is easy to verify that the only two dimensional parameters at hand, $\sigma$ and
$c$, are not sufficient to build the quantity ${\cal S}$, and this fact alone automatically invalidates expansion (\ref{hSexp}).

To get an insight in the behaviour of the chiral angle in the classical limit, let us proceed to the dimensionless mass-gap equation
obtained from (\ref{mgh}) with the help of the substitution $\vp=\mu c{\bm \xi}$ and $\vk=\mu c{\bm \eta}$, so that all dimensional parameters
in the equation produce a single mass scale $\mu=\sqrt{\sigma\hbar c}/c^2$. It easy to see that the scale $\mu$ depends on the Planck
constant. Then, the small-momentum expansion of the chiral angle reads
$$
\vpp_p\mathop{\approx}\limits_{p\to 0}\frac{\pi}{2}-{\rm
const}\frac{p c}{\mu c^2}+ \ldots=\frac{\pi}{2}- {\rm
const}\frac{pc}{\sqrt{\sigma\hbar c}}+\ldots,
$$
and, therefore, in the formal limit $\hbar\to 0$, the chiral angle becomes steeper at the origin thus approaching the chirally symmetric solution
$\vpp_p=0$. In other words, one witnesses the collapse of the chiral angle similar to the one which happens to the quantum mechanical wave function in
the classical limit. Indeed, the chiral angle can be viewed as the radial wave function of the quark-antiquark pairs in the vacuum --- see, for example,
formula (\ref{S00}). Besides that, the chiral angle defines the wave function of the chiral pion. Thus, the chiral angle --- solution of the mass-gap
equation --- depends on the Planck constant essentially nonperturbatively.

Beyond the chiral limit, if the quark mass is introduced as a perturbation, the chiral angle can be presented as a series in powers
of the small dimensionless parameter $\frac{mc^2}{\sqrt{\sigma\hbar c}}$,
\begin{equation}
\vpp_p=\sum_{n=0}^\infty\left(\frac{mc^2}{\sqrt{\sigma\hbar c}}\right)^nf_n
\left(\frac{pc}{\sqrt{\sigma\hbar c}}\right),
\label{vp1}
\end{equation}
where the plot of the leading term $f_0(p)$ is shown in Fig.~\ref{figvp}.

On the other hand, as was mentioned above, beyond the chiral limit, the quark mass $m$,
as furnishing an additional dimensional parameter, allows one to
build both the classical dimension of the action, ${\cal S}\sim\frac{m^2c^3}{\sigma}$,
together with the classical dimension of the momentum, $mc$. Then, expansion (\ref{hSexp}) becomes possible
and takes the form
\begin{equation}
\vpp_p=\sum_{n=0}^\infty\left(\frac{\sigma\hbar c}{(mc^2)^2}\right)^n\tilde{f}_n\left(\frac{p}{mc}\right),
\label{vp2}
\end{equation}
that is, the dimensionless expansion parameter is given by $\frac{\sigma\hbar c}{(mc^2)^2}$. The leading term in series
(\ref{vp2}) is known analytically and it is given by the free chiral angle $\tilde{f}_0 = \arctan\frac{mc}{p}$. In other words, perturbative
solution (\ref{vp2}) is given by expansion (\ref{Sser}).

Both expansions (\ref{vp1}) and (\ref{vp2}) reproduce the same solution for the chiral angle. However, expansion
(\ref{vp1}) converges fast near the chiral limit with $m=0$ and beyond the classical limit with $\hbar\neq 0$ (expansion (\ref{vp2}) blows up
in this limit). On the contrary, for $m\gg\frac{\sqrt{\sigma\hbar c}}{c^2}$ expansion (\ref{vp2}) converges much better than expansion
(\ref{vp1}). Meanwhile, there is no one-to-one correspondence between the functions $\{f_n\}$ and $\{\tilde{f}_n\}$, and
each function from one set is given by an infinite series in terms of the functions from the other set and vice versa.
For example, for asymptotically large momenta, the function $f_0$, depicted in Fig.~\ref{figvp}, tends to zero as $1/p^5$ (see
equation (\ref{asym})) while the asymptotic behaviour of the function $\tilde{f}_0$ is much slower, as $1/p$.

As a final remark, expansions (\ref{vp1}) and (\ref{vp2}) define two dynamical regimes of the system depending on the value of the
parameter $m/\sqrt{\sigma}$. Spontaneous breaking of chiral symmetry takes place in the limit $m\ll\sqrt{\sigma}$ --- regime
(\ref{vp1}) --- while, in the opposite limit of $m\gg\sqrt{\sigma}$ one deals with the ``heavy-quark'' physics --- regime (\ref{vp2}).

\subsection{Effective chiral symmetry restoration in the spectrum of excited mesons}

As was mentioned in Subsect.~\ref{subsec:gencom}, the spectrum of highly excited hadrons is expected to
show the phenomenon of an effective
restoration of chiral symmetry.
In Subsect.~\ref{subsec:quasi}, general qualitative arguments were given in favour of such a restoration in the Generalised Nambu--Jona-Lasinio model.
Below, we study this phenomenon quantitatively \cite{Kalashnikova:2005tr}.

\begin{figure}[t]
\centerline{\epsfig{file=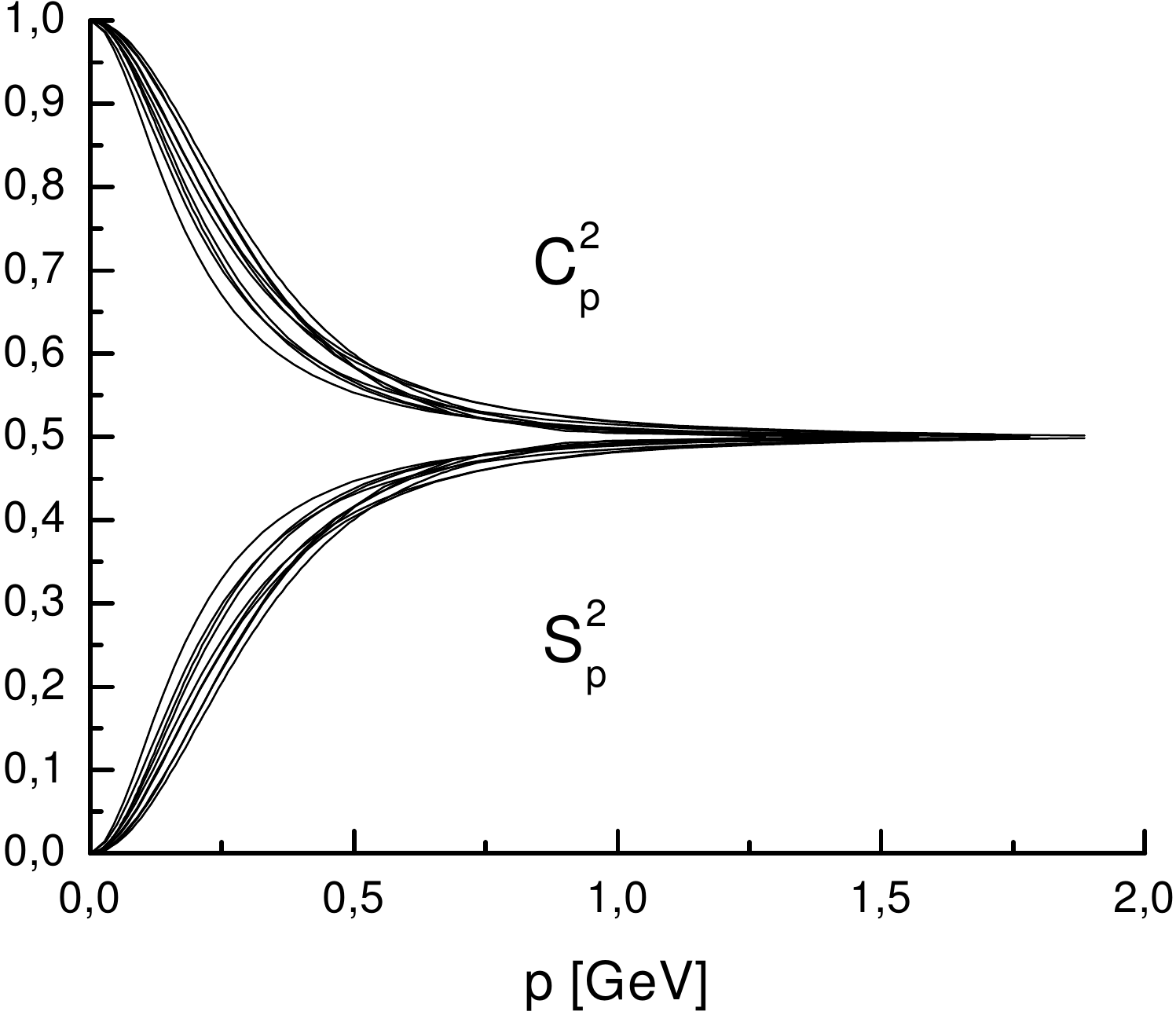,width=0.5\textwidth}}
\caption{The momentum dependence of the coefficients $C_p^2$ and $S_p^2$ for the potential $V(r)=K_0^{\alpha+1}r^\alpha$ with
$\alpha=0.3$, 0.5, 0.7, 0.9, 1.0, 1.1, 1.3, 1.7, and 2.0. For each potential, the parameter $K_0$ is tuned to provide the
same value of the chiral condensate equal to $-(250~\mbox{MeV})^3$.}\label{fig:cs}
\end{figure}

We start from Schr{\"o}dinger equation (\ref{FW4}) describing
the mass spectrum of heavy-light quarkonia. Multiplying this equation by
$(\vesig\vp)$ from the left, we can re-write the result in the form of
the bound-state equation for the wave function
\begin{equation}
\psi'({\bm p})=({\bm \sigma}\hat{\vp})\psi({\bm p})
\label{difpar}
\end{equation}
which, by construction, possesses the opposite parity, as compared with $\psi(\vp)$.
The resulting equations,
\begin{equation}
E_p\psi'({\bm p})+\int\frac{d^3k}{(2\pi)^3}V(\vp-\vk)\left[S_pS_k+
({\bm \sigma}\hat{\vp})({\bm \sigma}\hat{\vk})C_pC_k\right]\psi'({\bm k})=E\psi'({\bm p}),
\label{FW4pr}
\end{equation}
differs from equation (\ref{FW4}) by the permutation of the quantities $C_p$ and $S_p$ defined in (\ref{SandC}). It is easy to see then that,
in the limit of a large relative momenta, $\vpp_p\mathop{\to}\limits_{p\to\infty}0$
(see Fig.~\ref{figvp}), so that $C_p=S_p=\frac{1}{\sqrt{2}}$ and equations (\ref{FW4}) and (\ref{FW4pr}) coincide to take the form
\begin{equation}
E_p\psi^{(\prime)}({\bm p})+\frac12\int\frac{d^3k}{(2\pi)^3}V(\vp-\vk)\left[1+({\bm \sigma}\hat{\vp})({\bm
\sigma}\hat{\vk})\right]\psi^{(\prime)}({\bm k})=E\psi^{(\prime)}({\bm p}).
\label{FW4pr2}
\end{equation}
Therefore, the opposite-parity states $\psi({\bm p})$ and $({\bm \sigma}\hat{\vp})\psi({\bm p})$ become degenerate.
Notice that the Fourier transform of the potential in equations (\ref{FW4}) and (\ref{FW4pr}) picks up the region
$\vp\approx\vk$, so that, approximately, one can speak of the effective restoration of chiral symmetry if
$C_p^2\approx S_p^2$. Then it is easy to find that, in the chiral limit,
\begin{equation}
C_p^2-S_p^2=\sin\vpp_p={\cal N}_\pi^{-1}\vpp_\pi^\pm(p),
\end{equation}
where $\vpp_\pi^\pm(p)$ are the components of the wave function of the chiral pion (see equation (\ref{vppm})). This relation emphasises
the connection between the parity degeneracy observed in the spectrum of highly excited mesons and chiral symmetry.
In Fig.~\ref{fig:cs} one can see the dependences of the quantities $C_p^2$ and $S_p^2$ on the momentum for the potentials of the form
$V(r)=K_0^{\alpha+1}r^\alpha$ with different values of the power parameter $\alpha$. It is easy to see from the plot that the above functions, indeed,
reach the asymptotic value 1/2 fast.

As was many times mentioned above, qualitative predictions of the model do not depend on the power $\alpha$.
Moreover, the quantitative predictions also demonstrate only a very weak dependence on $\alpha$ --- see Fig.~\ref{fig:cs}.
Thus, for a detailed quantitative study of the problem of the chiral symmetry restoration in the spectrum of highly excited mesons it is
sufficient to choose the power $\alpha$ which provides the simplest form of the equations, that it, $\alpha=2$. Then,
\begin{equation}
V(\vp-\vk)=-K_0^3\Delta_k\delta^{(3)}(\vp-\vk),
\end{equation}
and the mass-gap equation reduces to the second-order differential equation (\ref{diffmge}). Chiral condensate (\ref{chircond})
equals $-(0.51K_0)^3$ and takes the standard value $-(250~\mbox{MeV})^3$ for
\begin{equation}
K_0=490~\mbox{MeV}.
\label{K0osc}
\end{equation}

It has to be noticed that the most adequate basis to deal with the spectrum of highly
excited mesons and, in particular, to study the prevalence of the chiral symmetry
restoration is the chiral basis \cite{Glozman:2007at}.
For example, in the work \cite{Wagenbrunn:2007ie}, this basis was successfully used
for numerical studies on the mass spectrum of excited mesons in the
Generalised Nambu--Jona-Lasinio model.
Nevertheless, to get a better contact with the calculations done in the framework of the Generalised Nambu--Jona-Lasinio model and for the simple
Salpeter equation, we adhere to the standard basis $\{J,L,S\}$.
Then, the wave function of the light quark $\psi(\vp)$
is decomposed in the basis of spherical spinors,
\begin{equation}
\Omega_{jlm}(\hat{{\bm p}})=\sum_{\mu_1\mu_2}C^{jm}_{l\mu_1\frac12\mu_2}
Y_{l\mu_1}(\hat{{\bm p}})\chi_{\mu_2},
\end{equation}
in the form
\begin{equation}
\psi(\vp)=\Omega_{jlm}(\hat{{\bm p}})\frac{u(p)}{p},
\end{equation}
where $u(p)$ is the radial wave function in the momentum representation
for which the following equation can be derived from (\ref{FW4}):
\begin{equation}
u''=[E_p-E]u+K_0^3\left[\frac{\varphi_p^{\prime 2}}{4}+\frac{\kappa}{p^2}\left(\kappa+\sin\varphi_p\right)\right]u,
\label{udd}
\end{equation}
and where the spin-orbit interaction for the central potential
is introduced in the standard way,
$$
\kappa=\left\{
\begin{array}{ccc}
l,&\mbox{for}&j=l-\frac12\\[1mm]
-(l+1),&\mbox{for}&j=l+\frac12
\end{array}
=\pm\left(j+\frac12\right).
\right.
$$

Equation (\ref{udd}) can now be re-written in the
form of the Schr{\"o}dinger equation,
\begin{equation}
-K_0^3u''+V_{[j,l]}(p)u=Eu,
\label{Se}
\end{equation}
with the effective potential
\begin{equation}
V_{[j,l]}(p)=E_p+K_0^3\left[\frac{\varphi_p^{\prime 2}}{4}+\frac{\kappa}{p^2}\left(\kappa+\sin\varphi_p\right)\right].
\label{Vnu}
\end{equation}

Notice that the well-known property of the spherical spinors,
\begin{equation}
({\bm \sigma}\hat{\vp})\Omega_{jlm}(\hat{{\bm p}})=-\Omega_{jl'm}(\hat{{\bm p}}),\quad
l+l'=2j,
\end{equation}
guarantees the opposite parity of the states with $j=l\pm\frac12$.
With the help of the explicit form of effective potential (\ref{Vnu})
it is easy to find the difference of the potential for the
states with $\kappa=\pm(j+\frac12)$,
\begin{equation}
\Delta V=-\frac{(2j+1)K_0^3}{p^2}\sin\vpp_p,
\label{DV}
\end{equation}
that demonstrates explicitly the relation between the splitting in mass for
the opposite-parity states and chiral symmetry which was
already discussed above in general terms. Obviously, for highly excited states with
larger mean values of the relative momentum,
the chiral angle decreases (see Fig.~\ref{figvp}) and so does the potential responsible for the
splitting of the opposite-parity energy levels.

\begin{table}[t]
\begin{center}
\begin{tabular}{|c|c|c|c|c|c|c|c|c|c|c|}
\cline{1-5}\cline{7-11}
$j$&1/2&3/2&5/2&7/2& \hspace*{3mm}& $j$&1/2&3/2&5/2&7/2 \\
\cline{1-5}\cline{7-11}
$E_{l=j-\frac12}$&2.04&3.51&4.51&5.35 && $E_{l=j-\frac12}^{\rm Salp}$&2.34&3.36&4.24&5.05 \\
\cline{1-5}\cline{7-11}
$E_{l=j+\frac12}$&2.66&3.69&4.57&5.36 && $E_{l=j+\frac12}^{\rm Salp}$&3.36&4.24&5.05&5.79 \\
\cline{1-5}\cline{7-11}
$\Delta E_j$&0.62&0.18&0.06&0.01 && $\Delta E_j^{\rm Salp}$&1.02&0.88&0.81&0.74 \\
\cline{1-5}\cline{7-11}
\end{tabular}
\end{center}
\caption{The masses of the orbitally excited states and their splittings for the radial quantum number $n=0$ as they come out from the solution of the
exact equation (\ref{Se}) with potential (\ref{Vnu}) and from the approximate Salpeter equation (\ref{Salp0}). All energies are
given in the units of the parameter $K_0$.}\label{tab:n0}
\end{table}

\begin{table}[t]
\begin{center}
\begin{tabular}{|c|c|c|c|c|c|c|c|c|c|c|}
\cline{1-5}\cline{7-11}
$j$&1/2&3/2&5/2&7/2 &\hspace*{3mm}& $j$&1/2&3/2&5/2&7/2 \\
\cline{1-5}\cline{7-11}
$E_{l=j-\frac12}$&3.91&5.03&5.87&6.60 &&$E_{l=j-\frac12}^{\rm Salp}$&4.09&4.88&5.63&6.33 \\
\cline{1-5}\cline{7-11}
$E_{l=j+\frac12}$&4.39&5.17&5.92&6.61 && $E_{l=j+\frac12}^{\rm Salp}$&4.88&5.63&6.33&7.00 \\
\cline{1-5}\cline{7-11}
$\Delta E_j$&0.48&0.14&0.05&0.01 && $\Delta E_j^{\rm Salp}$&0.79&0.75&0.70&0.67 \\
\cline{1-5}\cline{7-11}
\end{tabular}
\end{center}
\caption{The same as in Table~\ref{tab:n0} but for $n=1$.}\label{tab:n1}
\end{table}

\begin{figure}[t]
\centerline{\epsfig{file=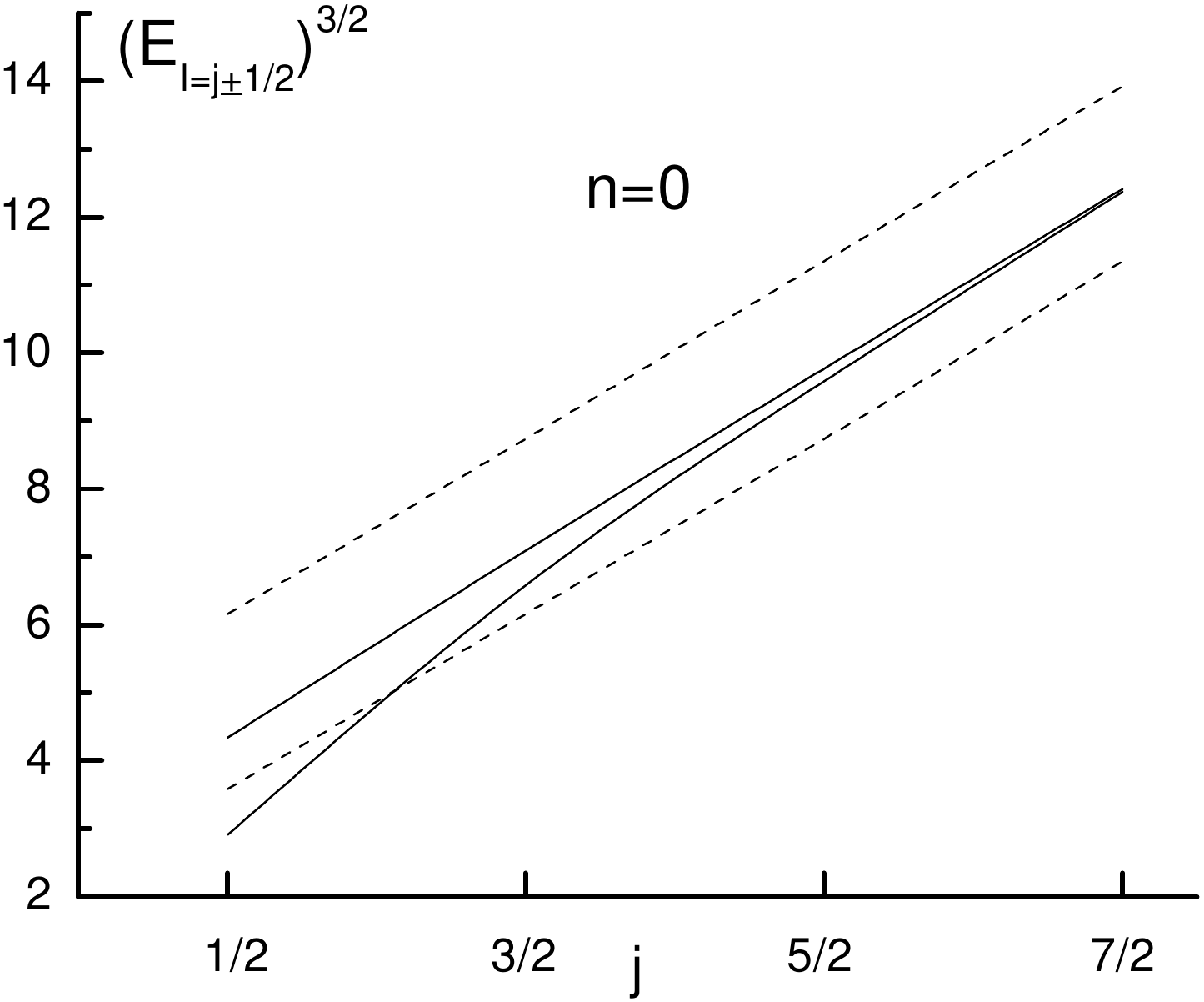,width=7cm}\hspace*{10mm}\epsfig{file=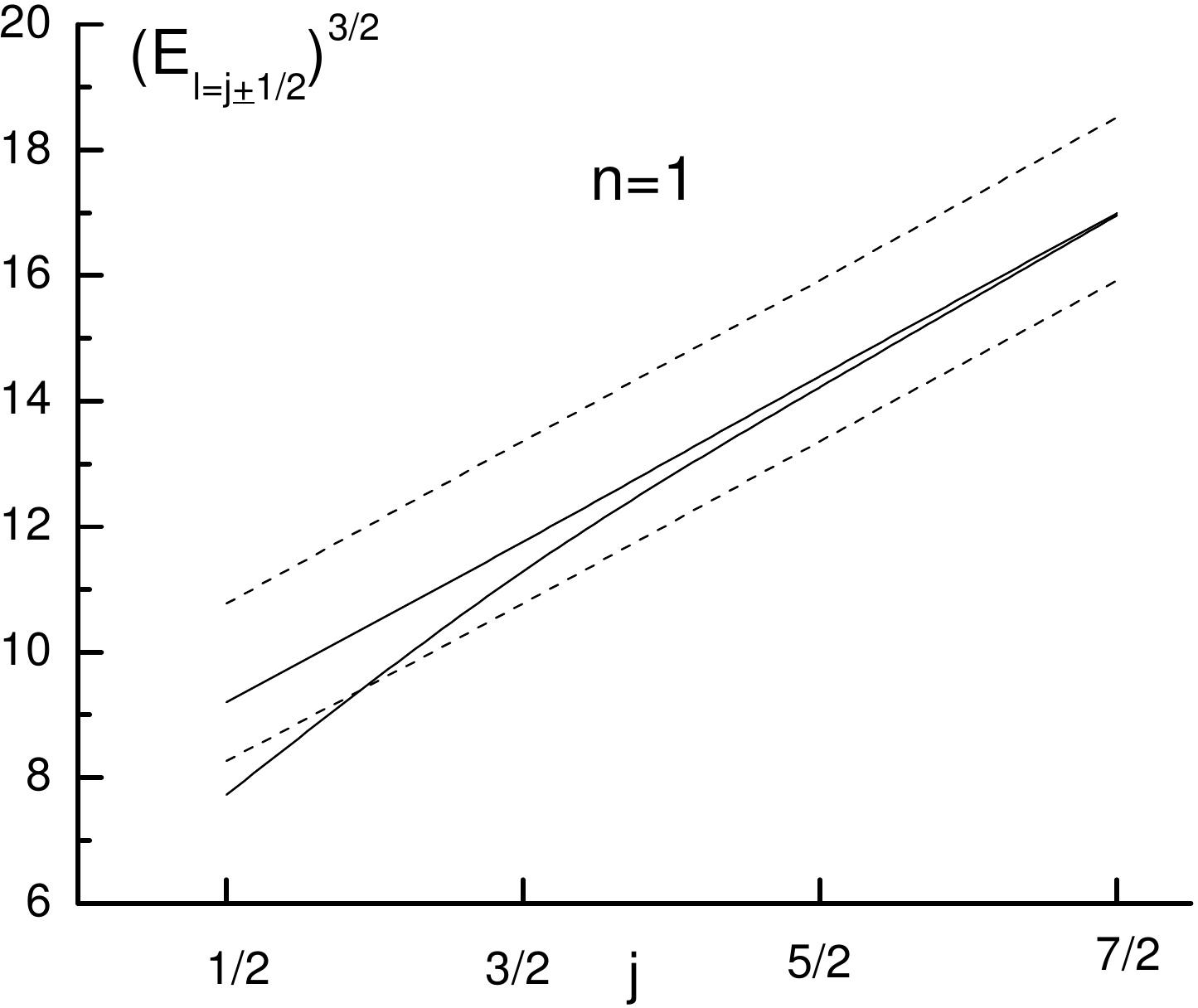,width=7cm}}
\caption{The Regge trajectories for equation (\ref{Se}) with potential (\ref{Vnu}) (solid line) and for
Salpeter equation (\ref{Salp0}) (dashed line). The lower and the upper line correspond to $l=j-\frac12$ and $l=j+\frac12$,
respectively.}\label{fig:n01}
\end{figure}

This type of behaviour is clearly seen from the explicit solution of
equation (\ref{Se}) quoted in Tables~\ref{tab:n0} and \ref{tab:n1}
and shown in Fig.~\ref{fig:n01}. For the sake of clarity, we compare the obtained solutions
with those for the na{\" i}ve Salpeter equation,
\begin{equation}
[\sqrt{\vp^2+m^2}+K_0^3\vx^2]\psi({\bm x})=E\psi({\bm x}),
\label{Salp0}
\end{equation}
as derived from equation (\ref{Se}) through the substitution $E_p=\sqrt{p^2+m^2}$ and $\vpp_p\equiv\pi/2$ in potential (\ref{Vnu}),
\begin{equation}
V_{[j,l]}^{\rm Salp}(p)=\sqrt{p^2+m^2}+K_0^3\frac{\kappa(\kappa+1)}{p^2}=\sqrt{p^2+m^2}+K_0^3\frac{l(l+1)}{p^2}.
\end{equation}
Once the opposite-parity states correspond to the angular momenta different by one unit then, in analogy with equation
(\ref{DV}), one can find that
\begin{equation}
\Delta V^{\rm Salp}=-\frac{2(l+1)K_0^3}{p^2}.
\label{DVSalp}
\end{equation}

The quasiclassical spectrum of equation (\ref{Salp0}) demonstrates a linear dependence of
$E^{3/2}$ ($E^{(\alpha+1)/\alpha}$ with $\alpha=2$) on the angular momentum $l$, so that, for a given radial quantum number $n$, equation
(\ref{Salp0}) produces two parallel trajectories $E^{3/2}(j)$ with $l=j\pm\frac12$. Similar trajectories for equation (\ref{Se}) with potential
(\ref{Vnu}) have a comparable level splitting for small $j$'s which, however, decreases fast with the growth of the angular momentum $l$.

This calculation explicitly demonstrates the phenomenon of the effective
restoration of chiral symmetry in the spectrum of highly excited
mesons in the Generalised Nambu--Jona-Lasinio model. As one can see by just comparing
potentials (\ref{DV}) and (\ref{DVSalp}),
\begin{equation}
E-E'\propto\lan\sin\vpp_p\ran,
\label{EEprgpi}
\end{equation}
where $E$ and $E'$ are the energies of the opposite-parity states, and averaging over the radial wave function is assumed on the r.h.s.

\subsection{Pion decoupling from excited mesons}

One of specific predictions for highly excited hadrons with effectively restored chiral symmetry is the decoupling of the chiral pion from them which
manifests itself through the decrease of the corresponding coupling constant with the increase of the hadron excitation number
\cite{Glozman:2003bt,Jaffe:2006aq,Jaffe:2005sq,Jaffe:2006jy,Cohen:2005am}.
This behaviour of the coupling can be readily established with the
help of the Goldberger--Treiman relation for the transitions $n\to n'+\pi$, where $n$ and $n'$ are the chiral partners, that is, the
opposite-parity hadronic states which become degenerate in mass if chiral
symmetry is restored in the spectrum.\footnote{Strictly speaking,
Goldberger--Treiman relation connects the pion-nucleon constant with the
nucleon axial constant; the derivation of this relation can be found
in any textbook in strong interactions.
Notwithstanding that, hereafter we shall be denoting by the name of
Goldberger--Treiman relation the one for the pion-hadron coupling
constant $g_{nn'\pi}$.}

Let us stick to the BCS approximation first and show
that the pion coupling to excited hadrons is defined by the effective mass of the
dressed quark. To this end, we consider the axial-vector current (for simplicity,
we consider the single-flavour case and the chiral anomaly is
omitted),
\begin{equation}
J_\mu^5(x)=\bar q(x)\gamma_\mu \gamma_5q(x),
\label{qax}
\end{equation}
which, due to the hypothesis of the Partial Conservation of the Axial-vector
Current (PCAC), is related to the wave function of the chiral pion
$\phi_\pi$,
\begin{equation}
J_\mu^5(x)=f_\pi\partial_\mu \phi_\pi(x).
\label{piax}
\end{equation}
Then, with the help of equation (\ref{piax}), it is easy to average the
divergence of this current, $\partial_\mu J_\mu^5$,
between the states of the dressed quarks,
\begin{equation}
\langle q(p)|\partial_\mu J_\mu^5(x)|q(p')\rangle=f_\pi m_\pi^2 \langle q(p)|\phi_\pi(x)|q(p')\rangle
\propto f_\pi g_\pi(q^2) (\bar{u}_p\gamma_5 u_{p'}),\quad q=p-p',
\label{eq1qq}
\end{equation}
where we have introduced the pion-quark-quark form factor $g_\pi(q^2)$.

On the other hand, if chiral symmetry is spontaneously broken and the
quark wave functions obey the effective Dirac equation with the
dynamically generated mass $m_q^{\rm eff}$ then,
with the help of equation (\ref{qax}), it is easy to arrive at
\begin{equation}
\langle q(p)|\partial_\mu J_\mu^5(x)|q(p')\rangle\propto m_q^{\rm eff}(\bar{u}_p\gamma_5 u_{p'}).
\label{eq22}
\end{equation}

Equating the r.h.s.'s of equations (\ref{eq1qq}) and (\ref{eq22}) one finds that
\begin{equation}
f_\pi g_\pi=m_q^{\rm eff},\quad g_\pi\equiv g_\pi(m_\pi^2),
\label{GTr}
\end{equation}
where, for simplicity, all numerical coefficients are absorbed into the definition
of the coupling constant $g_\pi$.
From equation (\ref{ABp}) one can see that the effective mass of the quark
is described by the quantity $A_p$. Then, with the help of
relation (\ref{GTr}), it is straightforward to find finally that \cite{Glozman:2006xq}
\begin{equation}
f_\pi g_\pi(p)\simeq A_p.
\label{gpiA}
\end{equation}

Beyond the BCS level, the Goldberger--Treiman relation connects the pion coupling constant
to an excited hadron with the mass splitting between
the two hadronic chiral partners. For definiteness, let us consider the transition ${\bar D}(J^P=0^+)\to{\bar D}'(J^P=0^-)\pi$,
where the quark contents of the $\bar{D}^{(\prime)}$ meson is $\bar{c}q$ with the light quark $q=u,d$.

From PCAC condition (\ref{piax}), generalised to the isospin
group $SU(2)$, one has
\begin{equation}
\langle 0|J_\mu^{5a}(0)|\pi^b(\vq)\rangle=if_\pi q_\mu\delta^{ab},
\end{equation}
so that for the transition matrix element $\langle n'|J_\mu^{5a}|n\rangle$ ($n^{(\prime)}=\bar{D}^{(\prime)}$) it is easy to find
\begin{equation}
\langle n'|J_\mu^{5a}|n\rangle=\langle n'|J_\mu^{5a}|n\rangle_{\mbox{nonpion}}-
\frac{2Mq_\mu f_\pi g_{nn'\pi}}{q^2-M_\pi^2+i0} D^{\prime\dag}\tau^aD,
\label{Ann}
\end{equation}
where we introduced the pion coupling constant $g_{nn'\pi}$ and the
isospin doublets $D$ and $D'$.

On the other hand, it is easy to establish the most general
form of the l.h.s. of equation (\ref{Ann}) compatible with Lorentz invariance,
\begin{equation}
\langle
n'|J_\mu^{5a}|n\rangle=\left[(P_\mu+P'_\mu)G_A(q^2)-(P_\mu-P'_\mu)G_S(q^2)\right]D^{\prime\dag}\left(\frac{\tau^a}{2}
\right)D,
\end{equation}
where $P_\mu$ and $P_\mu'$ are the momenta of the initial and final $D$ meson, respectively, and $q_\mu=P_\mu-P'_\mu$. Then, in the leading order in
the heavy-quark mass, conservation of the axial-vector current leads to the condition
\begin{equation}
2M(M-M')G_A-q^2G_S=0.
\label{GT}
\end{equation}

In the meantime, from relation (\ref{Ann}) one can see that, in the limit $q^2\to 0$,
one has $G_A(0)\equiv G_A\ne 0$ if
$G_S$ is identifies with the pion pole, that is,
\begin{equation}
\lim_{q^2\to 0}G_S(q^2)\to\frac{4Mf_\pi g_{n n'\pi}}{q^2}.
\end{equation}
The resulting equation,
\begin{equation}
\frac12(M-M')G_A=f_\pi g_{n n'\pi},
\label{gpiEEpr}
\end{equation}
is nothing but the sought Goldberger--Treiman relation for the heavy-light mesons.
This relation implies that, as the excitation of the
$D$ grows and, therefore, as its degeneracy with its chiral partner $D'$ becomes
more manifest, the pion decouples from this meson.

\begin{figure}[t]
\begin{center}
\epsfig{file=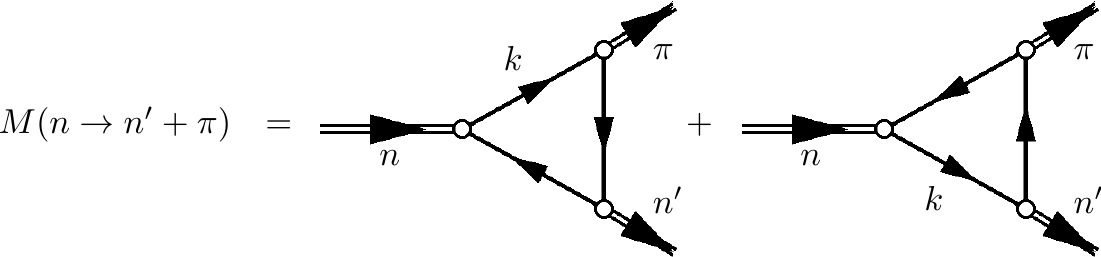,width=12cm}
\caption{The $n\to n'+\pi$ transition amplitude.}\label{hhpi}
\end{center}
\end{figure}

Let us now derive relations (\ref{gpiA}) and (\ref{gpiEEpr}) microscopically.
Consider the pion emission process
by the hadron $n$ (here hadrons $n$ and $n'$ are mesons; the case of the baryons is studied in the next chapter), $n\to n'+\pi$. The
corresponding diagrams are depicted in Fig.~\ref{hhpi} and the matrix element reads
\begin{equation}
\begin{array}{c}
\ds M(n\to n'+\pi)=\int\frac{d^4k}{(2\pi)^4}\Tr\left[\chi_n(\vk,\vP)S(k-P)\bar{\chi}_{n'}(\vk-\vP,\vP')
S(k-q)\bar{\chi}_\pi(\vk,\vq)S(k)\right]\\[3mm]
\ds +\int\frac{d^4k}{(2\pi)^4}\Tr\left[\chi_n(\vk,\vP)S(k-P)\bar{\chi}_\pi(\vk,\vq)S(k-q)\bar{\chi}_{n'}(\vk-\vP,\vP')
S(k)\right],
\end{array}
\label{Mhhpi}
\end{equation}
where $q=P-P'$ and each hadronic vertex contains the amplitude $\chi$ ($\bar{\chi}$
for the outgoing meson) which obeys
Bethe-Salpeter equation (\ref{GenericSal}). Thus, the pion emission vertex is given
by the overlap of the three vertex functions. The maximal overlap
is achieved if the wave functions of all three mesons are localised in the same region
in momentum. As it happens, the pion wave
function (\ref{Psiqq})
is localised at small relative momenta between the quark and
the antiquark and it decreases
fast with increased momentum. The wave functions
of the
mesons $n$ and $n'$ are localised at approximately the same momenta which grow with
the excitation number. This implies that the overlap of the
vertex functions decreases with the growth of the excitation of the meson $n$
and so does the pion coupling constant.

In order to describe this effect quantitatively we write the matrix
vertex $\chi(\vp,{\bm P})$ through the matrix wave function defined in equation (\ref{Psiqq}),
\begin{equation}
\chi(\vp,{\bm P})=\int\frac{d^3k}{(2\pi)^3}V(\vp-\vk)\gamma_0\tilde{\phi}(\vk,{\bm P})\gamma_0.
\label{bsp2}
\end{equation}

The matrix vertex for the incoming meson is simply related to the vertex for the outgoing meson,
\begin{equation}
\bar{\chi}(\vp,{\bm P})=\gamma_0\chi^\dagger(\vp,{\bm P})\gamma_0.
\label{chib}
\end{equation}

The explicit form of the matrix wave function $\tilde{\phi}_\pi$ for
the pion at rest (${\bm P}_\pi\equiv \vq\to 0$) is given by equation
(\ref{exp}) while the components $\vpp_\pi^\pm$ are quoted in equation (\ref{vppm}).
Then it is easy to find for the pion emission vertex
$\bar{\chi}_\pi(\vp,\vq=0)\equiv \bar{\chi}_\pi(\vp)$ that
\begin{equation}
f_\pi\bar{\chi}_\pi(\vp)=\sqrt{\frac{2\pi N_C}{m_\pi}}\gamma_5\int\frac{d^3k}{(2\pi)^3}V(\vp-\vk)\sin\vpp_k=
{\rm const}\times\gamma_5 A_p,
\label{Ap}
\end{equation}
where definition (\ref{ABp}) for the function $A_p$ was used.
If the pion form factor $g_\pi(p)$ is defined with the same constant as in
equation (\ref{Ap}),
\begin{equation}
\bar{u}_p\bar{\chi}_\pi(\vp)u_{p}={\rm const}\times g_\pi(p)(\bar{u}_p\gamma_5u_{p}),
\end{equation}
then we finally arrive at the Goldberger--Treiman relation in the form
\begin{equation}
f_\pi g_\pi(p)=A_p,
\label{GTr2}
\end{equation}
that coincides with formula (\ref{gpiA}), but now this relation is derived rigorously.

It is important to emphasise an essential difference between equations (\ref{GTr2}) and (\ref{GTr}), with the latter taken na{\"i}vely.
Indeed, na{\" i}vely, one could conclude that the r.h.s. of equation (\ref{GTr}) contains the
quantity which depends only on the momentum transfer in
the pion emission vertex, that is, on the pion momentum $\vq=0$. Then $m_q^{\rm eff}$ has to be treated as a constant, independent of the
excitation number of the hadron which emits the pion. The same would be true for the
pion coupling $g_\pi$. However, the microscopic
treatment performed above demonstrates that the pion emission vertex is a function of
two variables: the pion momentum and the loop momentum
floating through the pion emission vertex. The latter quantity also plays the role of
the momentum of the quark interacting with the pion. Therefore,
even in the limit $\vq=0$, the r.h.s of equation (\ref{GTr2}) is a decreasing function
of the momentum rather than a constant. It is easy to estimate
its decrease rate. Indeed, in the chiral limit, $A_p=E_p\sin\vpp_p$, while, for large
momenta, $E_p\approx p$ with the chiral angle behaving as
$\vpp_p\propto 1/p^{4+\alpha}$, where $\alpha$ is the parameter of the
power-like potential (see equation (\ref{asym})). Thus,
\begin{equation}
g_\pi(p)\mathop{\sim}_{p\to\infty}\frac{1}{p^{\alpha+3}}.
\end{equation}

\begin{figure}[t]
\begin{center}
\epsfig{file=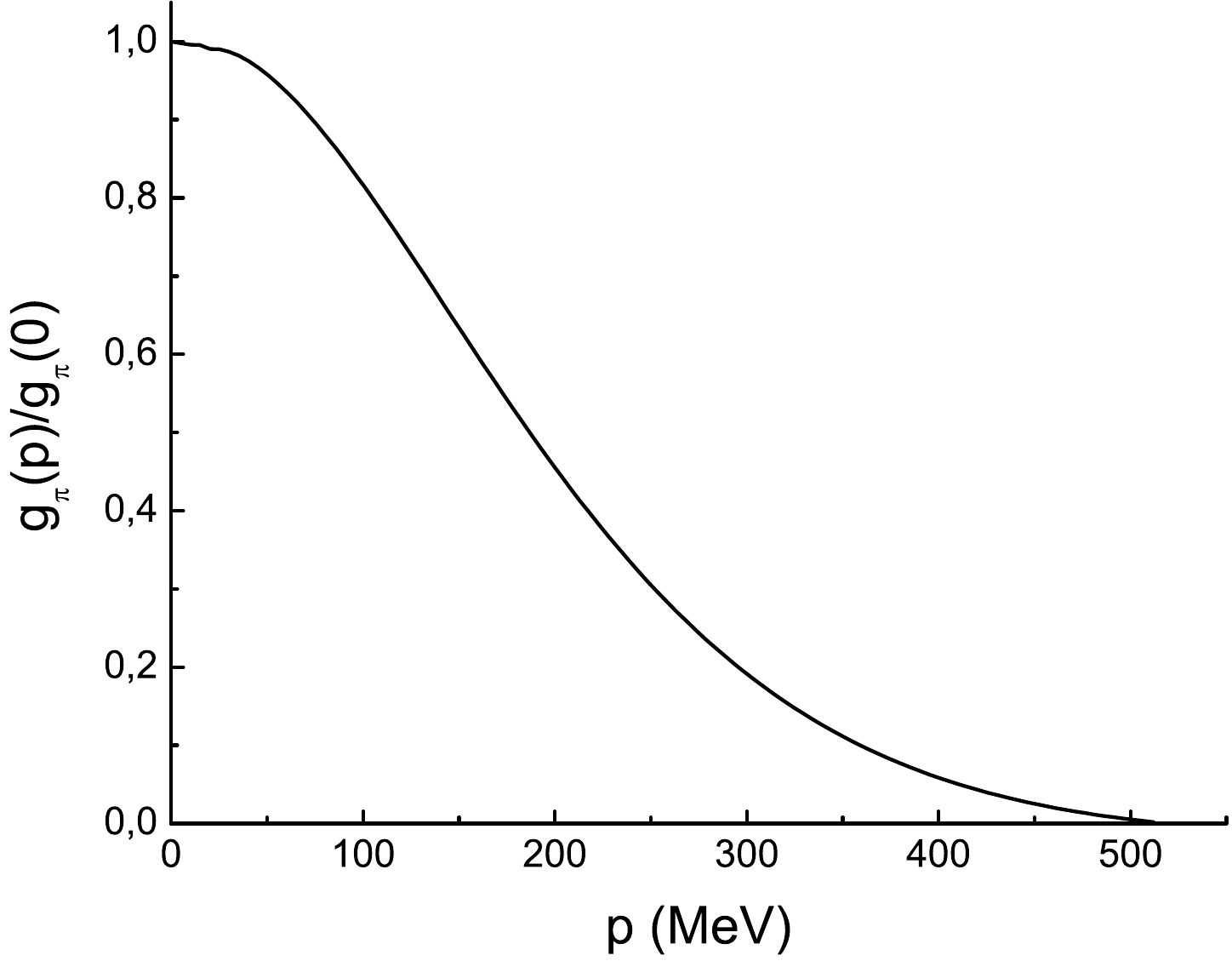,width=0.5\textwidth}
\caption{Dependence of the ratio $g_\pi(p)/g_\pi(0)$ on the momentum evaluated for the Harmonic Oscillator potential.
For definiteness, the parameter $K_0$ is fixed as in equation (\ref{K0osc}).}\label{gpiplot}
\end{center}
\end{figure}

The result of the numerical calculation of the ratio $g_\pi(p)/g_\pi(0)$
for the Harmonic Oscillator potential ($\alpha=2$), given
in Fig.~\ref{gpiplot}, shows that, indeed, the pion coupling decreases
with the quark momentum growth.

To finalise this chapter, let us derive microscopically Goldberger--Treiman relation (\ref{gpiEEpr}) for the pion emission by a
heavy-light meson \cite{Nefediev:2006bm}.

First, we proceed from the nonrelativistic normalisation (\ref{norm1}) for the pion wave functions (\ref{vppm}) to the relativistic one, so that we
define the wave functions
\begin{equation}
X_p=\frac{\sqrt{N_C}}{f_\pi}\left[\sin\vpp_p+M_\pi\Delta_p\right],\quad
Y_p=\frac{\sqrt{N_C}}{f_\pi}\left[\sin\vpp_p-M_\pi\Delta_p\right].
\label{vppm2}
\end{equation}
Then one can find that
\begin{equation}
\int\frac{d^3p}{(2\pi)^3}\Bigl(X_p^2-Y_p^2\Bigr)=2M_\pi,
\end{equation}
and the pionic field with the isospin projection $a$ in its rest frame can be written in the form
\begin{eqnarray}
&&|\pi^a\rangle=\frac{1}{\sqrt{N_C}}\sum_{\alpha=1}^{N_C}\sum_{s_1,s_2=\uparrow,\downarrow}
(\sigma_2)_{s_1s_2}\sum_{i_1,i_2=\pm1/2}\left(\frac{\tau^a}{2}\right)^{i_1i_2}
\int\frac{d^3p}{(2\pi)^3}
\left[b_{\alpha, s_1 i_1}^\dag(\vp)d_{s_2 i_2}^{\alpha\dag}(-\vp) X_p+\right.\nonumber\\[-2mm]
\label{pia0}\\[-2mm]
&&\left.\hspace*{70mm}+d_{s_2 i_2}^\alpha(-\vp)b_{\alpha, s_1 i_1}(\vp) Y_p\right]|0\rangle,\nonumber
\end{eqnarray}
where $|0\ran$ is the BCS vacuum.

The wave functions of the pseudo-scalar and scalar heavy-light mesons which obey equations (\ref{FW4}) and (\ref{FW4pr})
can be written in the form
\begin{equation}
\psi(\vp)=\frac{i}{\sqrt{2}}\sigma_2\phi_p, \quad
\psi'(\vp)=(\vesig\hat{\vp})\psi(\vp)=\frac{i}{\sqrt{2}}(\vesig\hat{\vp})\sigma_2\phi'_p
\label{normnn}
\end{equation}
and normalised by the relativistic conditions,
\begin{equation}
\Tr\int\frac{d^3p}{(2\pi)^3}|\psi(\vp)|^2=\int\frac{d^3p}{(2\pi)^3}\phi_p^2=2M,\quad
\Tr\int\frac{d^3p}{(2\pi)^3}|\psi'(\vp)|^2=\int\frac{d^3p}{(2\pi)^3}\phi_p^{\prime 2}=2M',
\end{equation}
where the trace is taken in the spin indices. Besides, in the above normalisation conditions one can set
$M=M'$.

The pion coupling constant $g_{nn'\pi}$ is defined through the relation
\begin{equation}
\langle {\bar D}'\pi^a |V|{\bar D} \rangle=2M i g_{nn'\pi}(D^{'\dag} \tau^a D)(2\pi)^3
\delta^{(3)}({\bm P}'+\vq-{\bm P}),
\label{Vnn}
\end{equation}
where $V$ is the interaction responsible for the pion emission. In what follows,
we shall evaluate the matrix element on the l.h.s of equation (\ref{Vnn})
in the framework of the Generalised Nambu--Jona-Lasinio model.

\begin{figure}[t]
\begin{center}
\includegraphics[width=0.6\textwidth]{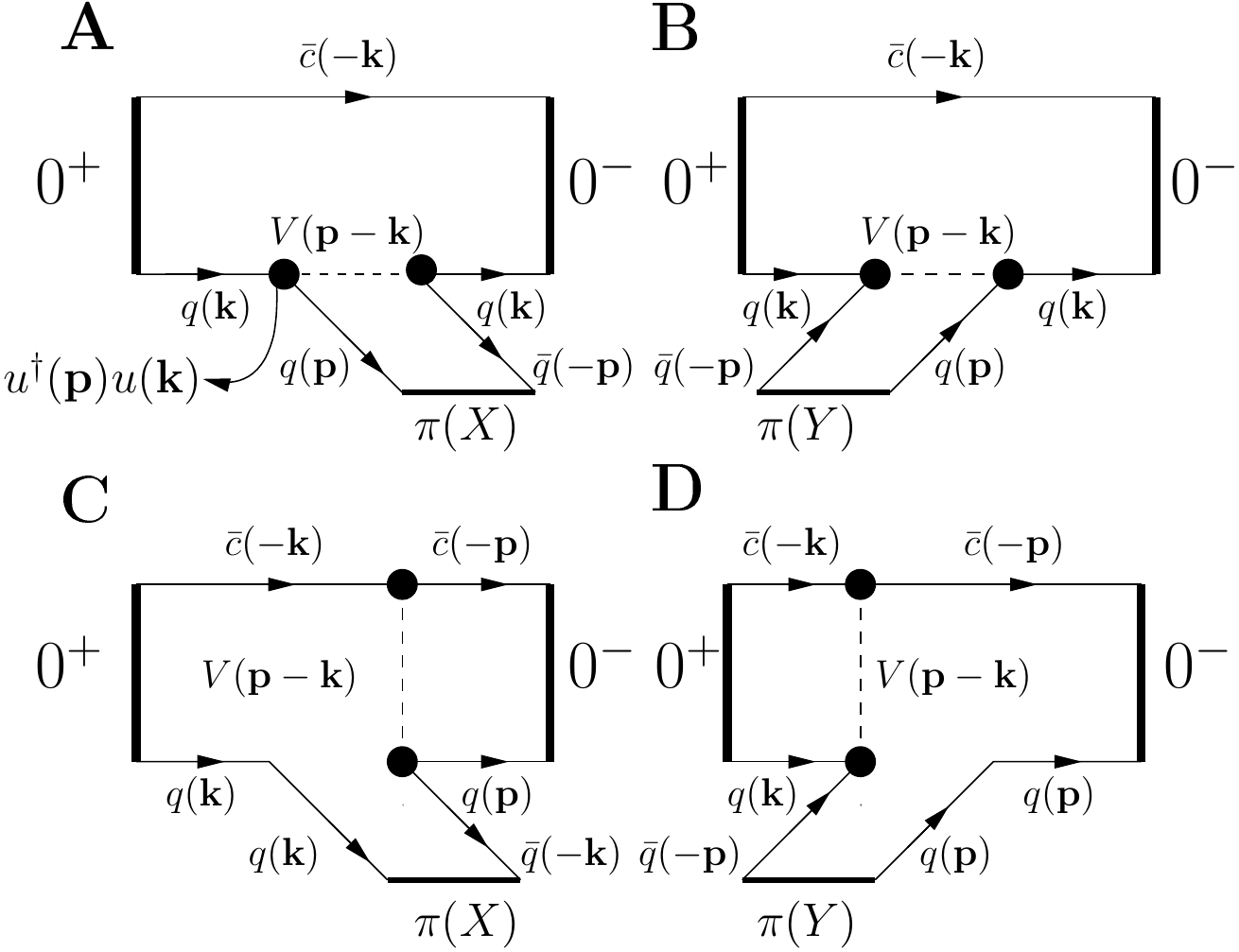}
\end{center}
\caption{The diagrams contributing to the matrix
element $\langle {\bar D}'\pi |V|{\bar D}\rangle$.}\label{ABCD}
\end{figure}

In Fig.~\ref{ABCD} we draw the four diagrams contributing to the matrix element
$\langle {\bar D}'\pi |V|{\bar D}\rangle$ for which we can write then
\begin{equation}
\langle {\bar D}'\pi^a |V|{\bar D} \rangle=2M\left[A^a_X+B^a_Y+C^a_X+D^a_Y\right](2\pi)^3\delta^{(3)}({\bf
P}'+\vq-{\bf P}),
\label{Vnn2}
\end{equation}
where the contributions $A^a_X$ and $B^a_Y$ cancel against each other while the amplitudes $C^a_X$ and $D^a_Y$ take the form
\cite{Nefediev:2006bm}
\begin{eqnarray}
C^a_X&=&\frac{(D'\tau^a D)}{4M\sqrt{N_C}}\int\frac{d^3k}{(2\pi)^3}\frac{d^3p}{(2\pi)^3}
\phi'_pV(\vp-\vk)X_k\phi_k[S_pC_k-(\hat{\vp}\hat{\vk})C_pS_k],\nonumber \\[-2mm]
\label{CD}\\[-2mm]
D^a_Y&=&\frac{(D'\tau^a D)}{4M\sqrt{N_C}}\int\frac{d^3k}{(2\pi)^3}\frac{d^3p}{(2\pi)^3}
\phi'_pV(\vp-\vk)Y_p\phi_k[S_pC_k-(\hat{\vp}\hat{\vk})C_pS_k],\nonumber
\end{eqnarray}
where $C_p$ and $S_p$ are defined in (\ref{SandC}) and their products stem from the 
vertices shown in the diagrams as black dots and given by various products of the dressed quark bispinors (\ref{uandv}).

For instance, in the amplitude $A_X^a$ (see Fig.~\ref{ABCD}) one has the following combination:
\begin{equation}
u^\dag(\vp)u(\vk)=C_pC_k+({\bm \sigma}\cdot\hat{\vp})({\bm \sigma}\cdot\hat{\vk})S_pS_k.
\end{equation}

Now, taking relations (\ref{Vnn}), (\ref{Vnn2}), and (\ref{CD}) together and using the explicit form of the pion wave functions
(\ref{vppm2}), it is easy to find for the coupling constant $g_{nn'\pi}$ in the leading order in $M_\pi$ \cite{Nefediev:2006bm},
\begin{equation}
f_\pi g_{nn'\pi}=\frac{1}{2M}\int\frac{d^3k}{(2\pi)^3}\frac{d^3p}{(2\pi)^3}
\phi_p'V(\vp-\vk)\phi_k(\sin\vpp_p+\sin\vpp_k)[S_pC_k-(\hat{\vp}\hat{\vk})C_pS_k].
\label{gp}
\end{equation}

The nonpion contribution to the off-diagonal axial charge $G_A$ can be evaluated with the help of the explicit expression for the temporal component
of the axial-vector current and it takes the form \cite{Nefediev:2006bm}
\begin{equation}
G_A=\frac{1}{2M}\int\frac{d^3k}{(2\pi)^3}\phi'_k\phi_k\cos\vpp_k.
\label{GAnn}
\end{equation}

To proceed, we multiply equation (\ref{FW4}) for the wave function $\psi(\vp)$ by $\psi'(\vp)\cos\vpp_p$, take trace in the
spinor indices, and integrate both sides of the resulting equation in the momentum $d^3p/(2\pi)^3$. Then we repeat the above procedures for
equation (\ref{FW4pr}) for the wave function $\psi'(\vp)$,
now multiplied by $\psi(\vp)\cos\vpp_p$.
Subtracting one resulting equation from the other, we arrive at the relation
\begin{eqnarray}
\frac12(E-E')G_A=\frac{1}{2M}\int\frac{d^3k}{(2\pi)^3}\frac{d^3p}{(2\pi)^3}
\phi_p'V(\vp-\vk)\phi_k[\cos\vpp_p(C_pC_k+(\hat{\vp}\hat{\vk})S_pS_k)-\nonumber \\[-2mm]
\label{gp2}\\[-2mm]
-\cos\vpp_k((\hat{\vp}\hat{\vk})C_pC_k+S_pS_k)].\nonumber
\end{eqnarray}
After simple trigonometric transformations the r.h.s. of the last equation reduces to that of relation (\ref{gp}).
Therefore, equating the l.h.s.'s of equations (\ref{gp}) and (\ref{gp2}) and proceeding from the energies
$E$ and $E'$ to the corresponding masses (that is, adding the mass of the heavy antiquark), we finally arrive at
Goldberger--Treiman relation (\ref{gpiEEpr}).

Two comments are in order here. On the one hand, it is easy to see that the
role played by the $Y$ ($\vpp_\pi^-$) component of the pion wave
function in the derivation of relation (\ref{gpiEEpr}) is
as important as the role of the component $X$
($\vpp_\pi^+$). This emphasises once more the Goldstone nature of the chiral pion which as
a matter of principle cannot be described in
na{\" i}ve (constituent) quark models.

On the other hand, it follows from equation (\ref{GAnn}), from the properties of
the chiral angle, and from normalisation condition
(\ref{normnn}) that for excited mesons the axial charge approaches unity, $G_A\toon 1$.
Therefore, as it was stated above, for highly excited mesons, the pion
coupling constant decreases,
\begin{equation}
g_{nn'\pi}=\frac{G_A\Delta M_\pm}{2f_\pi}\propto \Delta M_\pm\toon 0,
\end{equation}
and that implies that the Goldstone boson decouples from the spectrum of excited heavy-light quarkonia.

\subsection{Effective chiral symmetry restoration in the spectrum of excited baryons}

In the previous chapters we studied in detail the problem of the effective restoration
of chiral symmetry in the spectrum of highly excited hadrons
and the related question of the chiral pion (Goldstone boson) decoupling from the
spectrum of highly excited mesons. A similar situation takes place
in the spectrum of excited baryons. We start from a general
symmetry-based discussion.

Consider a chiral doublet $B$ built from the effective baryonic fields of the opposite parity $B_+$ and $B_-$ \cite{Glozman:2007ek},
\begin{equation}
B=\genfrac{(}{)}{0pt}{0}{B_+}{B_-}.
\label{doub}
\end{equation}
The states $B_+$ and $B_-$ are mixed by the axial transformation,
\begin{equation}
B\to\exp\left(i\frac{\theta^a_A\tau^a}{2}\sigma_1\right)B,
\label{VAD}
\end{equation}
where $\sigma_1$ is the Pauli matrix in the space of the doublet $B$.
It is easy to establish the form of the Lagrangian invariant
with respect to the above transformation (for the alternative forms of this
Lagrangian see papers \cite{Detar:1988kn,Jido:2001nt}),
\begin{equation}
\mathcal{L}_0 =i\bar{B}\gamma^\mu\partial_\mu B-m_0\bar{B}B
=i\bar{B}_+ \gamma^\mu \partial_\mu B_+ +i\bar{B}_-\gamma^\mu\partial_\mu B_-
-m_0\bar{B}_+B_+-m_0\bar{B}_-B_-.
\label{lagBB}
\end{equation}

An important property of the given Lagrangian is the presence of a nonvanishing
chirally invariant mass $m_0$, the same for the opposite-parity fields.
This implies the Wigner-Weyl realisation of chiral symmetry with massive fermions. Chiral doublets are inevitable in this scenario. Therefore,
in the spectrum of baryons, in addition to the ``standard'' scenario when the fermion mass emerges as a result of the spontaneous breaking of chiral
symmetry (Nambu--Goldstone realisation), an alternative realisation is possible which is consistent with chiral doublets. It is straightforward to
build the Noether current which corresponds to symmetry (\ref{VAD}) of Lagrangian (\ref{lagBB}),
\begin{equation}
j^a_{5\mu}=\bar{B}_+\gamma_\mu\frac{\tau^a}{2}B_-+\bar{B}_-\gamma_\mu\frac{\tau^a}{2}B_+,
\label{ac}
\end{equation}
which does not contain diagonal terms of the form either $\bar{B}_+\gamma_\mu \gamma_5\frac{\tau^a}{2}B_+$ or
$\bar{B}_-\gamma_\mu \gamma_5 \frac{\tau^a}{2}B_-$. Therefore, the diagonal axial charges of the baryons $B_+$ and
$B_-$ which form the chiral doublet vanish while the off-diagonal axial charges related to the transitions between the opposite-parity states
equal to unity \cite{Glozman:2007ek},
\begin{equation}
G_A^{+}=G_A^{-}=0,\quad G_A^{+-}=G_A^{-+}=1.
\label{ach}
\end{equation}
It has to be noticed that the axial charge of the standard Dirac fermion equals to 1.

It is easy to trace the consequences of property (\ref{ach}). Firstly, the diagonal pion couplings to baryons must vanish in unison with
the diagonal axial charges of the baryons, that is, $g_{\pi B_\pm B_\pm}=\frac{G_A^\pm m_\pm}{ f_\pi}=0$.

Below, we derive the formulae for the off-diagonal
constant $g_{\pi B_+B_-}$. To this end, consider the matrix element of the axial-vector current
between two arbitrary opposite-parity baryonic states, $1/2^+$ and $1/2^-$,
\begin{equation}
\langle B_-(p_f)|j^a_{5\mu}|B_+(p_i)\rangle=\bar{u}(p_f)\left[\gamma_\mu H_1(q^2)+
\sigma_{\mu \nu}q^\nu H_2(q^2)+q_\mu H_3(q^2)\right]\frac{\tau^a}{2}u(p_i),
\label{fff}
\end{equation}
where $p_i$ and $p_f$ are the initial-state and the final-state momenta,
respectively, ($q=p_f-p_i$) and we introduced the form factors $H_1$, $H_2$, and $H_3$.
Then, for the matrix element of the divergence of the axial-vector current
we arrive at
\begin{equation}
\langle B_-(p_f)|\partial^\mu
j^a_{5\mu}|B_+(p_i)\rangle=i\left[(m_+-m_-)H_1(q^2)+q^2H_3(q^2)\right]\bar{u}(p_f)\frac{\tau^a}{2}u(p_i).
\label{df}
\end{equation}
Once the l.h.s. of equation (\ref{df}) vanishes in the chiral limit due to the
PCAC condition, then, in the limit $q\to 0$, the form factors must obey
the following condition:
\begin{equation}
(m_+-m_-)H_1(0)+\lim_{q\to 0}q^2H_3(q^2)=0,
\end{equation}
which can be easily recognised as the Goldberger--Treiman relation,
\begin{equation}
g_{\pi B_+B_-}=\frac{G_A^{+-}(m_+ - m_-)}{2 f_\pi},\quad G_A^{+-}=H_1(0).
\label{ggt}
\end{equation}
Indeed, PCAC requires that the contribution of the term proportional to $H_1$ is compensated by the term proportional to
$H_3$, and the latter has to develop a pole at $q^2\to 0$ which is identified naturally with the Goldstone pole.
Thus, if the states $B^\pm$ belong to the same chiral doublet then they become degenerate in the mass that ensures that
$g_{\pi B_+B_-}=0$, and this
condition, being a consequence of PCAC, is independent of the particular degeneracy mechanism for the states $B^\pm$.

\begin{figure}[t]
\centerline{\epsfig{file=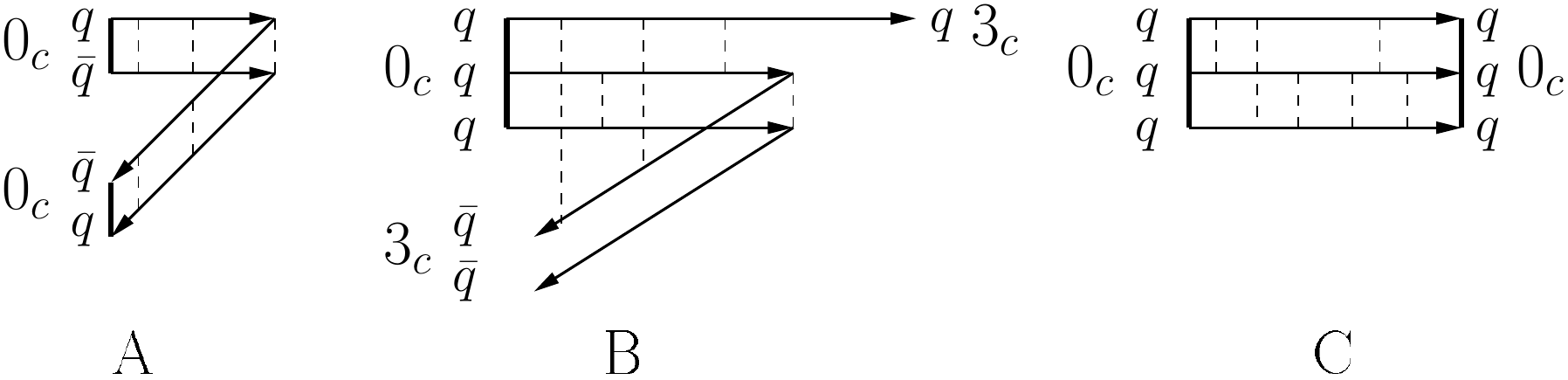,width=0.8\textwidth}}
\caption{Diagram A: a typical colour-allowed (singlet --- singlet, $0_c\rightarrow 0_c$)
transition between the positive-energy and negative-energy
amplitudes for the $q\bar{q}$ pair. Diagram B: a similar transition is forbidden in the
baryon since it results in the state with an open colour ($3_c$).
Diagram C: a typical colour-allowed diagram contributing to the Dyson equation for the
baryon in the ladder approximation.}
\label{baryonversusmeson}
\end{figure}

Similarly to mesons, the general symmetry-based arguments given above possess a particular microscopic realisation in the framework of the
Generalised Nambu--Jona-Lasinio model. However, it is important to make a comment on the baryonic states in this model. In spite of the fact that
the model is considered in the formal limit $N_C\to\infty$, properties of the baryons can be studied qualitatively (in many cases also quantitatively)
if $N_C=3$ is substituted. Then the baryon can be built by acting the three dressed-quark operators on the BCS vacuum and
contracting the result with the relevant wave function,
\begin{equation}
\Psi_B=\Psi_{\rm colour}\otimes\Psi_{\rm flavour}\otimes\Psi_{\rm spin}\otimes\Psi_{\rm space},
\quad \Psi_{\rm colour}=\frac{1}{3!}\varepsilon_{\alpha\beta\gamma}q^\alpha q^\beta q^\gamma,
\label{PsiB}
\end{equation}
where $\varepsilon_{\alpha\beta\gamma}$ is the antisymmetric Levi-Civita tensor.
Baryons do not bring any new effects to the model and, even more, the Dyson equations for
baryons turn out to be much simpler than the similar equations
for mesons. The simplification comes from the fact that for baryons the positive-energy component of the amplitude does not couple to the negative-energy
component, for such transitions would imply the existence of the states with open colour (see Fig.~\ref{baryonversusmeson}).

Since, by construction, the colour wave function of the baryon is antisymmetric
with respect to the permutation of the quarks, then it
is sufficient to study only symmetric combinations $\Psi_{\rm flavour}\otimes\Psi_{\rm spin}\otimes\Psi_{\rm space}$, so that in the generic case
the spatial wave function $\Psi_{\rm space}^{{\cal Y}}({\bm\rho}, {\bm\lambda)}$
(here $\bm\rho$ and $\bm\lambda$ are the standard Jacobi coordinates)
should contain all possible permutations ${\cal Y}$: antisymmetric (${\cal A}$),
symmetric (${\cal S}$), and mixed $F$ (${\cal D}_F$) or $D$
(${\cal D}_D$). So far, all considerations were quite general while below
in this chapter a particular form of some baryonic wave functions will be
quoted and used to calculate the axial charges of these baryons.

Consider the operator of the axial charge,
\begin{equation}
Q_5^a=\int d^{3}x\;\bar{\psi}_i\gamma_{0}\gamma_{5}\left(\frac{\tau^a}{2}\right)^{ij}\psi_j,
\label{Q5}
\end{equation}
and evaluate it for the dressed quarks using equations (\ref{psi})
and (\ref{uandv}). The result reads \cite{Nefediev:2008dv}
\begin{eqnarray}
&&Q_5^a=\sum_{i,j}\sum_{\alpha=1}^{N_C}\sum_{s,s'=\uparrow,\downarrow}\left(\frac{\tau^a}{2}\right)^{ij}\int\frac{d^3p}
{(2\pi)^3}
\Bigl[\cos\vpp_p({\bm\sigma} \hat{\bm p})_{ss'}\Bigl(b_{i\alpha s}^\dagger(\vp) b_{js'}^\alpha(-\vp)
+d_{js}^{\alpha\dagger}(-\vp) d_{i\alpha s'}(\vp)\Bigr)\nonumber\\[-2mm]
\label{ChiralChar}\\[-2mm]
&&\hspace*{50mm}+\sin\vpp_p(i\sigma_2)_{ss'}\Bigl(b_{i\alpha s}^\dagger(\vp) d^{\alpha\dagger}_{js'}(-\vp)+d_{i\alpha
s}(\vp)b_{js'}^\alpha(-\vp)\Bigr)\Bigr],\nonumber
\end{eqnarray}
where the two contributions in the square brackets have different physical interpretation.
The first term is diagonal in the quark creation and annihilation
operators (nonanomalous term) and because it contains
the operator $({\bm\sigma} \hat{{\bm p}})$ it is responsible for the transition from the baryonic state
with a given parity to the baryonic state with the opposite
parity, that is, to the chiral partner. The second term in equation (\ref{ChiralChar})
has the content of a Bogoliubov
anomalous term. Since the axial current is a component of the conserved Noether current
and, therefore, it commutes
with the Hamiltonian, $[Q_5^a,H]=0$, then the state $Q_5^a|0\rangle=|\pi^a\rangle$
is degenerate with the vacuum and it is the Goldstone
boson, that is, it is nothing but the chiral pion. Indeed, the quantity
$\sin\vpp_p(i\sigma_2)$ is the wave function of the pion in its rest frame --- see equations (\ref{vppm}) and (\ref{po}).

Consider the diagonal part of the axial charge operator of a baryon
defined as the sum of operators
(\ref{ChiralChar}) over all quarks in the baryon,
\begin{equation}
{\cal Q}_5\equiv {\cal Q}_5^3=\sum_{n=1}^3Q_{5n}^3.
\label{Q5bar}
\end{equation}
From the above consideration, it is easy to see that the result of such an operator acting on the baryonic state can be
schematically presented in the form
\begin{equation}
{\cal Q}_5|B\rangle=|B'\rangle+|B\pi\rangle,
\label{hh}
\end{equation}
where the first term on the r.h.s. describes the chiral partner of the state $|B\ran$
and the second term contains the neutral pion.
The relative weight of the above two terms is defined by
the chiral angle $\vpp_p$. In particular, in case of the maximal symmetry breaking
$\vpp_p=\pi/2$ and, therefore, only the second term
survives on the r.h.s. of equation (\ref{hh}). In the opposite limit of unbroken chiral symmetry
$\vpp_p=0$, so that only the first term survives. In this case it is
easy to arrive at the following (obvious) properties
of the operator
${\cal Q}_5$:
\begin{equation}
{\cal Q}_5^\dagger={\cal Q}_5,\quad \langle B_2|{\cal Q}_5^2|B_1\rangle\propto\langle B_2|B_1\rangle=\delta_{B_1B_2}.
\label{paritypairing}
\end{equation}
Therefore,
\begin{equation}
{\cal Q}_5|B^{\pm}\rangle=G_A^{\pm\raisebox{-0.5mm}{\scriptsize $\mp$}}|B^{\mp}\rangle,
\end{equation}
where $B^{\pm}$ stand for the opposite-parity baryons and $G_A^{\pm\raisebox{-0.5mm}{\scriptsize $\mp$}}$ stand for the axial charges.
Then, for the baryon spectrum in the limit of the exact chiral symmetry
restoration the following relations
between various axial charges hold true:
\begin{equation}
G_A^{+-}=G_A^{-+}= 1,\quad G_A^{++}\equiv G_A^+=0,\quad G_A^{--}\equiv G_A^-=0,
\label{Gbeh}
\end{equation}
which are approached asymptotically with the growth of the excitation number of the baryon.
Once, as was mentioned above, the axial charge operator
commutes with the Hamiltonian, then the states $|B^+\rangle$
and $|B^-\rangle$ should become degenerate in this limit.

\subsection{Axial charges of baryons in the nonrelativistic quark model}

In the previous chapter, the effective restoration of chiral symmetry in the spectrum of excited baryons was described in detail in the
framework of the microscopical framework provided by the Generalised
Nambu--Jona-Lasinio model. In particular, predictions were made for the
axial charges of baryons. For comparison, let us evaluate the axial
charges of some baryons in a different approach. In particular, a popular
alternative approach to baryons is provided by the well-known nonrelativistic $SU(6)$ quark model which includes the $SU(3)$ isospin and
$SU(2)$ spin group and which is quite successful in describing the ground states of baryons \cite{Gursey:1992dc,Kokkedee:1966wx}.
All ground-state baryons with the quantum numbers $1/2^+$ and $3/2^+$,
which form the octet and the decuplet, respectively,
of the isospin $SU(3)$ group enter the {\bf 56}-plet of the group $SU(6)$. Then, for example, magnetic moments of baryons (in fact, their ratios)
are reproduced in the $SU(6)$ model with accuracy about 10-15\%. One of the most well-known predictions of this model, the nucleon axial charge
$G_A=5/3$, coincides with the experimental value $G_A=1.26$ quite well, the discrepancy is caused by the neglected relativistic effects, by
the pionic cloud, by the effects of the $SU(6)$ symmetry breaking because of the different quark masses, and so on.

In the large-$N_C$ limit, the ground states in the spectrum of baryons indeed obey the $SU(6)$ algebra \cite{Gervais:1983wq,Dashen:1993as}
that is a consequence of the spontaneous breaking of chiral symmetry which results in the emergence of large constituent quark masses.
In order to predict the masses of the excited baryons, the symmetry group $SU(6)$ needs to be supplied with the dynamical assumptions about
the structure of the radial and angular excitations of the quarks in the baryons. In the simplest case of the Harmonic Oscillator confining potential
the energy of the quarks is fully fixed by the principal quantum number, and the masses of the excited negative-parity states of the nucleon and the
$\Delta$ agree well with the predictions of such a $SU(6)\times O(3)$ classification scheme with $N=1$. However, for the positive-parity states, the given
scheme meets certain difficulties such as an overestimated splitting with the negative-parity states and, for the Roper resonance, with the wrong ordering
of the opposite-parity levels. In the literature, there exist successful attempts to resolve the aforementioned difficulties through
a particular
symmetry breaking mechanism \cite{Glozman:1995fu}. However,
a systematic mass degeneracy of opposite-parity excited baryons
looks quite unnatural in the
framework of this quark model and there is no explanation for this phenomenon.
In particular, in the $SU(6)\times O(3)$ scheme there are no reasons
whatsoever for the axial charges of the baryons to obey formula (\ref{Gbeh}).
Meanwhile, once the axial charges of baryons can be evaluated on the
lattice (see, for example,
papers \cite{Kokkedee:1966wx,Takahashi:2009zz,Takahashi:2009zzb,Erkol:2009ev,Maurer:2012sf})
then it would be natural to confront the lattice results with the predictions
of the quark model $SU(6)\times O(3)$ in which the effective
restoration of chiral symmetry is not possible. Then, deviations of the lattice data from
the predictions of the quark model can be interpreted as
an argument in favour of the chiral symmetry restoration in the spectrum of excited baryons.

In order to evaluate the axial charges of baryons in the nonrelativistic quark model
one needs to work out the averages over the baryons' wave functions
of the form
\begin{equation}
G_A^{fi}=\langle\Psi_f(1,2,3)|\sum_{n=1}^3Q^A_n|\Psi_i(1,2,3)\rangle,
\label{me}
\end{equation}
where $Q^A_n$ is the operator of the axial charge of the $n$-th quark which in the
leading order is given by the Gamov--Teller formula
$\sigma_3\tau_3$ (here $\vesig$ and ${\bm \tau}$ are the spin and the isospin operators
of the Dirac fermion, respectively).
This operator admits relativistic corrections (which, however, will not be considered below) of the form
\begin{equation}
\frac{1}{2M}\vesig({\bm p}_i + {\bm p}_f)\tau^a e^{i \bm q \bm r},
\label{corr}
\end{equation}
where $\vq={\bm p}_f-{\bm p}_i$.
In addition, the operator of the axial charge of a nonrelativistic quark in the leading
order also contains a dependence on the
spatial coordinate in form of the exponent $\exp{[i\vq\ver]}$.
However, evaluation of both the diagonal and off-diagonal axial charges amounts to taking the limit
$\vq\to 0$ \cite{Glozman:2008vg}. Then, in the leading order, the operator of the
axial charge does not depend on the coordinate and, therefore, matrix element (\ref{me}) is nonzero only if the spatial wave functions of the
initial and the final baryons coincide. This prediction of the $SU(6) \times O(3)$ scheme appears to be at odds with the predictions of the
chiral symmetry restoration in the spectrum and, in particular, with its microscopic realisation in the framework of the
Generalised Nambu--Jona-Lasinio model --- see equation (\ref{Gbeh}).

Evaluation of the diagonal matrix elements (\ref{me}) requires one to know the wave
functions of the baryons in the $SU(6)
\times O(3)$ scheme which are well-known in the literature and are quoted
in Table~\ref{tabt2}. Each wave function is characterised by
several quantum numbers. First, there is the multiplet
of the spin-flavour group $SU(6)$ to which the given state belongs
and which is
encoded in the Young symbol $[f]_{FS}$. In each such multiplet the baryon's wave
function possesses a particular symmetry
in the flavour space (symbol $[f]_F$ with $f=3,21,111$) and a particular spin symmetry
(symbol $[f]_S$ with $f=3,21$ for $S=3/2$ and
$S=1/2$, respectively). Finally, the spatial part of the wave function is fixed
by the angular momentum $L$ and by the permutation symmetry
$[f]_X$ which is fixed by the Pauli principle as $[f]_X= [f]_{FS}$. For a particular basis used, additional quantum numbers may arise like
the principal quantum number $N$ in the Harmonic Oscillator basis or the spatial symmetry of the angular wave function $(\lambda \mu)$.

\begin{table}[t]
\begin{center}
\begin{tabular}{|l|l|}
\hline
$N(\lambda\mu)L[f]_X[f]_{FS}[f]_F[f]_S$& $J^P$, нуклон \\
\hline
$0(00)0[3]_X[3]_{FS}[21]_F[21]_S$ & $\frac12^+, N$ \\
\hline
$2(20)0[3]_X[3]_{FS}[21]_F[21]_S$ & $\frac12^+, N(1440)$\\
\hline
$1(10)1[21]_X[21]_{FS}[21]_F[21]_S$ & $\frac12^-, N(1535)\quad\frac32^-,N(1520)$\\
\hline
$1(10)1[21]_X[21]_{FS}[21]_F[3]_S$ & $\frac12^-, N(1650)\quad\frac32^-,N(1700)\quad\frac52^-,N(1675)$ \\
\hline
$2(20)2[3]_X[3]_{FS}[21]_F[21]_S$&$\frac32^+,N(1720)\quad \frac52^+,N(1680)$\\
\hline
$2(20)0[21]_X[21]_{FS}[21]_F[21]_S$ & $\frac12^+, N(1710)$ \\
\hline
\end{tabular}
\end{center}
\caption{Wave functions of some nucleons in the mass region below 2~GeV in the $SU(6)$ scheme (see, for example,
reference \cite{Glozman:1995fu}).}\label{tabt2}
\end{table}

The diagonal axial charges of some baryons evaluated with the help of the wave functions quoted in Table~\ref{tabt2} are given in
Table~\ref{GAs} (the details of the calculations can be found in paper \cite{Glozman:2008vg}).
The charges found allow one to compare the predictions of the $SU(6)$ model with the predictions of the chiral restoration model.
In particular, the states $N(1440)$ and $N(1535)$ form a chiral doublet and, therefore, according to the model of chiral restoration,
their axial charges should be small. From Table~\ref{GAs} one can see that the $SU(6)$ model also predicts a small axial charge for the
$N(1535)$, however, for the state $N(1440)$ the prediction of this model is rather large --- larger than unity. A similar situation takes place for
another pair of chiral partners --- for $N(1710)$ and $N(1650)$: the $SU(6)$ model predicts rather large values of both axial charges.
Other examples of model calculations of the axial charges of excited baryons can be found in papers
\cite{Choi:2010ty,An:2008tz,An:2009zza,Gallas:2009qp,Yuan:2009st}.

\begin{table}[t]
\begin{center}
\begin{tabular}{|c|c|c|c|c|}
\hline
Baryon&$N(1440)$&$N(1710)$&$N(1535)$&$N(1650)$\\
\hline
$J^P$&$1/2^+$&$1/2^+$&$1/2^-$&$1/2^-$\\
\hline
$G_A$&$5/3$&$1/3$&$-1/9$&$5/9$\\
\hline
\end{tabular}
\end{center}
\caption{Diagonal axial charges of baryons evaluated in the framework of the $SU(6)$ quark model.}\label{GAs}
\end{table}

\section{Conclusions}

In this review we discussed some aspects of the phenomenon of chiral symmetry breaking and properties of hadrons in the framework of
the Generalised Nambu--Jona-Lasinio model. An important feature of this model is
its microscopic approach to the chiral symmetry breaking in
the vacuum and the presence of confinement allowing one
to employ the model to address a wide class of problems related not only with the
low-lying states in the spectrum of hadrons but also with various properties
of excited hadrons. In particular, the phenomenon of the effective
restoration of chiral symmetry in the spectrum of excited mesons and baryons
is described microscopically.

The main problems discussed in the review are as follows.
\begin{itemize}
\item An explicit microscopic description of the phenomenon of chiral symmetry breaking in the vacuum is given in terms of the dressed quark
fields and the wave function of the chirally broken vacuum which has the form of a coherent-like state formed by the condensed
$^3P_0$ quark-antiquark pairs.
\item The bosonic Bogoliubov transformation is generalised to the case of compound mesonic operators and the equivalence of the given method to the
approach based on the Bethe-Salpeter equation for mesonic amplitudes is proved.
\item The problem of the interrelation between the Lorentz nature of confinement and the spontaneous breaking of chiral symmetry is addressed.
The connection of the spontaneous chiral symmetry breaking with the dynamically generated scalar interquark potential in quarkonium
is traced at the microscopic level.
\item The existence of two different dynamical regimes in the mass-gap equation
is established from the most general arguments and only one of them
is shown to be realised
in the chiral limit. For the latter regime the chiral angle is shown to collapse in the classical limit, that is,
the quantum nature of the spontaneous breaking of chiral symmetry in the vacuum is demonstrated directly.
\item A detailed microscopic description of the phenomenon of the effective chiral symmetry restoration in the spectrum of excited hadrons is
presented in the framework of the Generalised Nambu--Jona-Lasinio model. In particular, both quali\-ta\-tive and quantita\-tive analysis of the mass-gap
equation for the given model with an arbitrary power-like confining potential is presented and the existence of chirally nonsymmetric
solutions is demonstrated for all such power-like potentials.
\item The connection is traced between the interquark potential in quarkonium responsible for the splitting between the opposite-parity states and
the chiral angle, that is, the quantity which describes the effect of the spontaneous breaking of chiral symmetry. This potential is
demonstrated to decrease fast with the increase of the excitation number of the meson.
\item A microscopic derivation of the Goldberger--Treiman relation for the pion
coupling constant with a heavy-light quarkonium is presented and
the pion is explicitly shown to decouple from the excited hadrons which form approximate chiral multiplets.
\item A microscopic derivation of the behaviour of the diagonal and off-diagonal axial charges of baryons which form approximate chiral
multiplets is presented and the results are confronted with the predictions of the $SU(6)\times O(3)$ quark model.
\end{itemize}

In conclusion, let us mention a few questions and problems of the phenomenology of strong interactions which can be addressed using
Generalised Nambu--Jona-Lasinio model. First of all, it has to be noticed that highly excited mesons made of light quarks demonstrate a higher
level of degeneracy of the spectrum than just the restored chiral symmetry. In particular, the slopes of the Regge trajectories in the total spin
$J$ and in the radial quantum number $n$ coincide with a high accuracy (see, for example, a recent paper \cite{Pang:2015eha}) that complies well
with the idea of the existence of the principle quantum number $n+J$ \cite{Glozman:2007at}. Besides, in a series of recent papers it was conjectured that
highly excited hadrons form multiplets of the $SU(4)$ group which includes chiral symmetry as a subgroup \cite{Glozman:2014mka,Glozman:2015qva}.
This hypothesis finds support on the lattice if a cute trick is employed \cite{Glozman:2012fj,Denissenya:2015mqa}, namely, it was suggested to
investigate the properties of hadrons using the field configurations after the artificial removal of the near-zero modes of the Dirac operator.
Once the chiral condensate in QCD vacuum is defined by the density of such near-zero modes \cite{Casher:1979vw}, then their removal
should result in the chiral symmetry restoration and, therefore, all results obtained with the help of such special lattice configurations should
demonstrate all implications of the restored chiral symmetry. Indeed, the result demonstrates the emergence of a rather high degeneracy in the
spectrum which is consistent with the $SU(4)$ group \cite{Glozman:2012fj,Denissenya:2015mqa}. Building a dynamical model of QCD string
which possesses the above property is an important problem of the theoretical high-energy physics which can be addressed using, in particular,
the experience gained from the microscopic calculations in the framework of the Generalised Nambu--Jona-Lasinio model.
\bigskip

Work of Yu. K. and A. N. was performed within the Institute of Nuclear Physics and Engineering supported by 
MEPhI Academic Excellence Project (contract No 02.a03.21.0005, 27.08.2013). They also acknowledge support from the
Russian Foundation for Basic Research (Grant No. 17-02-00485).

\end{document}